%% file: main.tex
\documentclass[twoside,openright]{report}

\usepackage[b5paper,layout=b5paper,bmargin=1in,lmargin=2.5cm,rmargin=1.5cm]{geometry}

\usepackage[english]{babel}

\usepackage{setspace}
\usepackage{scrextend}
\KOMAoption{fontsize}{11pt}

\usepackage{fancyhdr,xcolor}
\makeatletter
\definecolor{azuloscuro}{rgb}{0.04,0.05,0.5}
\def\headrule{{\color{azuloscuro}\if@fancyplain\let\headrulewidth\plainheadrulewidth\fi\hrule\@height\headrulewidth\@width\headwidth\vskip-\headrulewidth}}
\makeatother
\usepackage{amsmath}
\usepackage{graphicx}
\usepackage{float}
\usepackage{soul}
\usepackage{amsfonts}
\usepackage{pdfpages}
\usepackage{hyperref}
\usepackage[fleqn]{nccmath}
\usepackage{array,multirow}

\usepackage[section]{placeins}

\hypersetup{
    colorlinks,
    citecolor=[rgb]{0.04,0.05,0.5},
    filecolor=black,
    linkcolor=[rgb]{0.04,0.05,0.5},
    urlcolor=[rgb]{0.04,0.05,0.5}
}

\usepackage[none]{hyphenat}
\usepackage{caption,setspace}
\begin{document}

\includepdf[pages={1-}]{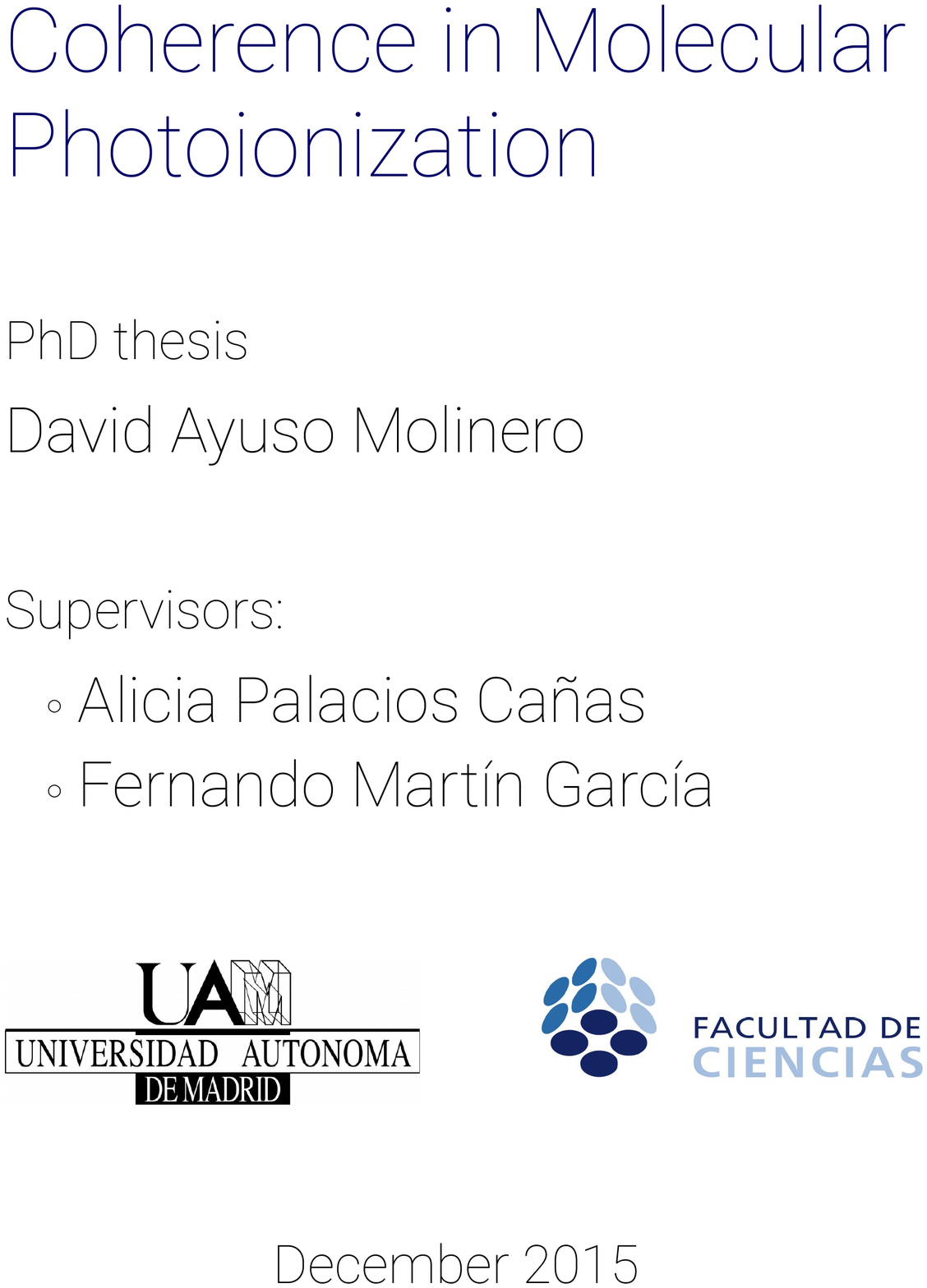}
\includepdf[pages={1-}]{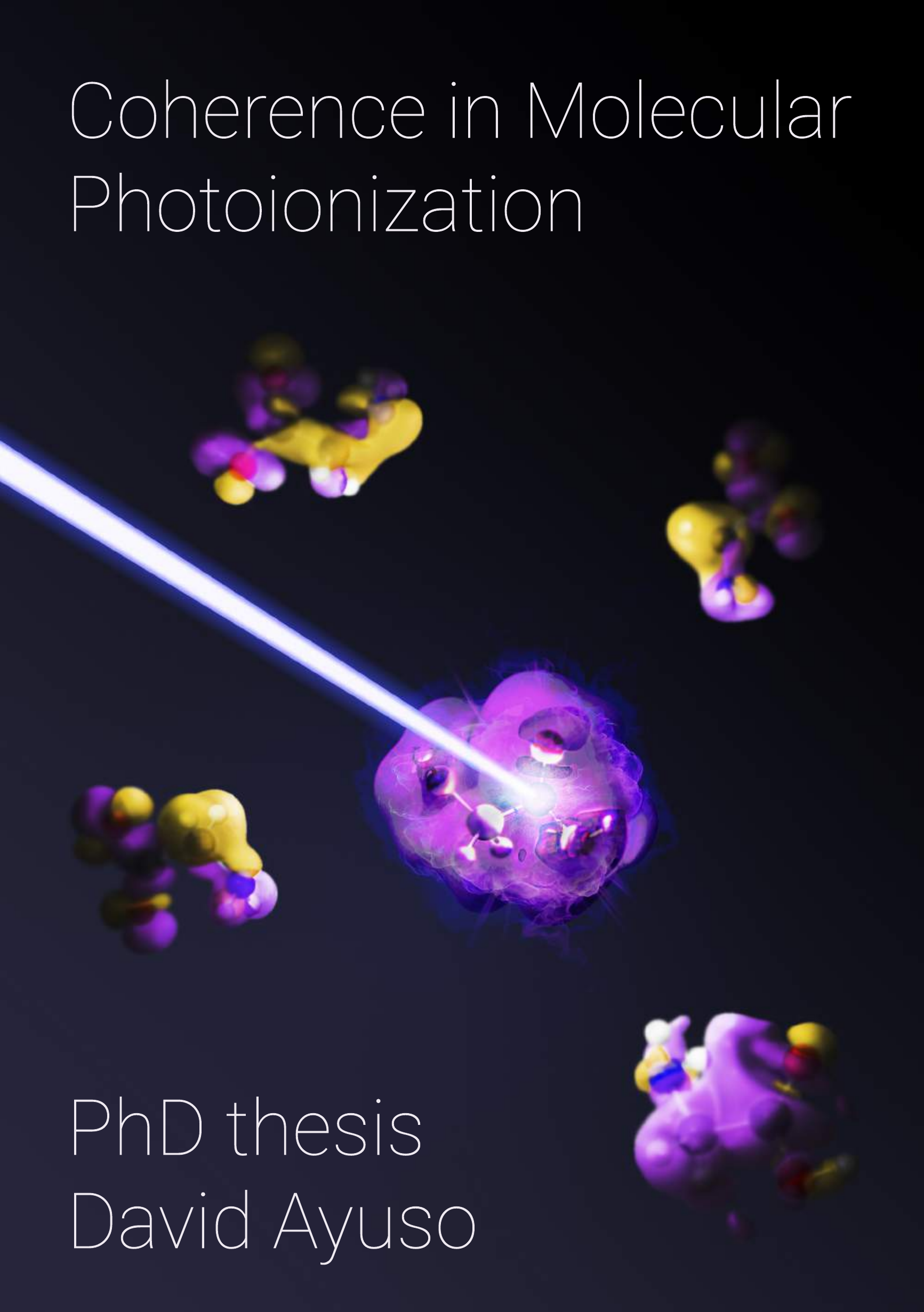}

\setcounter{page}{1}

\input{./Chapters/Abstract}
\input{./Chapters/Resumen}
\input{./Chapters/Publications}

\input{./Chapters/Acknowledgements}

\tableofcontents

\onehalfspacing

\input{./Chapters/Introduction}
\part*{Theory}\addcontentsline{toc}{part}{Theory}
\input{./Chapters/Chapter1}
\input{./Chapters/Chapter2}

\input{./Chapters/Chapter3}
\part*{Results}\addcontentsline{toc}{part}{Results}
\input{./Chapters/Chapter4}
\input{./Chapters/Chapter5}
\input{./Chapters/Conclusions}
\input{./Chapters/Conclusiones}

\part*{Appendices}\addcontentsline{toc}{part}{Appendices}
\appendix
\chapter{\textnormal{Intramolecular photoelectron diffraction in the gas phase}}
\label{appendix1}
\includepdf[pages={1-}]{blank.pdf}
\includepdf[pages={1-}]{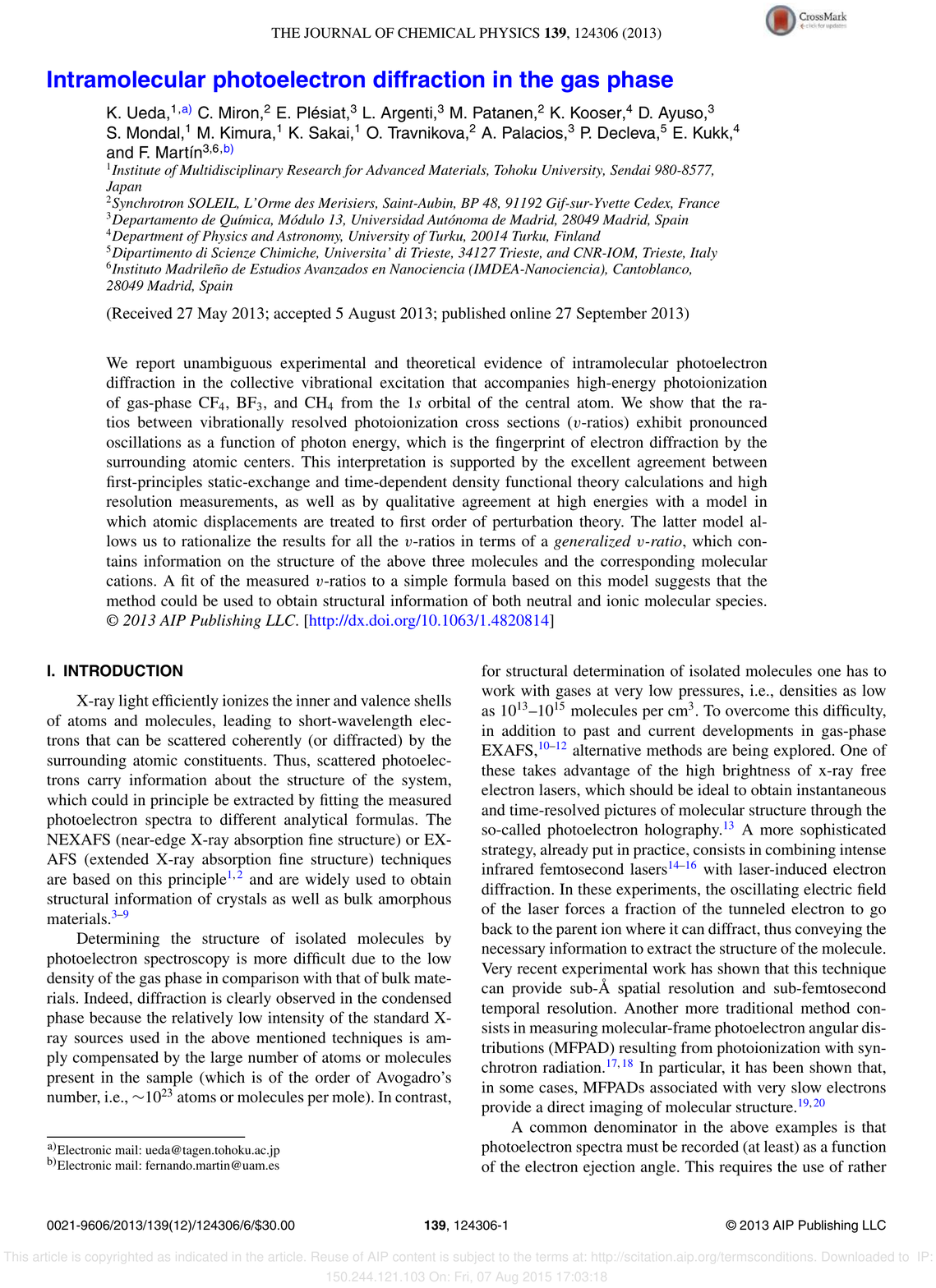}

\chapter{\textnormal{Vibrationally resolved B1s photoionization cross section of BF$_3$}}
\label{appendix2}
\includepdf[pages={1-}]{blank.pdf}
\includepdf[pages={1-}]{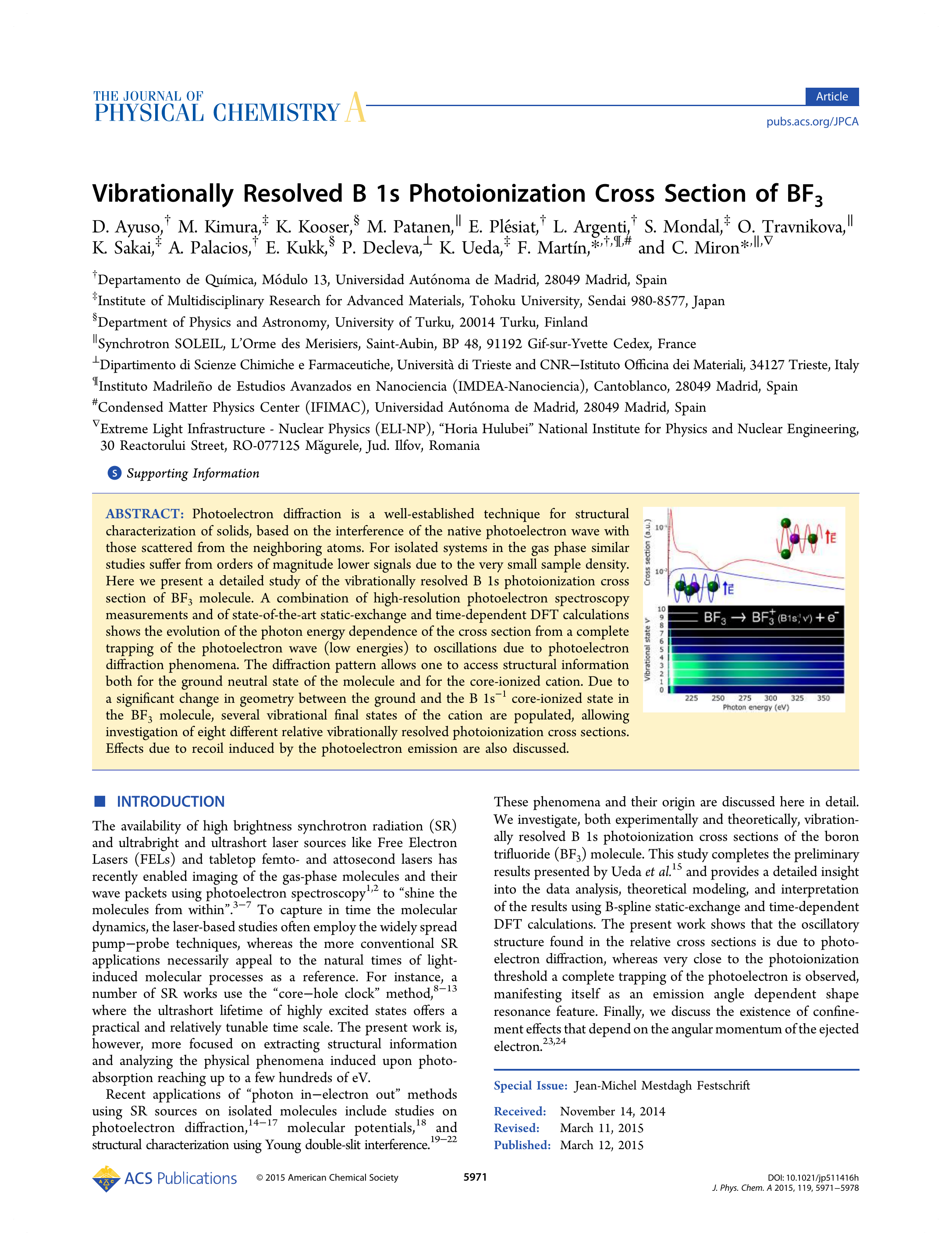}
\includepdf[pages={1-}]{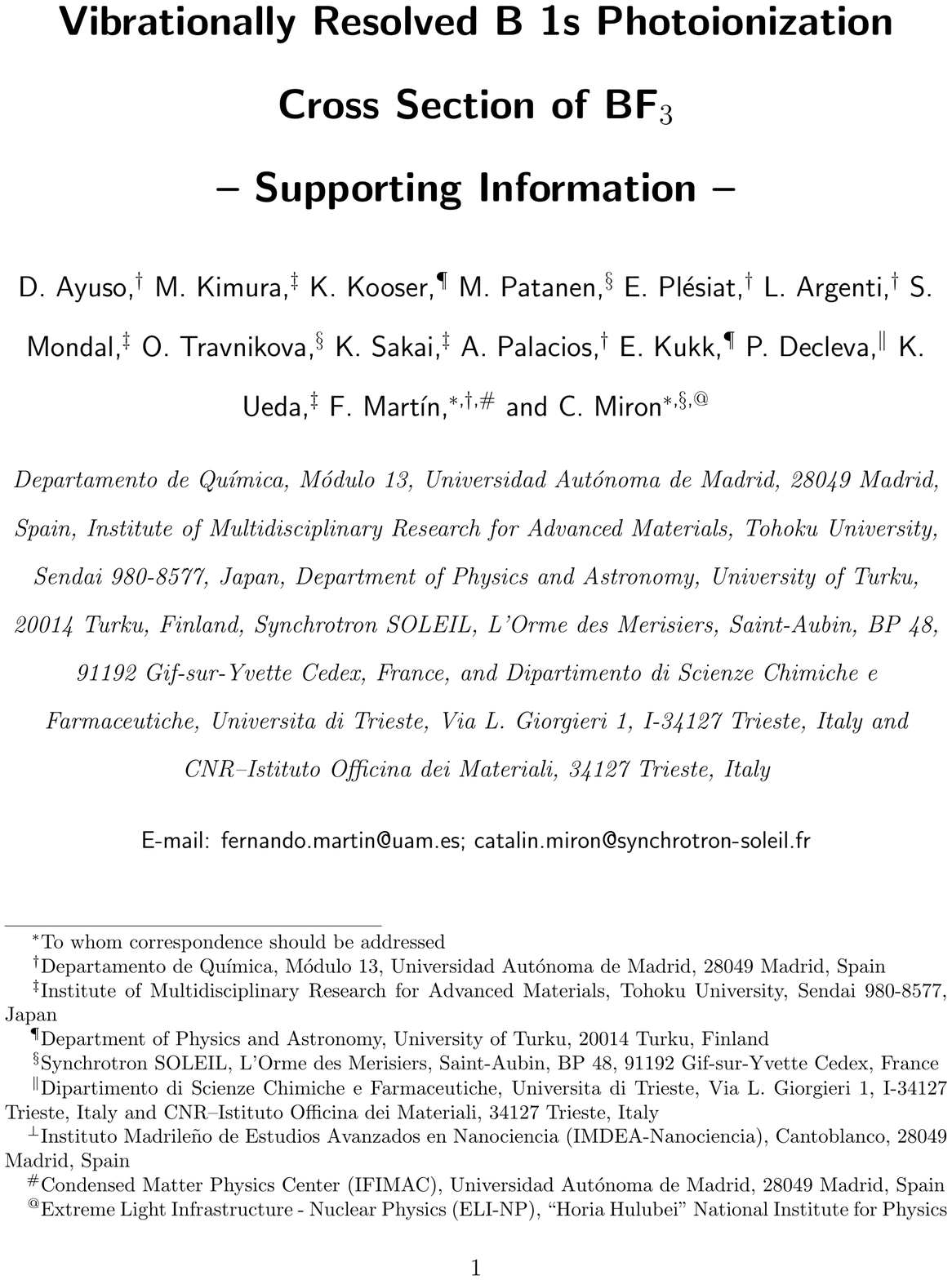}

\chapter{\textnormal{Vibrationally resolved C1s photoionization cross section of CF$_4$}}
\label{appendix3}
\includepdf[pages={1-}]{blank.pdf}
\includepdf[pages={1-}]{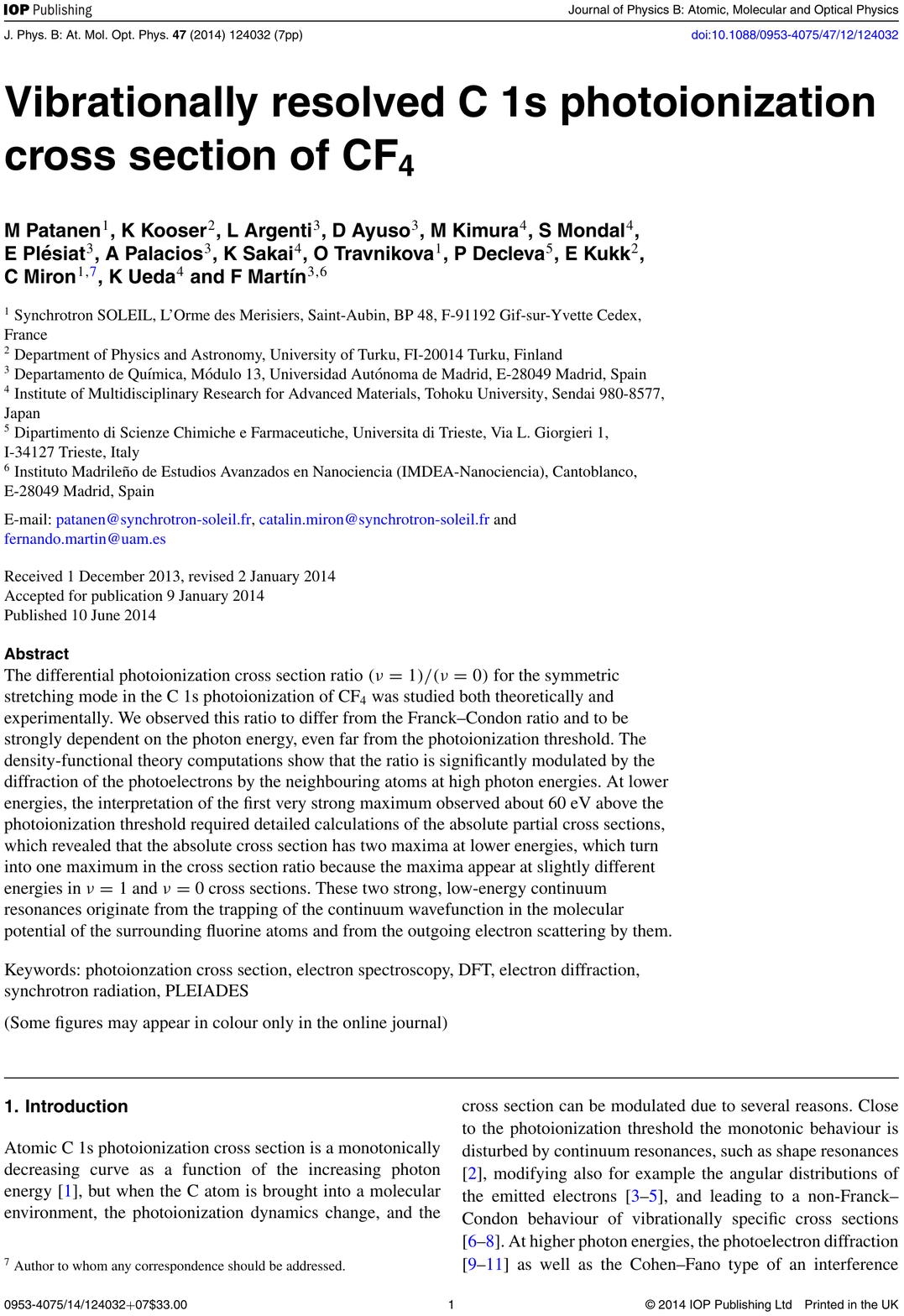}

\chapter{\textnormal{Effects of molecular potential and geometry on atomic core-level photoemission over an extended energy range: The case study of the CO molecule}}
\label{appendix4}
\includepdf[pages={1-}]{blank.pdf}
\includepdf[pages={1-}]{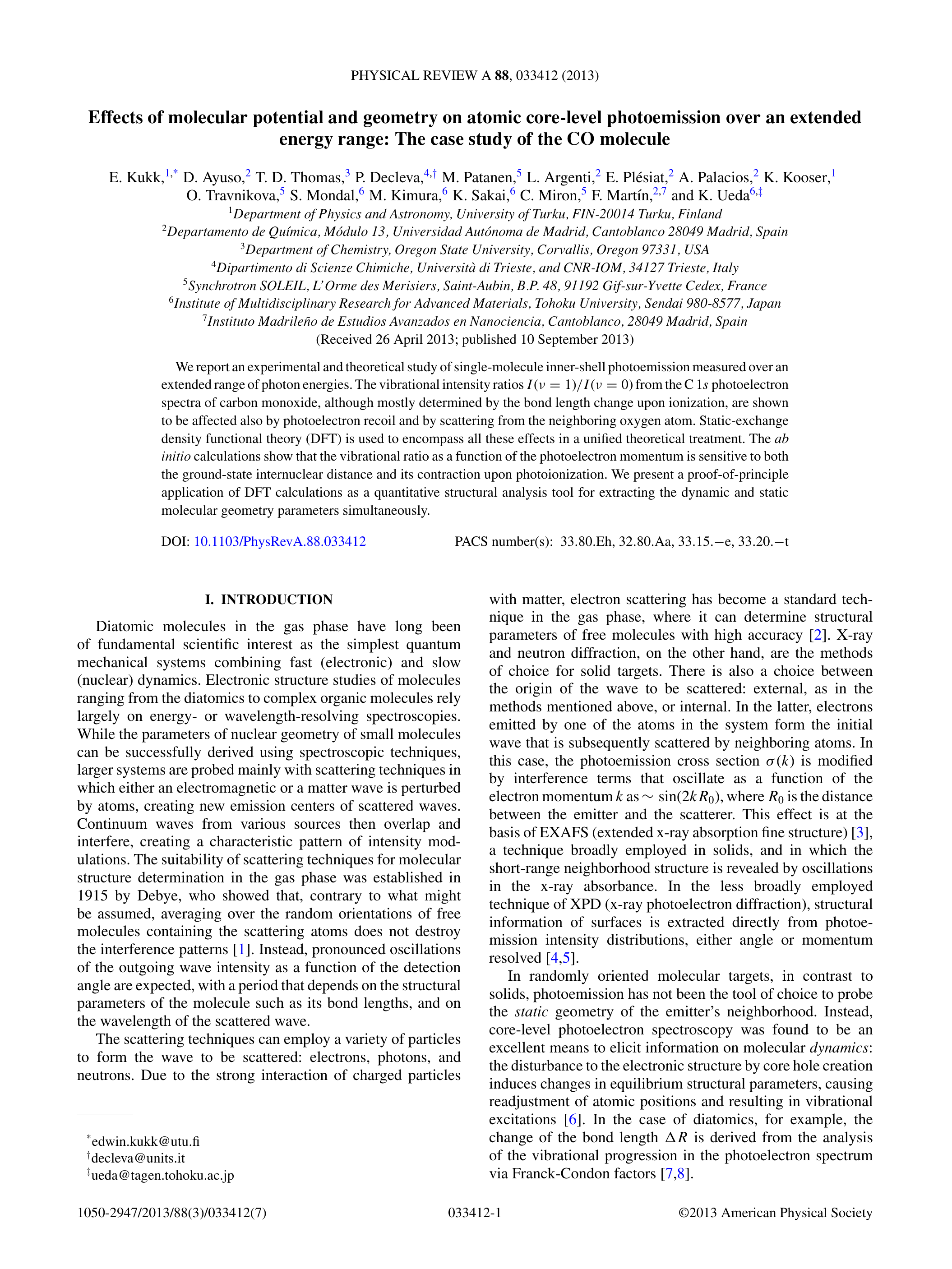}

\chapter{\textnormal{Dissociative and non-dissociative photoionization of molecular fluorine from inner and valence shells}}
\label{appendix5}
\includepdf[pages={1-}]{blank.pdf}
\includepdf[pages={1-}]{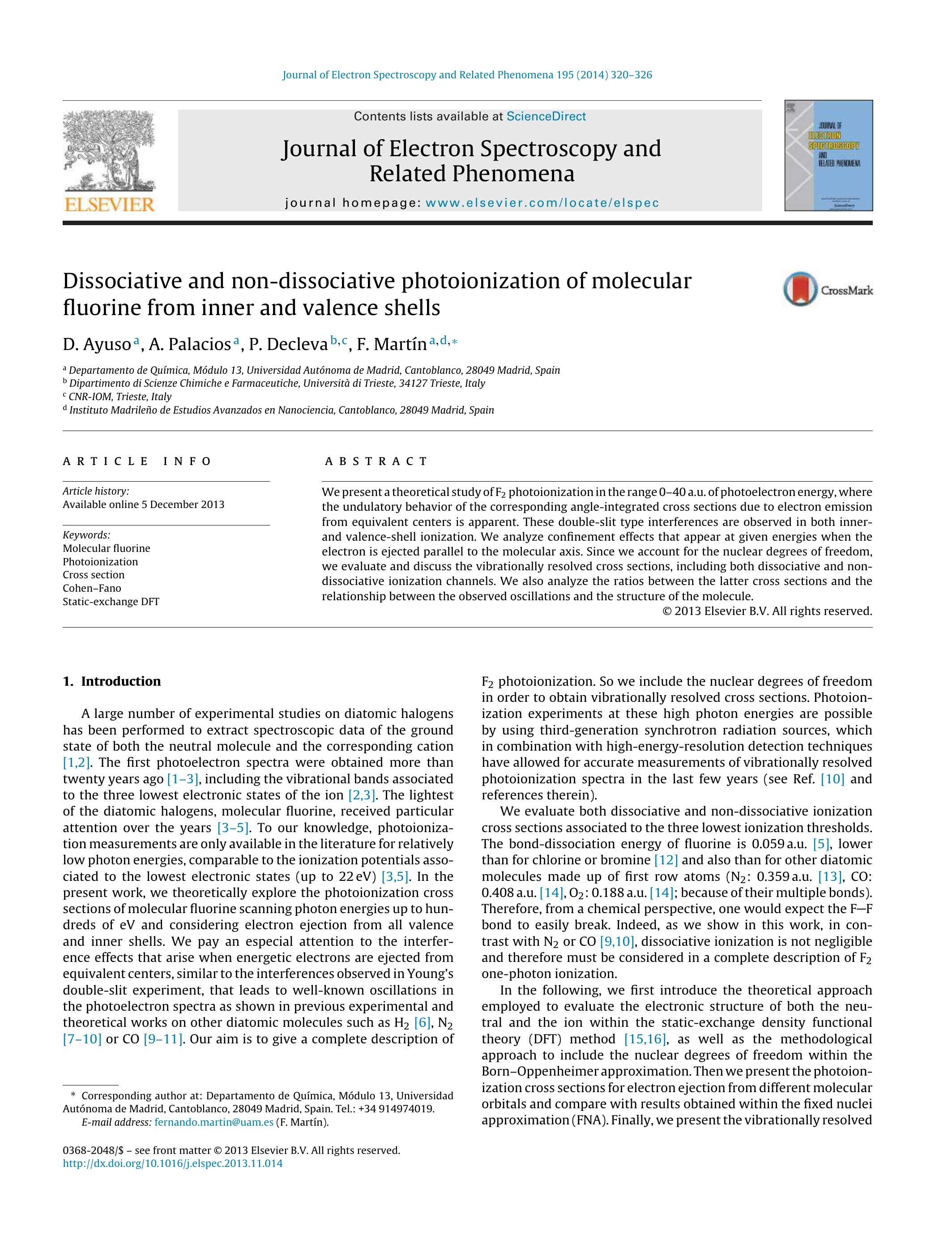}

\chapter{\textnormal{Ultrafast electron dynamics in phenylalanine initiated by attosecond pulses}}
\label{appendix6}
\includepdf[pages={1-}]{blank.pdf}
\includepdf[pages={1-}]{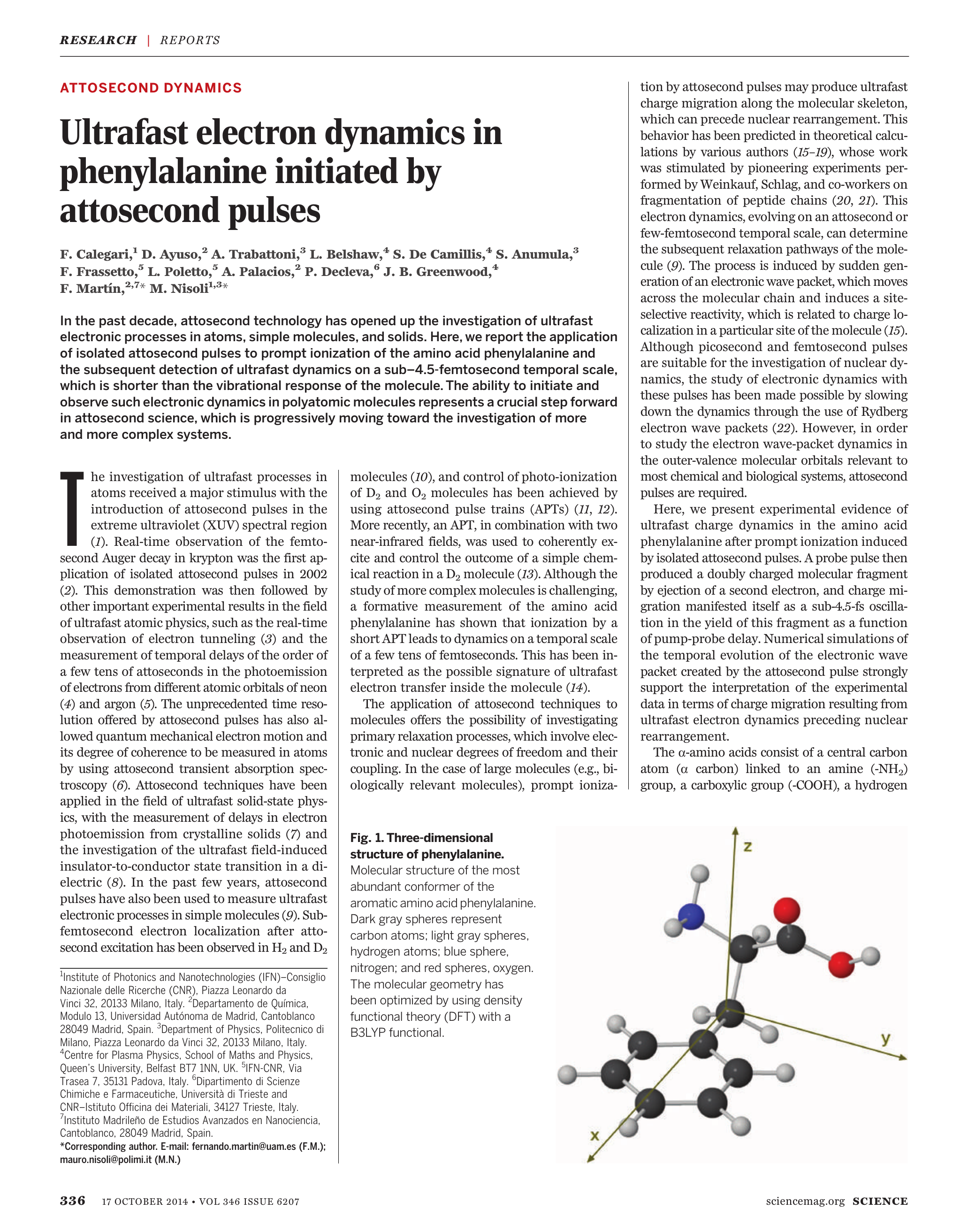}
\includepdf[pages={1-}]{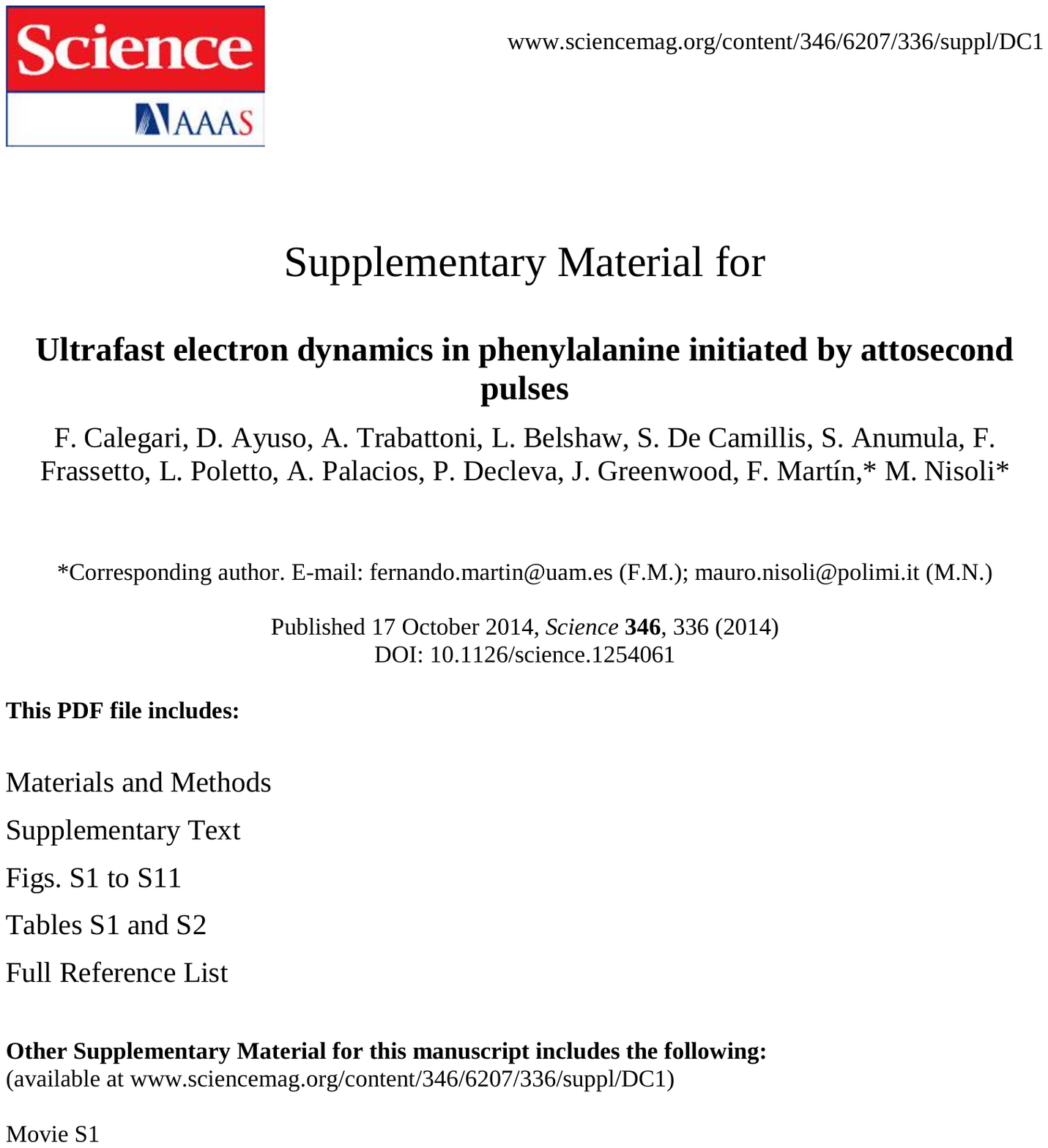}

\chapter{\textnormal{Ultrafast charge dynamics in an amino acid induced by attosecond pulses $\quad$ }}
\label{appendix7}
\includepdf[pages={1-}]{blank.pdf}
\includepdf[pages={1-}]{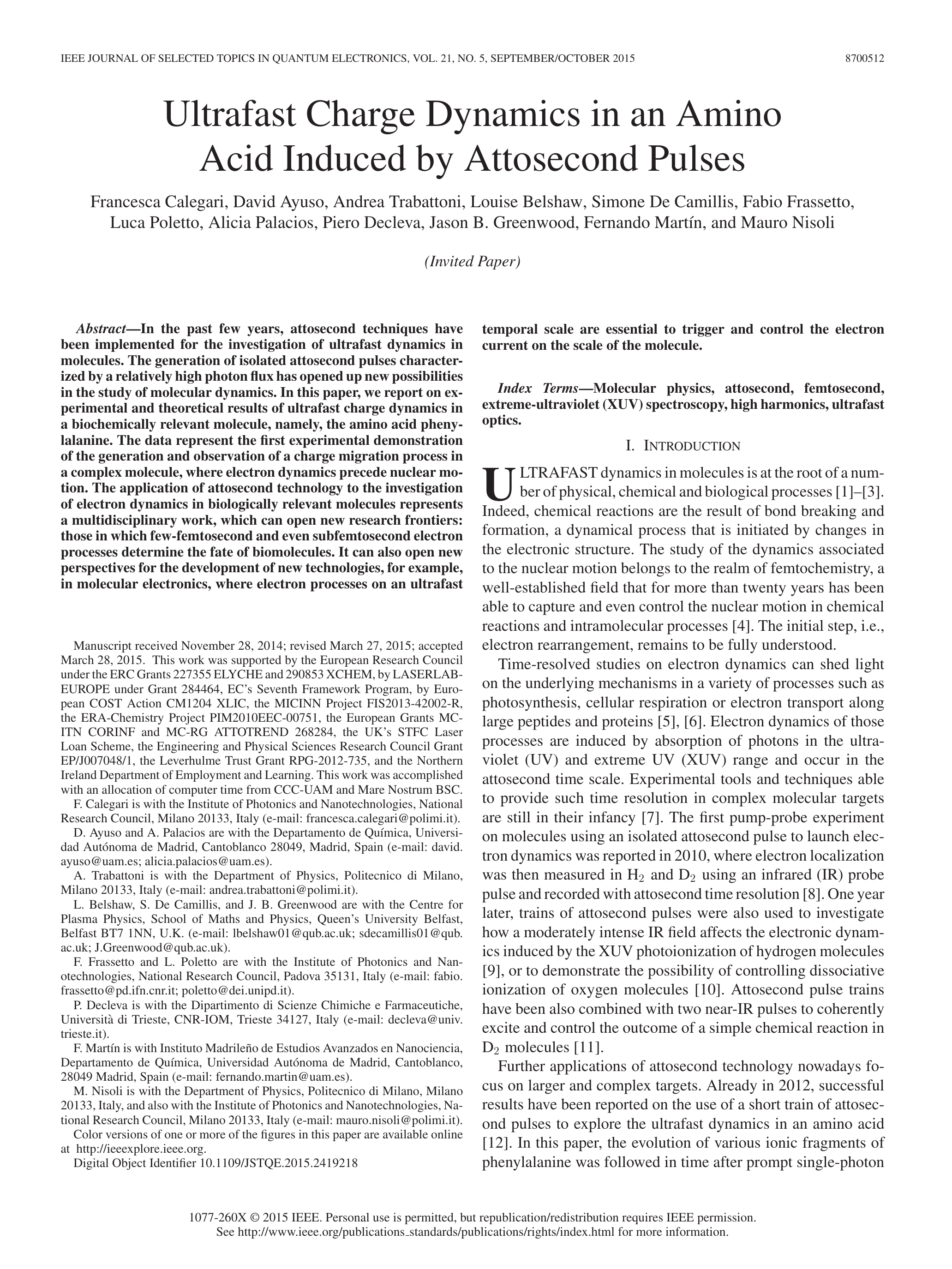}

\addcontentsline{toc}{part}{Bibliography}
\bibliographystyle{unsrt}
\bibliography{mybib}
\end{document}

%% file: Chapters/Abstract.tex
\chapter*{Abstract}
\label{Abstract}
\fancyhead[LE]{}
\fancyhead[RO]{\fontsize{11pt}{11pt}\selectfont Abstract}

We present a theoretical study of light-induced phenomena in gas-phase molecules, exploring the physical phenomena arising in two distinct contexts: using synchrotron radiation and ultrashort laser pulses.
This work has been done in close collaboration with Piero Decleva (Universit\`a degli Studi di Trieste).\vspace{2 mm}

We first present our work on inner-shell photoionization of diatomic (CO) and small polyatomic (CF$_4$, BF$_3$) molecules at high photoelectron energies performed in collaboration with the experimental the groups of Edwin Kukk (Turku University), Catalin Miron (Synchrotron SOLEIL), Kiyosi Ueda (Synchrotron SPring-8) and Thomas Darrah Thomas (Oregon State University).
The combination of state-of-the-art Density Functional Theory (DFT)-like calculations, capable to describe photoionization accounting for the nuclear degrees of freedom, together with high-resolution third-generation synchrotron facilities, has enabled the investigation of non-Franck-Condon effects observable in vibrationally resolved photoionization measurements.
We demonstrate that the nuclear response to intramolecular electron diffraction is observable and can be used to obtain structural information. 
As a proof-of-principle, by using the DFT calculations as an analysis tool to fit the experimental data, we have accurately determined the equilibrium distance of the CO molecule and the bond contraction that takes place upon C 1s ionization.
This is a surplus of photoelectron spectroscopy with respect to more conventional spectroscopic techniques, which usually can only provide structural information of neutral molecular species.
Furthermore, we have explored the different phenomenon arising when an electron is emitted from a delocalized orbital: multicenter emission.
The results on molecular fluorine coming from our numerical simulations are in good qualitative agreement with those provided by the simple formula proposed by Cohen and Fano in the sixties.\vspace{2 mm}

We have employed the same DFT-based methodology together with time\hyp{}dependent first-order perturbation theory and a reduced density matrix formalism to report the first demonstration of purely electron dynamics in a biological molecule: the amino acid phenylalanine, in collaboration with the experimental groups of Mauro Nisoli (Politecnico di Milano), Luca Poletto (Istituto Nazionale di Fotonica - Consiglio Nazionale delle Ricerche) and Jason Greenwood (Queen's University).
The use of attosecond pulses in combination with novel detection techniques has enabled the capture of purely electron motion at its intrinsic time scale.
Because of their wide energy bandwidth, attosecond pulses are ideal sources to generate coherent superpositions of states, triggering an ultrafast electronic response that can be later tracked with attosecond resolution.
Our theoretical study enabled to interpret the experimental findings in terms of charge migration, thus confirming the first observation of purely electron dynamics in a biomolecule.
The work presented here has been extended to treat the amino acids glycine and tryptophan, which has allowed the investigation of radical substitution effects in the charge migration mechanism.

%% file: Chapters/Resumen.tex
\chapter*{Resumen}
\label{Resumen}
\fancyhead[LE]{}
\fancyhead[RO]{\fontsize{11pt}{11pt}\selectfont Resumen}

Esta tesis doctoral constituye un estudio te\'orico de procesos ultrarr\'apidos que ocurren en mol\'eculas aisladas cuando son expuestas a radiaci\'on electromagn\'etica. En particular, hemos investigado los fen\'omenos f\'isicos que surgen en dos contextos diferentes: (i) cuando la energ\'ia de los fotones incidentes est\'a bien definida, como ocurre en experimentos que hacen uso de radiaci\'on sincrotr\'on, y (ii) cuando la duraci\'on de la interacci\'on radiaci\'on-materia es extremadamente corta, lo cual es posible gracias al desarrollo de pulsos l\'aser ultra-cortos, del orden de tan solo unos pocos cientos de attosegundos (1 as = 10$^{-18}$ s). Este trabajo ha sido supervisado por Fernando Mart\'in y Alicia Palacios (Universidad Aut\'onoma de Madrid), y ha sido realizado en estrecha colaboraci\'on con Piero Decleva (Universit\'a degli Studi di Trieste) y con varios grupos experimentales de distintos pa\'ises, como se indica a continuaci\'on.

En primer lugar, presentamos un estudio sobre fotoionizaci\'on de capa interna de mol\'eculas diat\'omicas (CO) y poliat\'omicas (CF4, BF3) con radiaci\'on sincrotr\'on a altas energ\'ias del fotoelectr\'on, que ha sido realizado en colaboraci\'on con los grupos experimentales de Edwin Kukk (Turku University), Catalin Miron (Synchrotron SOLEIL), Kiyosi Ueda (Synchrotron SPring-8) y Thomas Darrah Thomas (Oregon State University). El uso de instalaciones sincrotr\'on de tercera generaci\'on, en combinaci\'on con t\'ecnicas de detecci\'on avanzadas de alta resoluci\'on en energ\'ia, nos ha permitido observar claras violaciones del principio de Franck-Condon (FC) en espectros fotoelectr\'onicos resueltos vibracionalmente. Nuestro trabajo te\'orico, basado en la aplicaci\'on de la Teor\'ia del Funcional de la Densidad (DFT, del ingl\'es, “Density Functional Theory”) para describir procesos de fotoionizaci\'on en sistemas multielectr\'onicos, ha sido esencial para guiar las campañas experimentales, as\'i como para interpretar los resultados de las mismas. Para ello, ha sido necesario tener en cuenta en nuestras simulaciones computacionales los grados de libertad adicionales debidos al movimiento nuclear. Nuestro estudio conjunto demuestra que las violaciones del principio de FC observadas son consecuencia de un fen\'omeno de difracci\'on electr\'onica: cuando se emite un electr\'on desde una regi\'on bien localizada en el centro de una mol\'ecula poliat\'omica, como el orbital 1s del \'atomo de carbono en la mol\'ecula CF4, \'este puede ser difractado por los \'atomos circundantes, originando interferencias constructivas y destructivas en los espectros fotoelectr\'onicos.
La respuesta nuclear al fen\'omeno de difracci\'on electr\'onica es observable, y los espectros fotoelectr\'onicos contienen informaci\'on estructural del sistema. Sin embargo, la extracci\'on de esta informaci\'on puede llegar a ser todo un desaf\'io. En esta tesis doctoral proponemos un m\'etodo de determinaci\'on estructural, basado en el uso de c\'alculos DFT como herramienta de an\'alisis para el ajuste de datos experimentales. Como prueba de concepto, hemos determinado con precisi\'on, y simult\'aneamente, la distancia de equilibrio de la mol\'ecula de CO y la contracci\'on de enlace que tiene lugar tras la extracci\'on de un electr\'on del orbital 1s del \'atomo de carbono. Nuestro m\'etodo presenta una clara ventaja frente a t\'ecnicas espectrosc\'opicas m\'as convencionales que, en general, tan solo son capaces de proporcionar informaci\'on estructural de la especie neutra.

Cuando el electr\'on no se emite desde una regi\'on bien localizada en el centro de la mol\'ecula, sino desde un orbital deslocalizado entre varios \'atomos, tiene lugar un fen\'omeno f\'isico diferente: emisi\'on multic\'entrica. Nuestras simulaciones en la mol\'ecula de fl\'uor muestran que este fen\'omeno es experimentalmente observable en espectros fotoelectr\'onicos, y nuestros resultados coinciden, de forma cualitativa, con los predichos por la sencilla f\'ormula propuesta por Cohen y Fano en los años 60.

Poder observar y controlar el movimiento de los electrones en mol\'eculas biol\'ogicas es uno de los principales objetivos de la ciencia de attosegundos (attociencia). El uso de pulsos laser ultra-cortos permite inducir corrientes electr\'onicas ultra-r\'apidas en la materia mediante la creaci\'on de superposiciones coherentes de autoestados del sistema. Adem\'as, estos pulsos tienen la duraci\'on adecuada para visualizar las corrientes creadas con la resoluci\'on temporal requerida. En colaboraci\'on con los grupos experimentales de Mauro Nisoli (Politecnico di Milano), Luca Poletto (Instituto Nazionale di Fotonica – Consiglio Nazionale delle Ricerche) y Jason Greenwood (Queen's University), hemos reportado la primera observaci\'on de din\'amica puramente electr\'onica en una mol\'ecula biol\'ogica: el amino \'acido fenilalanina. Esto ha sido posible gracias a la creaci\'on y aplicaci\'on de pulsos laser de tan solo 300 attosegundos de duraci\'on, en combinaci\'on con novedosas t\'ecnicas de detecci\'on de fragmentos i\'onicos doblemente cargados (formados tras la ionizaci\'on de la biomol\'ecula). La migraci\'on de carga observada precede cualquier reordenamiento estructural y es la base de un gran n\'umero de procesos biol\'ogicos.
Nuestras simulaciones computacionales, basadas en DFT junto con teor\'ia de perturbaciones a primer orden y un formalismo de matriz de densidad reducida, han sido esenciales en la interpretaci\'on de los hallazgos experimentales en t\'erminos de migraci\'on de carga ultra-r\'apida. Para ello, ha sido necesario describir de forma precisa la interacci\'on de la mol\'ecula con el pulso laser empleado en los experimentos, as\'i como la respuesta molecular inducida. Nuestro trabajo ha revelado, de manera inequ\'ivoca, que las variaciones de carga observadas se deben \'unica y exclusivamente al movimiento ondulatorio de los electrones en la biomol\'ecula. Hemos descubierto que la migraci\'on de carga desde un extremo a otro del amino \'acido tarda entre 3 y 4 femtosegundos (1 fs = 10$^{-15}$ s). Merece la pena señalar que, hasta entonces, ning\'un trabajo te\'orico hab\'ia sido capaz de describir el proceso de ionizaci\'on por un pulso de attosegundos y el posterior reordenamiento electr\'onico en una mol\'ecula compleja, de relevancia biol\'ogica.

En esta tesis doctoral presentamos un estudio completo que incluye los amino \'acidos glicina y tript\'ofano. Esto nos ha permitido generalizar los resultados obtenidos a otras mol\'eculas biol\'ogicas, as\'i como investigar c\'omo afecta la sustituci\'on del radical al mecanismo de migraci\'on de carga ultra-r\'apida en un amino \'acido. Nuestros resultados demuestran que, modificando las caracter\'isticas el pulso laser (amplitud y fase de sus componentes espectrales) es posible modificar la respuesta electr\'onica inducida, permitiendo controlar a la carta el movimiento de los electrones en sistemas de relevancia biol\'ogica.

%% file: Chapters/Publications.tex
\chapter*{Publications}

The work presented in this PhD thesis has lead to the following publications peer reviewed journals (inverse chronological order):\vspace{2 mm}
\begin{enumerate}

\item[7.] \textbf{Ultrafast Charge Dynamics in an Amino Acid Induced by Attosecond Pulses}.
F. Calegari, D. Ayuso, A. Trabattoni, L. Belshaw, S. De Camillis, F. Frassetto, L. Poletto, A. Palacios, P. Decleva, J. B. Greenwood, F. Mart\'in, and M. Nisoli.
\emph{IEEE Journal of Selected Topics in Quantum Electronics} \textbf{21}, 5, 8700512 (2015).
Attached in appendix \ref{appendix7}.

\item[6.] \textbf{Vibrationally Resolved B 1s Photoionization Cross Section of BF$_3$}.
D. Ayuso, M. Kimura, K. Kooser, M. Patanen, E. Pl\'esiat, L. Argenti, S. Mondal, O. Travnikova, K. Sakai, A. Palacios, E. Kukk, P. Decleva, K. Ueda, F. Mart\'in and C. Miron.
\emph{The Journal of Physical Chemistry A} \textbf{119}, 5971-5978 (2015).
Attached in appendix \ref{appendix2}.

\item[5.] \textbf{Ultrafast electron dynamics in phenylalanine initiated by attosecond pulses}.
F. Calegari, D. Ayuso, A. Trabattoni, L. Belshaw, S. De Camillis, S. Anumula, F. Frassetto, L. Poletto, A. Palacios, P. Decleva, J. B. Greenwood, F. Mart\'in and M. Nisoli.
\emph{Science} \textbf{346}, 6207, 336 (2014).
Attached in appendix \ref{appendix6}.

\item[4.] \textbf{Vibrationally resolved C 1s photoionization cross section of CF$_4$}.
M. Patanen, K Kooser, L. Argenti, D. Ayuso, M. Kimura, S. Mondal, A. Palacios, K. Sakai, O. Travnikova, P. Decleva, E. Kukk, E. Pl\'esiat, C. Miron K. Ueda and F Mart\'in.
\emph{Journal of Physics B: Atomic, Molecular and Optical Physics} \textbf{47}, 124032 (2014).
Attached in appendix \ref{appendix3}.

\item[3.] \textbf{Dissociative and non-dissociative photoionization of molecular fluorine from inner and valence shells}.
D. Ayuso, A. Palacios, P. Decleva and F. Mart\'in.
\emph{Journal of Electron Spectroscopy and Related Phenomena} \textbf{195}, 320-326 (2014).
Attached in appendix \ref{appendix5}.

\item[2.] \textbf{Intramolecular photoelectron diffraction in the gas phase}.
K. Ueda, C. Miron, E. Pl\'esiat, L. Argenti, M. Patanen, K. Kooser, D. Ayuso, S. Mondal, M. Kimura, K. Sakai, O. Travnikova, A. Palacios, P. Decleva, E. Kukk, and F. Mart\'in. 
\emph{The Journal of Chemical Physics} \textbf{139}, 124306 (2013).
Attached in appendix \ref{appendix1}.

\item[1.] \textbf{Effects of molecular potential and geometry on atomic core-level photoemission over an extended energy range: The case study of the CO molecule}.
E. Kukk, D. Ayuso, T. D. Thomas, P. Decleva, M. Patanen, L. Argenti, E. Pl\'esiat, A. Palacios, K. Kooser, O. Travnikova, S. Mondal, M. Kimura, K. Sakai, C. Miron, F. Mart\'in, and K. Ueda.
\emph{Physical Review A} \textbf{88}, 033412 (2013).
Attached in appendix \ref{appendix4}.

\end{enumerate}

%% file: Chapters/Acknowledgements.tex
\chapter*{Acknowledgements}

I would like to express my gratitude to the people that have made this work possible:\vspace{2 mm}
\begin{itemize}
\item[$\circ$] To Alicia, for guiding me during these four years, for her constant support and her very useful advice. 
\item[$\circ$] To Fernando, for giving me the opportunity to join this wonderful group, for his support and for being a constant source of inspiration.
\item[$\circ$] To Piero, for being so accessible, for his clear explanations, for his support and for being the perfect host every time I had the opportunity to visit him in Trieste.
\item[$\circ$] To the experimental collaborators I had the opportunity to work with, in Milan, Belfast, Oregon, Turku, Sendai and Saint Malo, for allowing me to be part of amazing projects.
\item[$\circ$] To \'Alvaro, Oriana, Lara and Darek, for all the good moments shared during these four years and for their constant support.
\item[$\circ$] To In\'es, Jes\'us, Luca, Sergio and Selma, for their useful advice and for sharing a bit of their knowledge with me.
\item[$\circ$] To my office mates, and to everyone in the department, because it has been wonderful to spend these four years with them.
\item[$\circ$] And to my family and friends, and to Steven.
\end{itemize}

The research leading to these results has received funding from European Union's Seventh Framework Programme (FP7/2007-2013) projects ERC-AdG-XChem (GA 290853) and MC-IRG ATTOTREND (GA 268284).

%% file: Chapters/Introduction.tex
\chapter*{Introduction}
\label{Introduction}
\addcontentsline{toc}{part}{Introduction}
\fancyhead[LE]{}
\fancyhead[RO]{\fontsize{11pt}{11pt}\selectfont Introduction}

Chemical reactions occur as a result of bond breaking and formation, a dynamical process that, in general, is initiated by changes in the electronic structure of a molecule and followed by the subsequent nuclear rearrangement.
The study of the dynamics associated to the nuclei belongs to the realm of femtochemistry, a well\hyp{}established field that for more than twenty years has been able to capture and even control the nuclear motion in chemical reactions and intramolecular processes \cite{ZewailJPCA2000,ZewailPAC2000}.
The field obtained an important recognition in 1999, when Ahmed Zewail was awarded the Nobel prize in Chemistry ``for his studies of the transition states of chemical reactions using femtosecond spectroscopy'' \cite{ZewailNobelPrize}.
One of the most common techniques to investigate the dynamics of a chemical reaction constitutes the well-known pump-probe spectroscopy: a short pulse of light (pump) is used to induce a process in a molecular target and, after some time, the dynamical response of the system is monitored with a second pulse (probe).
By performing measurements with different time delays between the two pulses, it is possible to take ``snapshots'' of a chemical reaction.
Of course, the duration of the light pulses employed needs to be (at least) of the same order of magnitude (or shorter) than the dynamics to observe.
For instance, in order to monitor a process of charge transfer mediated by the nuclear motion in organic molecules \cite{LehrJPCA2005}, pulses in the femtosecond time domain were needed.
This is the reason why purely electron motion, occurring in the attosecond time domain, has remained hidden from direct experimental observation until very recently, when pulses with durations as short as a few tens of attoseconds became available.

\subsubsection*{Attosecond science}

\begin{figure}[h]
\centering
\includegraphics[width=\textwidth]{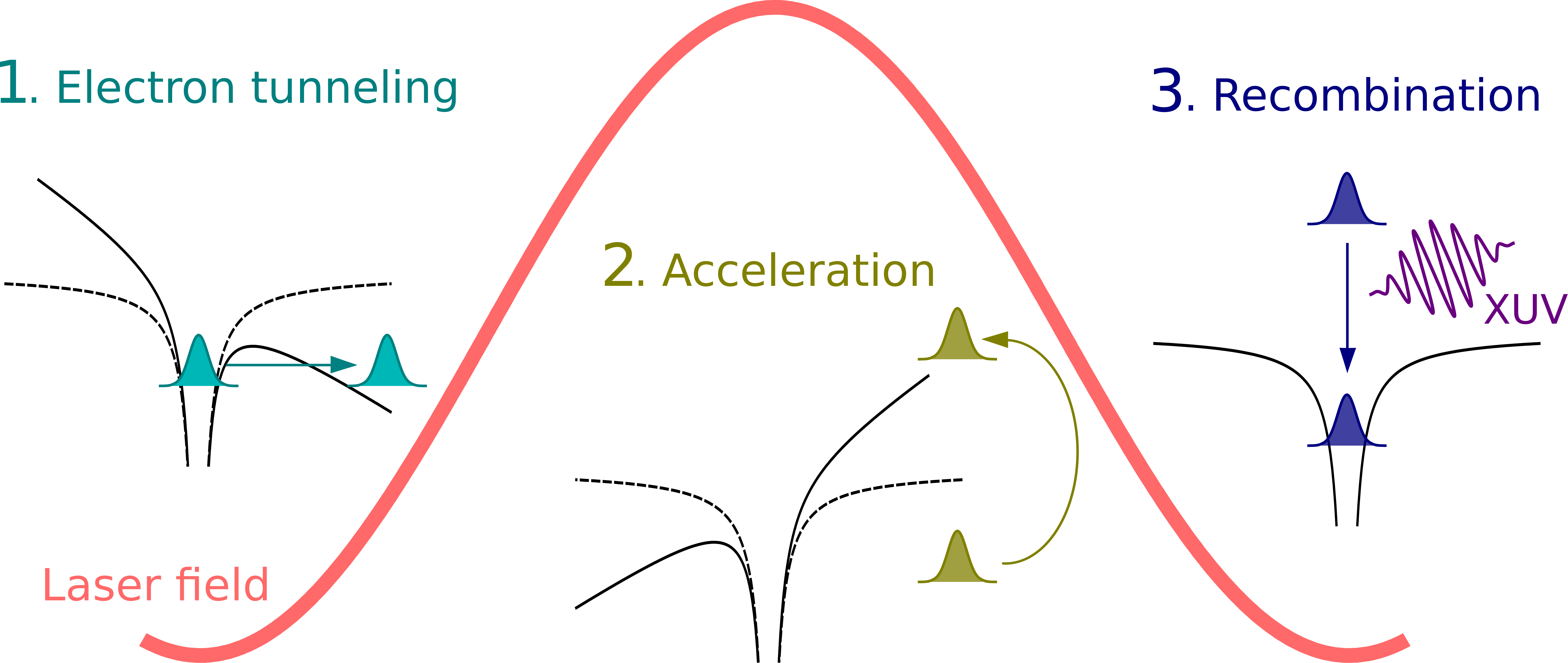}
\caption{Schematic representation of the three-step model \cite{CorkumPRL1993} for HHG: (1) tunnel ionization, (2) acceleration by the driving field and (3) recombination with the parent ion.}
\label{fig_ThreeStepModel}
\end{figure}

The experimental demonstration of attosecond pulses was achieved in 2001 \cite{HentschelNature2001,PaulScience2001} using high harmonic generation (HHG) techniques, which opened a new era of time resolved experiments.
HHG is a non-linear process occurring when an atomic or molecular gas is irradiated with an intense femtosecond laser, usually a Ti:sapphire laser with a central wavelength of 800 nm.
The target will then emit XUV light with frequencies that are high odd multiples of the driving field \cite{McPhersonJOSAB1987,FerrayJPB1988}.
An interpretation of this phenomenon was given in 1993 \cite{CorkumPRL1993} by Paul Corkum by means of a three-step model.
Due to the distortion of the Coulomb potential generated by the strong IR field, the electron is ionized by tunneling through the electric barrier (see fig. \ref{fig_ThreeStepModel}).
Then (second step), the electron is accelerated by the laser field and driven back towards the parent ion.
In the last step, the recombination, the energy that the electron has accumulated during its journey is released as an energetic XUV or X-ray photon \cite{CorkumPRL1993}.
Since the first step (tunnel ionization) is a non-linear process that requires the absorption of several photons, it is more likely to occur at the maxima of the laser field, when the tunneling picture is valid.
Thus, electrons are released and recombined in ultrashort intervals, leading to the formation attosecond laser pulses \cite{AgostiniJPB2004}.

According to the three-step model, an attosecond pulse is generated every half a cycle of the driving femtosecond pulse, which leads to the production of attosecond pulse trains (APTs) rather than single attosecond pulses (SAPs).
A lot of work has been directed towards the generation of SAPs \cite{KrauszRMP2009}.
The first SAP was produced in 2001 by the group of Ferenc Krausz by using the selection of cutoff harmonics generated with a 7 fs IR pulse \cite{HentschelNature2001}.
This short duration made that the highest harmonics were produced only during one half-cycle, thus generating a SAP.
In order to characterize a SAP, one can use a co-propagating IR laser pulse by means of the attosecond streak camera \cite{ItataniPRL2002}:
the photoelectron emitted by the XUV pulse is driven by the presence of the IR field and its momentum is modified in a extent that depends on the relative time delay between the two pulses.
The first fully-characterized few-cycle SAP was generated in 2006 in the group of Mauro Nisoli \cite{SansoneSciencie2006}, with a duration of 130 as and a central frequency of 35 eV.
Only two years later, an even shorter pulse was produced, with duration of 80 as and 80 eV of central energy \cite{GoulielmakisScience2008}.
To date, the shortest XUV pulses ever produced and characterized were 67 as-long and their central energy was 90 eV \cite{ZhaoOL2012}.\vspace{2 mm}

These impressive achievements have enabled the real-time observation and control of electron motion in atoms, molecules and condensed phases \cite{CorkumNatPhys2007,KrauszRMP2009,bookAttosecondAndXUVPhysics,PalaciosJPB2015}.
The observation of the femtosecond Auger decay in krypton in 2002 was the first application of isolated attosecond pulses \cite{DrescherNature2002}.
This demonstration was followed by other important experimental results in the field of ultrafast atomic physics, such as the real-time observation of electron tunneling \cite{UiberackerNature2007} and the measurement of temporal delays of the order of a few tens of attoseconds in the photoemission of electrons from different atomic orbitals of neon \cite{SchultzeScience2010} and argon \cite{KlunderPRL2011}.
The unprecedented time resolution offered by attosecond pulses has also allowed quantum mechanical electron motion and its degree of coherence to be measured in atoms by using attosecond transient absorption spectroscopy \cite{GoulielmakisNature2010}.
Attosecond techniques have also been applied in the field of ultrafast solid-state physics, with the measurement of delays in electron photoemission from crystalline solids \cite{CavalieriNature2007} and the investigation of the ultrafast field-induced insulator-to-conductor state transition in a dielectric \cite{SchiffrinNature2013}. \vspace{2 mm}

The application of attosecond techniques to molecules offers the possibility of investigating primary relaxation processes, which involve electronic and nuclear degrees of freedom and their coupling \cite{SansoneNature2010,KelkensbergPRL2011,SiuPRA2011,RanitovicPNAS2014,LepineNatPhot2014}. 
In the case of large molecules (e.g., biologically relevant molecules), prompt ionization by attosecond pulses may produce ultrafast charge migration along the molecular skeleton, preceding any nuclear rearrangements.
This phenomenon has been predicted by various authors \cite{RemaclePNAS2006,CederbaumCPL1999,BreidbachJCP2003,BreidbachPRL2005,HennigJPCA2005}, whose work was stimulated by pioneering experiments performed by R. Weinkauf, E. W. Schlag and collaborators on fragmentation of peptide chains \cite{WeinkaufJPC1994,WeinkaufJPC1995,WeinkaufJPC1996}.
Ultrafast electron dynamics evolving on an attosecond or few-femtosecond temporal scale can determine the subsequent relaxation pathways of a molecule \cite{LepineNatPhot2014}.
The process is induced by sudden generation of an electronic wave packet, which moves across the molecular chain and induces a site selective reactivity, which is related to charge localization in a particular site of the molecule \cite{RemaclePNAS2006}.
Although the study of complex molecules is challenging, a formative measurement of the amino acid phenylalanine has shown that ionization by a short APT leads to dynamics on a temporal scale of a few tens of femtoseconds.
This has been interpreted as the possible signature of ultrafast electron transfer inside the molecule \cite{BelshawJPCL2012}.
Even though picosecond and femtosecond pulses are suitable for the investigation of nuclear dynamics, the study of electronic dynamics with these pulses has been made possible by slowing down the dynamics through the use of Rydberg electron wave packets \cite{WalsPRL1994}.
However, in order to study the electron wave packet dynamics in the outer-valence molecular orbitals, relevant to most chemical and biological systems, attosecond pulses are required.\vspace{2 mm}

Despite the impressive progress made in the last decades, attosecond pulses generated via HHG still have two main constraints: their relatively low intensities, which can make some attosecond pump-probe experiments quite challenging, and the limited range of photon energies in which they can be produced.
These limitations are overcome in the recently operating XUV/X-ray free-electron laser (FEL) facilities, which can already produce bright few-femtosecond pulses based on synchrotron light.

\subsubsection*{Synchrotron light}

Synchrotron radiation is emitted when charged particles moving at velocities close to the speed of light are forced to change direction by a magnetic field, as dictated by the fundamental laws of electrodynamics \cite{bookJackson}.
The first observation of artificial synchrotron light occurred at the General Electric Research Laboratory in New York in 1947 \cite{ElderPR1947}, opening a new era of accelerator-based light sources.
Synchrotron light sources rapidly evolved ever since through up to four different generations \cite{BilderbackJPB2005}.\vspace{2 mm}

The so-called first-generation light sources were high-energy physics facilities were synchrotron radiation was generated as a byproduct.
The interest for the use of synchrotron light increased over the years, motivating a series of pioneering advances.
The most important was the development of storage rings \cite{ONeillPR1956}, the basis for all of today's synchrotrons.
A storage ring is a type of circular accelerator in which the beam of particles can circulate at a fixed energy during several hours, providing stable beam conditions and reducing the radiation hazard.
In 1975, the Synchrotron Radiation Source (SRS) at the Daresbury Laboratory in the UK, the first synchrotron exclusively designed to the production of light, began to be constructed \cite{HermanNS1979,HermanNS1980}.
It became operational in 1981 \cite{Holder2008}, the same year as the BESSY synchrotron \cite{Baumgaertel1981} in Berlin, giving birth to the second-generation light sources.
Over the years, some facilities that were originally developed to high-energy physics, such as the HASYLAB (Hamburger Synchrotronstrahlungslabor) at DESY, or the Stanford Synchrotron Radiation Laboratory at the SLAC National Accelerator Laboratory at Standford,
were upgraded to second-generation status and gradually started to dedicate more operating time to light production.\vspace{2 mm}

The development of insertion devices \cite{WinickPT1981} lead to the advent of third-generation light sources.
Insertion devices are arrays of magnets with alternating polarity that, located into the straight sections of the storage rings, generate very bright beams and allow to shift the light spectrum towards higher photon energies. 
These devices were quickly incorporated into existing the synchrotron facilities.
Third-generation facilities, designed with insertion devices in place from the beginning, saw the light in the 90s.
The first of them was the European Synchrotron Radiation Facility (ESRF), located in Grenoble, France, which started operating in 1994 \cite{KvickJPCS1991,AdamsJAS1995}.\vspace{2 mm}

To date, there are more than 50 dedicated second- and third-generation light sources all over the world serving many areas of science. 
The BESSY II in Berlin, the ELETTRA in Trieste, the National Synchrotron Light Source in the USA, Spring-8 in Japan or SOLEIL in France are just a few examples.
All of them present similar structures, with a storage ring having many ports connected to the beamlines, which are small stations where the experiments are performed.
The configuration of the different beamlines, however, can be more different, depending on the kind of experiments they have been designed for.\vspace{2 mm}

Synchrotron radiation spans a wide energy range, from infrared light up to hard X-rays.
It is characterized by its high brightness, being orders of magnitude brighter than that produced in conventional light sources.
Synchrotron light is also highly polarized (linearly, circularly or elliptically), tunable, collimated (consisting of almost parallel rays) and concentrated over a small area.
These properties make of synchrotron radiation one of the most important and universal research tools, with a steadily increasing number of applications \cite{BilderbackJPB2005,SaishoASL1996,LewisPMB1997}. \vspace{2 mm}

The brightness of the light beams produced these facilities and the use of efficient X-ray monochromators \cite{CaciuffoPR1987} have brought unprecedented energy resolution to X-ray spectroscopy.
This has lead to major advances in X-ray absorption spectroscopy (XAS) that include the development of the near edge X-ray absorption fine structure spectroscopy (NEXAFS) and the X-ray absorption fine-structure spectroscopy (EXAFS) \cite{StummAPF1989,SternJSR2001} techniques.
In addition to conventional XAS, ion-yield spectroscopy techniques, in which the yield of ionic fragments is recorded as a function of the photon energy, can provide useful information about photodissociation dynamics occurring in highly excited states \cite{Becker1996,GuilleminJPB2009}.
By recording fragmentation yields for different ejection angles one can also gain information about the symmetry of the states involved in the fragmentation dynamics \cite{ShigemasaPRA1992}.\vspace{2 mm}

Photoelectron spectroscopy has allowed to measure core-hole lifetime widths of molecular species \cite{CarrollPRA1999,CarrollPRA2000,JurvansuuPRA2001} as well as their corresponding vibrational structures \cite{BorvePRA2000}.
Recent work \cite{MironNatPhys2012} has demonstrated that high-resolution resonant photoelectron spectroscopy can be applied to image molecular potentials of excited states and to identify new states that are invisible in conventional photoelectron spectroscopy techniques.
The combination of high-resolution photoelectron spectroscopy with normal Auger electron spectroscopy has allowed to obtain information about potential energy curves of two-hole states in diatomic molecules \cite{PuttnerPRA2002}.
By recording photoelectrons, Auger electrons and ionic fragments in coincidence, it has been possible to get a deeper insight into the Auger decay process \cite{KammerlingPRL1991,UedaPRL1999,WehlitzPRA1999,LablanquiePRL2000,EberhardtPRA1988,LevinPRL1990}.\vspace{2 mm}

The application of angle-resolved photoelectron spectroscopy to gas-phase molecules can reveal structural details about the target.
By measuring the photoemission intensity over a hemispherical region, it has been possible to reconstruct molecular orbital densities of large $\pi-$conjugated molecules \cite{PuschnigScience2009}.
When an electron is emitted from an inner shell of a molecule, the outgoing wave can be coherently scattered by the surrounding atomic constituents.
Therefore, molecular-frame photoelectron angular distributions (MFPADs) are sensitive to molecular potentials and can thus be used to image molecular structures.
MFPADs can be measured by registering photoelectrons in coincidence with ions \cite{GolovinZPD1992} or with Auger electrons \cite{GuilleminRSI2000,GuilleminPRL2001}.
It has been shown that in the case of small polyatomic molecules in which the central atom is bonded to hydrogens, such as CH$_4$, photoelectrons with low kinetic energy are emitted along the chemical bonds \cite{WilliamsPRL2012,WilliamsJPB2012,PlesiatPRA2013}.
Although this assumption is no longer true when hydrogens are replaced by heavier fluorine atoms \cite{PlesiatPRA2013}, MFPADs can still retrieve information about the tridimensional structure of the molecular target.
Measuring MFPADs requires the use of rather sophisticated coincidence setups with high angular resolution and high detection efficiency \cite{DornerPR2000,MironNIMPRSA2009}, which are not always available or applicable.
Alternatively, recent work on the photoionization of diatomic and small polyatomic molecules by synchrotron radiation \cite{CantonPNAS2011,PlesiatPRA2012} has shown that, under some circumstances, photoelectron spectra integrated over electron ejection angle can also be a valuable tool for structural determination.\vspace{2 mm}

Free-electron lasers (FELs) constitute the fourth generation of light sources, being able to generate ultrashort and ultrabright pulses of coherent light by using powerful linear accelerators which are several kilometers long.
The underlying principle behind FELs is the self-amplified spontaneous emission (SASE).
As in third-generation facilities, FELs make use of insertion devices to generate synchrotron light.
This radiation interacts with the oscillating electrons, making them drift into microbunches, which are separated by a distance that is equal to one radiation wavelength.
Through this interaction, the electrons start to emit coherent light.
The emitted radiation can reinforce itself, leading to high beam intensities and laser-like properties.\vspace{2 mm}

The first FEL to become operational was the Tesla Test Facility in Hamburg in 2000, that was replaced by FLASH in 2005, the first soft X-ray FEL \cite{AckermannNatPhot2007}, where the first experiments on coherent diffractive imaging were performed \cite{ChapmanNatPhys2006,ChapmanNature2007,MarchesiniNature2008} by means of the so-called diffraction-before-destruction \cite{DoerrNatMeth2011} technique.
In 2009, the Linac Coherent Light Source (LCLS) in Stanford became the first hard X-ray FEL, improving the resolution of coherent diffractive imaging experiments \cite{ChapmanNature2011,RedeckeScience2013}.
Other FEL facilities such as FERMI \cite{AllariaNatPhot2012} (2010), at the Elettra Sincrotrone in Trieste, or SACLA (2011), embedded in the Spring-8 complex \cite{PileNatPhot2011}, have recently become operational,
and new facilities are being designed or under construction. 

\subsubsection*{Theoretical methods}

Advances in synchrotron- and HHG-based light sources have opened the door for structural determination of single isolated molecules, as well as for imaging and even controlling electron and nuclear dynamics, with exciting applications in physics, chemistry and biology.
In order to guide and to understand these new generation of experiments, theoretical calculations are crucial.
As we have seen, the development of high-resolution third-generation synchrotron light sources enabled the application of traditional photoelectron X-ray diffraction techniques to small molecules in the gas phase.
However, solid theoretical support combining state-of-the-art calculations with the use of simple models was needed to understand and interpret the diffraction patterns arising in the photoelectron spectra.
Moreover, theoretical predictions have guided and motivated a large number of attosecond experiments, such as the recent observation of ultrafast charge migration in a biomolecule \cite{CalegariScience2014}.
The first theoretical prediction of charge migration in complex systems came from Lorenz S. Cederbaum and J. Zobeley more than 15 years ago \cite{CederbaumCPL1999}, demonstrating that electron correlation can drive purely electron dynamics in a sub-femtosecond time scale, faster than the onset of the nuclear motion.
Over the years, they have investigated this phenomenon in a large number of organic molecules, ranging from small amino acids such as glycine \cite{KuleffJCP2005,KuleffCP2007} to larger systems containing aromatic rings \cite{LunnemannCPL2008}, using \emph{ab initio} approaches that accurately account for electron correlation.
Charge migration has become a hot topic, attracting the interest of a number of researchers and giving rise to exciting theoretical predictions \cite{KuleffJPB2014}.
In 2006, Fran\c{c}oise Remacle and Raphael D. Levine showed that a positive hole could migrate from the amino to the carboxyl terminal of a tetrapeptide in just one and a half femtosecond \cite{RemaclePNAS2006} by using Density Functional Theory.
These predictions have motivated recent experimental observations \cite{BelshawJPCL2012,CalegariScience2014,CalegariIEEE2015} with attosecond pulses generated via HHG as well as the design of nobel experiments based on FELs \cite{CooperFD2014}.
In these theoretical works, the initial wave packet was prepared by removing an electron from a given molecular orbital, creating a hole in the electronic structure of the molecule.
Then, since the ionized state is not a stationary state of the ionic Hamiltonian but a linear superposition of several of them, the hole moves through the molecular skeleton with a velocity that is dictated by the energy spacing between the interfering states.
Although approaches accounting for the interaction with experimentally realistic attosecond pulses have been recently developed \cite{MignoletJPB2014,GolubevPRA2015,MarciniakNatCom2015,CalegariScience2014,CalegariIEEE2015}, the evaluation of electronic wave packets in the continuum is still challenging, especially in the case of complex molecules.\vspace{2 mm}

The so-called fixed-nuclei approximation has been assumed in the works above mentioned.
Although it is a reasonable approach because the nuclear motion usually comes into play in a longer time scale, some influence cannot be completely excluded.
For instance, it has been shown that stretching a few picometers of carbon bonds can occur in a few femtoseconds and can modify the charge dynamics in organic molecules \cite{LunnemannCPL2008}.
Semi-classical approaches have been recently applied to the description of coupled electron-nuclear dynamics in complex molecules.
An implementation of the Ehrenfest method within the Complete Active Space Self Consistent Field (CASSCF) formalism has allowed to explore the damper effect of the nuclear dynamics in the charge migration mechanism \cite{MendiveTapiaJCP2013,VacherJCP2014,VacherTCA2014}.
Another recent work \cite{DespreJPCL2015} has shown that attosecond hole migration in a benzene cation can survive the nuclear motion for more than 10 fs.
However, despite these advances, a full description of charge migration in a complex molecule including the nuclear degrees of freedom in which the photoionization step is property accounted for is yet to be done.\vspace{2 mm}

The theoretical description of electron and nuclear dynamics in which all degrees of freedom are fully correlated is very difficult even for the case of small molecules.
In this context, the multi\hyp{}configurational time\hyp{}dependent Hartree\hyp{}Fock (MCTDHF) method \cite{CaillatPRA2005,KatoJCP2008,KatoJCP2009,AlonPRA2009,NestCPL2009,HaxtonPRA2011,HaxtonPRA2014,HaxtonPRA2015} has been proved to be a powerful tool.
Also, approaches based on Density Functional Theory (DFT) can be very useful because of their good compromise between accuracy and computational effort.
Our group, in collaboration with Piero Decleva in Trieste, is pioneer in the use of the static-exchange DFT method and a more elaborate time-dependent version together with new approaches to include the nuclear motion at the Born-Oppenheimer level in diatomic molecules \cite{CantonPNAS2011,PlesiatPRA2012,PlesiatPCCP2012,PlesiatJPB2012,KukkPRA2013} and for the symmetric stretching mode in molecules with a small number of atoms \cite{ArgentiNJP2012,UedaJCP2013,PlesiatCEJP2013,PatanenJPB2014,AyusoJESRP2014,AyusoJPCA2015}, finding very good agreement with recent experimental results obtained with synchrotron radiation.
These studies have been partially developed during this PhD thesis.
Nevertheless, it should be mentioned that the inclusion of more vibrational degrees of freedom while keeping a reliable description of electron correlation remains a challenge.

\subsubsection*{Outline}

The aim of this work is to investigate light-induced phenomena in simple isolated systems, with especial emphasis in the theoretical description of processes that can be traced and manipulated using synchrotron radiation and ultrafast laser pulses.
In particular, we have focused on the investigation of (1) photoionization of small molecules using methods that account for both nuclear and electronic degrees of freedom, and (2) ultrafast electron dynamics in larger molecules initiated by attosecond XUV pulses.
All the work presented in this PhD thesis has been done in collaboration with Piero Decleva, from Universit\`a degli Studi di Trieste.\vspace{2 mm}

The theoretical methods employed in this work are described in chapters \ref{chapter1}, \ref{chapter2} and \ref{chapter3}.
Chapter \ref{chapter1} reviews the general concepts to treat light-matter interaction.
How to apply these concepts to the special case of molecules interacting with synchrotron radiation and attosecond pulses is presented in chapters \ref{chapter2} and \ref{chapter3}.
In this work we have employed the static-exchange DFT method, developed by Mauro Stener, Piero Decleva and collaborators, briefly described in chapter \ref{chapter2},  to evaluate bound and continuum electronic stationary states.
The nuclear motion has been accounted for within the Born-Oppenheimer approximation in the case of diatomic and small polyatomic molecules.
This has allowed for the computation of vibrational excitations upon ionization and the investigation of non-Franck-Condon effects.
Results are presented in chapters \ref{chapter4} and \ref{chapter5}.
Chapter \ref{chapter4} is devoted to our study of small molecules with synchrotron radiation, while chapter \ref{chapter5} is focused on the investigation of electron dynamics upon photoionization with ultrashort pulses in amino acids.\vspace{2 mm}

We have first investigated the interferences arising in the photoionization of small molecules (BF$_3$, CF$_4$, CO and F$_2$) at high photoelectron energies by analyzing the role of the nuclear motion.
In collaboration with the experimental groups of Edwin Kukk (Turku University), Catalin Miron (Synchrotron SOLEIL), Kiyosi Ueda (Synchrotron SPring-8) and Thomas Darrah Thomas (Oregon State University), we have found evidence of intramolecular scattering occurring in the inner-shell photoionization of CO, CF$_4$ and BF$_3$ imprinted the collective vibrational excitation that accompanies 1s ionization from the C (CO, CF$_4$) or the B (BF$_3$) atom at high photoelectron energies.
The ratios between vibrationally resolved photoionization cross sections ($\nu$-ratios) show pronounced oscillations as a function of the photon energy which are the fingerprint of electron diffraction by the surrounding atomic centers and therefore carry information of the molecular target as well as of the ionization process.
As a proof of principle, we have illustrated how to retrieve the structural information encoded in the $\nu$-ratios by determining the internuclear distances of the CO molecule and of the core-hole species generated upon C 1s ionization.
A different scenario occurs when the electron is emitted not from a well confined region in the molecule but from a delocalized orbital. In these situations, double (triple...) slit-like interferences are expected to arise.
We have investigated this phenomenon in the F$_2$ molecule, where, due to symmetry, the orbitals are delocalized between the two atomic centers.
By analyzing the role of the nuclear motion upon photoabsorption, we have found experimentally measurable evidence of double slit-like interferences in the angle-integrated photoelectron spectra.
All these results are summarized in chapter \ref{chapter4} and the published manuscripts are attached in appendices \ref{appendix1}, \ref{appendix2}, \ref{appendix3}, \ref{appendix4} and \ref{appendix5}.\vspace{2 mm}

We then move to more complex molecules, where we restrict ourselves to frozen nuclei, although applying time-dependent treatments for electron dynamics.
Ultrashort light pulses can create coherent superpositions of electronic states, triggering electron motion at a speed that is determined by the energy spacing between the interfering states.
We have investigated the ultrafast electronic response of large biological systems to attosecond XUV pulses, in collaboration with the experimental groups of Mauro Nisoli (Politecnico di Milano), Luca Poletto (Istituto Nazionale di Fotonica - Consiglio Nazionale delle Ricerche) and Jason Greenwood (Queen's University).
Our study, presented in chapter \ref{chapter5} and in appendices \ref{appendix6} and \ref{appendix7}, includes the amino acids glycine, phenylalanine and tryptophan, with the aim of understanding the influence of the different radicals in the charge migration mechanism. 
For each molecule, we have evaluated the electronic wave packet generated by an attosecond XUV pulse by means of the static-exchange DFT method and time-dependent first-order perturbation theory.
The Fourier analysis of the hole density over different portions of the molecule reveals ultrafast beatings that are in very good agreement with the oscillations found in a XUV/NIR pump-probe experiment where the yield of different fragments is measured as a function of the pump-probe time delay.

%% file: Chapters/Chapter1.tex
\chapter{Light-matter interaction}
\label{chapter1}
\fancyhead[LE]{\fontsize{11pt}{11pt}\selectfont Chapter 1. Light-matter interaction}
\fancyhead[RO]{\fontsize{11pt}{11pt}\selectfont\nouppercase{\rightmark}}

The aim of this chapter is to review the general concepts on light-matter interaction and the expressions that we employ to describe the behavior of atomic and molecular systems in an electromagnetic field.
We focus on the interaction with ``weak'' radiation, which can be accurately described using perturbative approaches.

\section{Time-dependent Schr\"odinger equation}

The evolution of a quantum system is fully determined by the time-dependent Schr\"odinger equation (TDSE) \cite{bookCohenTannoudji}:
\begin{equation}\label{TDSE}
i\hbar\frac{\partial}{\partial t}\Phi(t)=\hat{H}(t)\Phi(t)
\end{equation}
where $\Phi(t)$ is the wave function of the system and $\hat{H}(t)$ is the Hamiltonian operator.
For the sake of simplicity, nor spatial or spin coordinates are explicitly indicated here.
The formal solution of the TDSE is given by
\begin{equation}\label{TDSE2}
\Phi(t)=U(t_0,t) \Phi(t_0)
\end{equation}
where $U(t_0,t)=e^{-\frac{i}{\hbar}\int_{t_0}^{t}\hat{H}(\tau)d\tau}$ is the so-called evolution operator, which propagates the wave function from an initial time $t_0$ to $t$, $\hbar$ being the reduced Planck constant.
Eqs. \ref{TDSE} and \ref{TDSE2} are general and, in principle, applicable to any system.
Our goal is to investigate the behavior of matter upon interaction with bright light.
Specifically, we are interested in exploring molecular targets subject to ultrashort laser pulses and synchrotron radiation.
These light sources are usually intense enough so that one can describe the flux of photons as a continuum variable, i.e., by means of Maxwell equations \cite{bookJackson,bookCohenTannoudji}.
Furthermore, magnetic interactions are usually weak in these contexts and light can be modeled as an oscillating electric field.
Neglecting spin orbit couplings, mass polarization and relativistic effects, the Hamiltonian operator of a set of $N$ charged particles may be written as
\begin{equation}\label{H}
\hat H(t) = \underbrace{\sum_{i=1}^{N} \Bigg[ \frac{\bold{p_i}^2}{2m_i} + \sum_{j=1}^{N} \frac{q_i q_j}{|\bold{r_i}-\bold{r_j}|^2}\Bigg]}_{\hat{H}_0} + \underbrace{\sum_{i=1}^{N} q_i \bold{r}_i \bold{E}(t)}_{\hat{V}(t)}
\end{equation}
where $\bold{r}_i$, $\bold{p}_i$, $m_i$ and $q_i$ are the position, momentum, mass and charge of the $i$-th particle, respectively, and $\bold{E}(t)$ is the electric field of the electromagnetic wave in the dipole approximation \cite{bookJackson,bookCohenTannoudji}, which neglects the spatial dependence of the field across the system.
This is usually a good approach for long and medium wavelength fields and for small atomic systems as long as the wavelength is significantly larger than the dimensions of the system.
For our purposes, it is convenient to split $\hat{H}$ into two parts (see eq. \ref{H}): the field-free Hamiltonian, $\hat{H}_{0}$, and a term accounting for the interaction with the radiation, $\hat{V}(t)$.
Note that the interaction term in eq. \ref{TDSE} has been written in the length gauge.

\subsubsection*{Spectral methods}
Even though eq. \ref{TDSE} does not have an exact analytical solution in most cases, it can be solved, for instance, by defining an initial wave function in a grid of points and then propagate it numerically.
However, grid-based methods are typically expensive since one has to use large numbers of grid points in order to get accurate results.
In quantum chemistry, it is more efficient to use spectral methods, in which the wave function is expanded onto a complete basis set of functions.
Of course, the efficiency and accuracy here depend on the adequate choice of the basis set for the particular problem.
It is usually a good approach to use the eigenstates of the field-free Hamiltonian, which are the solutions of the the eigenvalue problem given by
\begin{equation}\label{TISE}
\hat{H}_0 \phi_k = E_k \phi_k
\end{equation}
where $\phi_k$ are the eigenfunctions of $\hat{H}_0$, which form a complete basis set, and $E_k$ are the corresponding eigenvalues.
The total wave function can be expanded as
\begin{equation}\label{spectralExp}
\Phi(t) = \sum_k c_k(t) \phi_k
\end{equation}
where the time dependence is contained in the spectral coefficients $c_k(t)$, which satisfy
\begin{equation}
c_k(t) = \langle \phi_k | \Phi(t) \rangle
\end{equation}
due to the orthogonality of the basis.
Inserting the spectral expansion of the wave function (eq. \ref{spectralExp}) into the TDSE (eq. \ref{TDSE}), we obtain
\begin{equation}\label{spectral1}
i\hbar\frac{\partial}{\partial t}\sum_k c_k \phi_k = \sum_k c_k [E_k + \hat{V}(t)] \phi_k
\end{equation}
where we have made use of eq. \ref{TISE}. By left-side projecting onto $\langle \phi_n |$ and applying the orthogonality relation $\langle \phi_i | \phi_j \rangle = \delta_{ij}$, eq. \ref{spectral1} reads
\begin{equation}\label{spectral2}
i\hbar\frac{d}{dt} c_n(t) = c_n(t) E_n + \sum_k  c_k \langle\phi_n | \hat{V}(t) | \phi_k \rangle
\end{equation}
We have obtained a set of coupled equations that describe the time-evolution of the spectral coefficients.
A more compact version of eq. \ref{spectral2} can be obtained by performing the change of variables
\begin{equation}\label{coefPhase}
c_n(t) = \tilde{c}_n(t) e^{-iE_nt/\hbar}
\end{equation} 
where $\tilde{c}_n(t)$ are the coefficients in the interaction picture \cite{bookCohenTannoudji}, which are equivalent to those in the Schr\"odinger picture $c_n(t)$ except for the corresponding stationary phases.
Applying eq. \ref{coefPhase} to the coefficients in eq. \ref{spectral2} and multiplying both sides of eq. \ref{spectral2} by $e^{iE_nt/\hbar}$, we obtain
\begin{equation}\label{spectral3}
i\hbar \frac{d}{dt}\tilde{c}_n(t) =  \sum_k  \tilde{c}_k(t) e^{{i\omega_{nk}t}} \langle\phi_n | \hat{V}(t) | \phi_k \rangle
\end{equation}
where we have introduced the Bohr angular frequency $\omega_{nk} = \frac{En - E_k}{\hbar}$.
This set of coupled equations is completely general and rigorously equivalent to eq. \ref{TDSE}.
The coupling between different states arises from the existence of the external potential $\hat{V}(t)$, which relates the evolution of $\tilde{c}_n(t)$ to that of all the other coefficients.
In general, eq. \ref{spectral3} can be solved numerically by breaking the time domain into small steps $\{t_1, t_2, ...\}$ and the set $\{\tilde{c}_n(t_j)\}_n$ is obtained from $\{\tilde{c}_n(t_{j-1})\}_n$ through iterative procedures \cite{bookAtomsInIntenseLaserFields,bookAttosecondAndXUVPhysics,BandraukJTCC2013}.
However, this approach might become computationally expensive in some situations.
If the external potential $\hat{V}(t)$ is weak, the time-evolution of the wave function can be evaluated more efficiently making use of perturbation theory, as we explain in the next section.

\section{Time-dependent perturbation theory}\label{section_TDPT}

Perturbation theory provides a useful approach to solve the TDSE when the field applied to the system is weak and therefore $\hat{V}(t)$ can be treated as a perturbation.
Under this assumption, the set of coefficients $\tilde{c}_n(t)$ hardly vary in time and their zero-th order solution is given by their initial values:
\begin{equation}\label{TDPT0}
\tilde{c}_n^{(0)}(t) \simeq \tilde{c}_n(0)
\end{equation}
Solutions of higher order ($r>0$) can be evaluated using the recurrence relation
\begin{equation}\label{TDPTr}
i\hbar \frac{d}{dt}\tilde{c}_n^{(r)}(t) =  \sum_k  \tilde{c}_k^{(r-1)}(t) e^{{i\omega_{nk}t}} \langle\phi_n | \hat{V}(t) | \phi_k \rangle 
\end{equation}
which enables to obtain the $r$-th order solution from the $(r-1)$-th order one.
We are interested in situations in which non linear processes are negligible and can thus be accurately described using \textbf{first-order perturbation theory}, which approximates the exact wave function to its first-order solution.
If the system is assumed to be in the ground state at $t=0$, that is, $\tilde{c}_n(0) = \delta_{0n}$, the zero-th order solution is given by $\tilde{c}_n^{(0)}(t) = \delta_{0n}$.
Inserting it into the right side of eq. \ref{TDPTr}, we can evaluate the first-order solution
\begin{equation}
i\hbar \frac{d}{dt}\tilde{c}_n^{(1)}(t) \simeq e^{i\omega_{n0}t} \langle\phi_n | \hat{V}(t) | \phi_0 \rangle
\end{equation}
By integrating in time and making use of the initial condition $\tilde{c}_n(0) = \delta_{0n}$ we obtain:
\begin{equation}
\tilde{c}_n^{(1)}(t) \simeq  \delta_{0n} -\frac{i}{\hbar} \int_0^t e^{i\omega_{n0}\tau} \langle\phi_n | \hat{V}(\tau) | \phi_0 \rangle d\tau
\end{equation}
In our particular case $\hat{V}(t)=\boldsymbol\mu \bold{E}(t)$, where $\mu = \sum_n q_n \bold{r}_n$ is the dipole moment operator.
Then, for $n\ne 0$, we have
\begin{equation}\label{coefFOPT}
\tilde{c}_n^{(1)}(t)  = -\frac{i}{\hbar} \langle \phi_n| \boldsymbol\mu_{\boldsymbol\epsilon}|\phi_0\rangle \int_{0}^{t} e^{i\omega_{n0}\tau} \text{E}(\tau) d\tau
\end{equation}
where $\boldsymbol\mu_{\boldsymbol\epsilon}= \hat{\boldsymbol\epsilon} \boldsymbol\mu$ is the component of the dipole operator along $\hat{\boldsymbol\epsilon}$, the polarization direction of the field and $\bold{E}(t)=\text{E}(t)\hat{\boldsymbol\epsilon}$. 
Eq. \ref{coefFOPT} can provide accurate values of the time-dependent coefficients upon interaction with an ultrashort laser pulse provided only linear effects come into play.
We can retrieve the set of coefficients in the Schr\"odinger picture using using eq. \ref{coefPhase}.
The corresponding transition probabilities are given by the square of the spectral amplitudes:
\begin{equation}\label{transProb}
P_{n\leftarrow 0}(t) = |c_n(t)|^{2} = |\tilde{c}_n(t)|^{2}
\end{equation}
where one can use $c_n(t)$ or $\tilde{c}_n(t)$ since they are equal except for a stationary phase.

\subsubsection*{The special case of a sinusoidal perturbation}
Let us consider the case of monochromatic light in the dipole approximation, i.e., 
\begin{equation}\label{Esin}
\bold{E}(t) = \text{E}_0 \sin{(\omega t)} \hat{\boldsymbol\epsilon} = \frac{\text{E}_0}{2} \Big[ e^{i\omega t} - e^{-i\omega t}\Big] \hat{\boldsymbol\epsilon}
\end{equation}
where $\omega$ is the frequency of the radiation.
This is a reasonable approach to model an experiment with synchrotron radiation, where the photon energy is well defined \cite{BilderbackJPB2005}.
Inserting eq. \ref{Esin} into eq. \ref{coefFOPT}, we obtain:
\begin{align}
\tilde{c}_n^{(1)}(t) &= -\frac{i}{2\hbar} \langle \phi_n| \boldsymbol\mu_{\boldsymbol\epsilon}|\phi_0\rangle  \int_{0}^{t} \Bigg[ e^{i(\omega_{n0}+\omega)\tau} - e^{i(\omega_{n0}-\omega)\tau}\Bigg] d\tau \\
&= -\frac{i}{2\hbar} \langle \phi_n| \boldsymbol\mu_{\boldsymbol\epsilon}|\phi_0\rangle  \Bigg[ \frac{1-e^{i(\omega_{n0}+\omega)t}}{\omega_{n0}+\omega} - \frac{1-e^{i(\omega_{n0}-\omega)t}}{\omega_{n0}-\omega} \Bigg] 
\end{align}
By making use of eq. \ref{transProb} we can evaluate the transition probability:
\begin{equation}\label{transProb2}
P_{n\leftarrow 0}(t)=|\tilde{c}_n|^{2}(t) = \frac{^2}{4\hbar^2} |  \langle \phi_n| \boldsymbol\mu_{\boldsymbol\epsilon}|\phi_0\rangle |^2 \Bigg| \frac{1-e^{i(\omega_{n0}+\omega)t}}{\omega_{n0}+\omega} - \frac{1-e^{i(\omega_{n0}-\omega)t}}{\omega_{n0}-\omega} \Bigg|^2 
\end{equation}
For a fixed value of $t$, the transition probability is a function of $\omega$ having two pronounced maxima for $\omega=\omega_{n0}$ and $\omega=-\omega_{n0}$ due to the two terms inside the bracket.
The first term, which maximizes for $\omega=-\omega_{n0}$, accounts for transitions from the initial to lower energy states occurring through induced photoemission.
Here we seek to describe excitations that take place upon photoabsorption from the ground state.
These are accounted for in the second term, which maximizes for $\omega=\omega_{n0}$.
Removing the induced photoemission term in eq. \ref{transProb2} and making use of the identity $e^{i\alpha}-1 = 2i e^{i\alpha/2}\sin{(\alpha/2)}$, we obtain:
\begin{equation}
P_{n\leftarrow 0}(t) = \frac{^2}{\hbar^2} |  \langle \phi_n| \boldsymbol\mu_{\boldsymbol\epsilon}|\phi_0\rangle |^2 \frac{\sin^2{\big([\omega_{n0}-\omega]t/2\big)}}{(\omega_{n0}-\omega)^2} 
\end{equation}
We are interested in finding the transition probability upon a long-time interaction.
In the limit $t\rightarrow\infty$, the function $\frac{\sin^2(\alpha t/2)}{\alpha^2}$ can be approximated by $\frac{\pi t}{2}\delta(\alpha)$.
Then, in the long-time limit, we have:
\begin{equation}
P_{n\leftarrow 0}(t) = \frac{\pi ^2 t}{2\hbar^2} |  \langle \phi_n| \boldsymbol\mu_{\boldsymbol\epsilon}|\phi_0\rangle |^2 \delta(\omega_{n0}-\omega)
\end{equation}
Note that this limit corresponds to the case of perfectly monochromatic light, where the photon energy is well defined and given by $\hbar\omega$.
Therefore, a transition from the ground to an excited state $n$ will only occur if $\omega=\omega_{n0}$.
The transition rate, i.e., the transition probability per unit of time can be obtained by integrating over a range of frequencies $(\omega_1,\omega_2)$ containing $\omega_{n0}$ and derivating with respect to time:
\begin{equation}\label{transRate}
\Gamma_{n\leftarrow 0} = \frac{d}{dt} \int_{\omega_1}^{\omega_2} P_{n\leftarrow 0}(t) d\omega = \frac{\pi ^2}{2\hbar^2} |  \langle \phi_n| \boldsymbol\mu_{\boldsymbol\epsilon}|\phi_0\rangle |^2 
\end{equation}
Although transition rates are experimentally measurable quantities, in practice, it is more convenient to measure photoionization cross sections since they are independent of the experimental conditions, as we explain in the next section.

\section{Photoionization cross section}\label{section_CS}

The cross section is defined as the hypothetical surface of effective interaction between a flux of particles and their targets.
In the particular case of photoionization, it refers to the probability of an electron to be emitted from the target upon interaction with the field.
The cross section corresponding to a transition from the ground state $\Phi_{0}$ a to a final $\Phi_{n}$ state is given by \cite{bookCohenTannoudji}:
\begin{equation}\label{CS}
\sigma = \frac{\Gamma_{n \leftarrow 0}}{F}
\end{equation}
where $F$ is the flux of photons per unit of area and time, which is related to the amplitude of the electric field $\text{E}_{0}$ according to
\begin{equation}\label{flux}
F=\frac{\text{E}_{0}^{2}c}{8\pi \hbar \omega}
\end{equation}
and $c$ is the speed of light.
Inserting \ref{transRate} and \ref{flux} into \ref{CS}, we obtain the photoionization cross section for a given orientation of the system with respect to the field:
\begin{equation}\label{CSphotoionization}
\sigma_{\boldsymbol\epsilon}=\frac{4\pi^{2}\omega}{\hbar c}  |\langle \phi_n| \boldsymbol\mu_{\boldsymbol\epsilon}|\phi_0\rangle |^2 
\end{equation}
As can be seen in eq. \ref{CSphotoionization}, the cross section does not depend on the parameters of the field.
For this reason, it is a very useful quantity to compare results obtained under different experimental conditions.

\subsubsection*{Randomly oriented targets}

In the case of randomly oriented molecules, the total cross section can be retrieved by averaging incoherently over three orthogonal directions of $\boldsymbol\epsilon$, let us call them $\bold{x}$, $\bold{y}$ and $\bold{z}$:
\begin{equation}\label{CSrandom}
\sigma=\frac{ \sigma_{\bold{x}} + \sigma_{\bold{y}} + \sigma_{\bold{z}}}{3}
\end{equation}
Eq. \ref{CSrandom} allows to reproduce experimental results in which targets without spherical symmetry are are not aligned with the field.

%% file: Chapters/Chapter2.tex
\chapter{Molecular structure}
\label{chapter2}
\fancyhead[LE]{\fontsize{11pt}{11pt}\selectfont Chapter 2. Molecular structure}
\fancyhead[RO]{\fontsize{11pt}{11pt}\selectfont \nouppercase{\rightmark}}

The present chapter describes the methodology employed to include the electronic and nuclear degrees of freedom in the theoretical description of molecular photoionization.
Ionization of molecules is more complex than in the case of atoms because of the lack of spherical symmetry and due to the added degrees freedom of the nuclear motion.
Thus, one has to assume certain approximations that simplify the full problem which, as the number of degrees of freedom increases, becomes  computationally intractable.
In this work we have evaluated the electronic structure of the molecules we have investigated using methods based on the Density Functional Theory (DFT), which can account for electronic exchange and correlation effects in medium and large size systems using reasonable computational resources.
The nuclear motion has been included at the Born-Oppenheimer level in the case of diatomic and small polyatomic molecules, allowing to evaluate vibrationally resolved photoionization cross sections.

\section{The molecular Hamiltonian}

The Hamiltonian operator representing the energy of the electrons and the nuclei in a molecule, the field-free molecular Hamiltonian, can be written as \cite{bookSzabo}:
\begin{equation}\label{Hmolecular}
\hat{H} = \hat{T}_{e} + \hat{T}_{N} + \hat{V}_{ee} + \hat{V}_{eN} + \hat{V}_{NN}
\end{equation}
where:
\begin{align*}\label{Hmolecular2}
&\circ\quad T_{e} = -\frac{1}{2}\sum_{i=1}^{N}\frac{1}{m_e}\nabla_{i}^{2} \text{ is the kinetic energy of the $N$ electrons,}\\
&\circ\quad \hat{T}_{N} = -\frac{1}{2}\sum_{i=1}^{M}\frac{1}{M_{\alpha}}\nabla_{\alpha}^{2} \text{ is the kinetic energy of the $M$ nuclei,}\\
&\circ\quad \hat{V}_{ee} = \sum_{i=1}^{N}\sum_{j>i}^{N}\frac{e^{2}}{|\bold{r}_{i}-\bold{r}_{j}|} \text{ is the electrostatic repulsion between electrons,}\\
&\circ\quad \hat{V}_{eN} = -\sum_{i=1}^{N}\sum_{\alpha=1}^{M}\frac{Z_{\alpha}e^{2}}{|r_{i}-\bold{R}_{\alpha}|}\text{ is the attraction between electrons and nuclei,} \\
&\circ\quad \hat{V}_{NN} = \sum_{\alpha=1}^{M}\sum_{\beta>\alpha}^{M}\frac{Z_{\alpha}Z_{\beta}e^{2}}{|\bold{R}_{\alpha}-\bold{R}_{\beta}|} \text{ is the repulsion energy term between nuclei,}
\end{align*}
$\bold{r}_{i}$ and $\bold{R}_{\alpha}$ stand for the coordinates of the electron $i$ and nuclei $\alpha$, respectively, $m_e$ and $e$ are the mass and the absolute value of the charge of the electron, and $M_\alpha$ and $Z_\alpha$ are the mass and the atomic number of the nuclei $\alpha$.
In order to find the eigenstates of $\hat{H}$, which constitute the set of stationary solutions of the molecular system, one can take advantage of the fact that since nuclei are more massive than the light electrons, their motion is slower, as we explain in the following.

\subsection{The Born-Oppenheimer approximation}
\label{section_BO}

Since the electromagnetic forces acting on electrons and nuclei have similar intensity, one might assume their momenta to be of the same magnitude.
Then, as the nuclei are significantly heavier, they must accordingly have much smaller velocities.
Based on this idea, Max Born and J. Robert Oppenheimer proposed a way to decouple electron and nuclear dynamics by splitting the total wave function into two parts \cite{Born1927}.
Within the Born-Oppenheimer approximation, the stationary states of the full-system can written as product of an electronic stationary state $\Psi_{n}(\bold{\bar{x}},\bold{\bar{R}})$, depending on both the electronic and the nuclear coordinates, and a nuclear stationary state $\chi_{n\nu}(\bold{\bar{R}})$, which only depends on the nuclear degrees of freedom:
\begin{equation}\label{totalWF}
\Phi_{n\nu}(\bold{\bar{x}},\bold{\bar{R}})=\Psi_{n}(\bold{\bar{x}},\bold{\bar{R}})\chi_{n\nu}(\bold{\bar{R}})
\end{equation}
where $n$ and $\nu$ are indexes (in general, sets of indexes) over the electronic and nuclear eigenstates labeling the vibronic state $\Phi_{n\nu}(\bold{\bar{x}},\bold{\bar{R}})$, where $\bar{\bold{x}}=(\bold{x}_1,...,\bold{x}_N)$ is a vector containing the spin and spatial coordinates of all electrons and $\bar{\bold{R}}=(\bold{R}_1,...,\bold{R}_M)$ contains all nuclear spatial coordinates (nuclear spin coordinates have been dropped).
Electronic stationary states satisfy the electronic time\hyp{}independent Schr\"odinger equation:
\begin{equation}\label{Eevp}
\Big[ \underbrace{ \hat{T}_{e} +\hat{V}_{ee}+\hat{V}_{eN}+\hat{V}_{NN} }_{\hat{H}_e} \Big] \Psi_{n}(\bold{\bar{x}},\bold{\bar{R}}) = E_{n}(\bold{\bar{R}})     \Psi_{n}(\bold{\bar{x}},\bold{\bar{R}})
\end{equation}
where $\hat{H}_e$ is the electronic Hamiltonian and $E_{n}(\bold{\bar{R}})$ is the energy of the electronic state $n$, which depends on the nuclear coordinates $\bold{\bar{R}}$.
Eq. \ref{Eevp} can be solved parametrically in a grid of nuclear geometries.
By doing so, one obtains the potential energy surfaces $E_{n}(\bold{\bar{R}})$ in which the nuclei move.
Note that, in each electronic calculation, the nuclear repulsion energy term ($\hat{V}_{NN}$ in eq. \ref{Eevp}) is just a constant value and thus its only effect is increasing the electronic eigenvalue.
The nuclear stationary states $\chi_{n\nu}$ associated to a given electronic state $n$ can be obtained by solving nuclear time-independent Schr\"odinger equation:
\begin{equation}\label{Nevp}
\Big[\underbrace{ T_N + E_{n}(\bold{\bar{R}})}_{\hat{H}_N}\Big] \chi_{n\nu}(\bold{\bar{R}}) = E_{n\nu} \chi_{n\nu}(\bold{\bar{R}})
\end{equation}
where $E_{n\nu}$ is energy of the vibronic state defined by the quantum numbers $n$ and $\nu$.
The Born-Oppenheimer approximation assumes that $\Phi_{n\nu}(\bold{\bar{x}},\bold{\bar{R}})$ varies very smoothly with $\bold{\bar{R}}$ and therefore that the electrons rearrange instantaneously as the nuclei move.
This assumption is valid as long as the energy spacing between electronic states, i.e., $E_n(\bold{\bar{R}})-E_{n-1}(\bold{\bar{R}})$, is sufficiently large and, in a photoionization process, as long as the photoelectron is not emitted very slowly, that is, with very low kinetic energy.
The Born-Oppenheimer approximation provides a powerful tool for the accurate evaluation of vibronic stationary states of diatomic and small polyatomic molecules, and also of larger systems in situations in which reduced-dimensionality models are applicable.

\subsubsection*{Potential energy curves}

As already indicated, the eigenvalues of eq. \ref{Eevp}, when solved in a grid of molecular geometries, constitute a set of potential energy surfaces (PESs) or curves (PECs) in the monodimensional case.
In this work we have employed the static-exchange DFT method and also the more elaborate time-dependent DFT to evaluate the electronic stationary states of the molecules we have investigated, as we explain in section \ref{section_electronic}.
Although these methods can accurately describe transitions to the electronic continuum, essential in order to describe photoionization, the energy values they provide might not be accurate enough in some situations.
In general, \emph{ab initio} multi-reference methods can produce accurate PESs of medium-size systems \cite{bookSzabo}.
In the case of a core-hole species, the situation is more challenging since one needs to develop specific approaches to avoid the variational collapse of the wave function while keeping a good description of electron correlation \cite{JensenJCP1987,RochaJCP2011}.
We have investigated the role of the nuclear motion in the photoionization of diatomic (CO, F$_2$) and small polyatomic (BF$_3$, CF$_4$) molecules under conditions in which only one vibrational mode is active.
In these situations, the harmonic and the Morse approximations provide a good alternative for the evaluation of the PECs:

\begin{itemize}
\item[$\circ$] \textbf{The harmonic oscillator} models an ideal system that when taken away from the equilibrium position $R_{eq}$ experiences a restoring force that is proportional to the extent of the displacement.
It allows to write the potential energy as
\begin{equation}\label{harmonicPotential}
E(R) = \frac{1}{2}m\omega^{2}(R-R_{eq})^{2}
\end{equation}
where $m$ is the mass of the system, $\omega$ is the angular frequency and $R$ is the nuclear coordinate.
This simple formula can provide a good representation of the PEC around the equilibrium geometry, but it cannot describe molecular dissociation since it does not take into account the anharmonicity of the chemical bonds.
Consequently, it allows to evaluate low-energy stationary eigenstates with accuracy \cite{UedaJCP2013,PatanenJPB2014}, but it should not be used to describe the high-energy region.

\item[$\circ$] \textbf{The Morse potential} \cite{MorsePR1929} provides a valid description of the PEC in a larger range of internuclear distances in terms of a simple analytical formula that takes into account the anharmonicity of the chemical bonds:
\begin{equation}\label{morse}
V(R)=V(R_{eq})+D_{e}\Big[1-e^{\alpha(R-R_{eq})}\Big]^2
\end{equation}
where $R_{eq}$ is the equilibrium distance, $D_{e}$ is the depth of the potential energy well and $\alpha$ is a parameter controlling its width.
The Morse parameters are related to the usual spectroscopic ones (the oscillator strength, $\omega_{e}$, and the anharmonicity parameter, $\omega_{e}x_{e}$) by the formulas
\begin{align}
D_{e}=\frac{\omega_{e}^2}{4\omega_{e}x_{e}}&-\frac{\omega_{e}x_{e}}{4} \simeq \frac{\omega_{e}^2}{4\omega_{e}x_{e}} \label{eq:morse_De}\\
\alpha&=\sqrt{\frac{k_{e}}{2D_{e}}}  \label{eq:morse_alpha}
\end{align}
Although eq. \ref{morse} was designed for studying diatomic molecules, it can still provide accurate results for the totally-symmetric stretching mode of small polyatomics \cite{PlesiatPRA2012,UedaJCP2013,PlesiatCEJP2013,AyusoJPCA2015}.

\end{itemize}

\subsubsection*{The fixed-nuclei approximation}

Some purely electronic processes can occur before the onset of the nuclear motion and can thus be described in the framework of the fixed-nuclei approximation (FNA), in which the nuclei are assumed to remain frozen at their equilibrium positions ($\bold{\bar{R}}=\bold{\bar{R}_{\text{eq}}}$).
Within the FNA, electronic stationary states satisfy:
\begin{equation}\label{EevpFNA}
\Big[ \underbrace{ \hat{T}_{e} +\hat{V}_{ee}+\hat{V}_{eN}+\hat{V}_{NN} }_{\hat{H}_{e}} \Big] \Psi_{n}(\bold{\bar{x}},\bold{\bar{R}_{\text{eq}}}) = E_{n}(\bold{\bar{R}_{\text{eq}}})     \Psi_{n}(\bold{\bar{x}},\bold{\bar{R}_{\text{eq}}})
\end{equation}
The fixed-nuclei approximation can provide accurate values of total photoionization cross sections (see, for instance \cite{LucchesePRA1982,FronzoniPCCP1999,StenerJCP2000}) when the variation of the electronic structure with the internuclear distances is smooth around the Franck-Condon region.
The FNA has also been successfully applied to time\hyp{}dependent problems in large systems.
For instance, most theoretical work on charge migration \cite{CederbaumCPL1999,KuleffJCP2005,RemaclePNAS2006,KuleffCP2007,MignoletJPB2014,KuleffJPB2014} relies on the validity of the FNA to propagate electronic wave packets. 
Of course, its applicability depends on the characteristics of each particular problem and, in general, the nuclear motion is expected to play a role in the femtosecond time domain.

\section{Evaluation of electronic states}
\label{section_electronic}

As we discussed, the electronic eigenvalue problem given in eq. \ref{Eevp} does not have an exact analytical solution in most cases.
A widely used approach is given by the Hartree$-$Fock (HF) method, that can provide a first approximation to the ``exact'' ground state solution in terms of a Slater determinant constructed from HF molecular orbitals, which are obtained within the mean field approximation through a self consistent field procedure.
However, it is well known that HF solutions are usually rather poor since the mean field approximation cannot describe electron correlation properly \cite{bookSzabo}.
Post-HF methods manage this problem by expanding the total wave function as a linear combination of Slater determinants (electronic configurations), being able to yield accurate solutions for the ground and for excited states of few-electron systems.
Yet, the number of configurations one might need to include in the expansion to reach the desired accuracy can make these methods extremely costly.
In this sense, DFT constitutes a useful alternative to \emph{ab initio} methods, providing an excellent compromise between accuracy and computational effort for medium and large size systems \cite{bookLevine}.

\subsection{Density functional theory}

Density functional theory (DFT) is widely used in physics, chemistry and materials science to investigate the ground state electronic structure of atoms, molecules and condensed phases.
According to DFT, the energy (or any other observable) of a many-electron system in the ground state can be determined by using functionals which solely depend on the electron density.
The most essential concepts of the method are given here; for a deeper insight, see, for instance \cite{bookLevine}, \cite{bookDreizler} or \cite{bookCramer}.\vspace{2 mm}

DFT is supported on the theorems proposed by Pierre Hohenberg and Walter Kohn in 1964 \cite{HohenbergPRB1964}, namely:
\begin{enumerate}
\item ``Any observable of a stationary non-degenerate ground state can be calculated, exactly in theory, from the electron density of the ground state''. 
\item ``The electron density of a non-degenerate ground state can be calculated, exactly in theory, determining the density that minimizes the energy of the ground state''.
\end{enumerate}
The use of the electron density $\rho(\bold{r})$ instead of the wave function $\Psi(\bold{x}_1,...,\bold{x}_N)$ is the foundation of DFT. 
Both entities are related through the equation:
\begin{equation}\label{elecDens}
 \rho(\bold{r})=N\int  \ldots \int  | \Psi(\bold{x},\bold{x}_2\ldots d\bold{x}_N)|^2 ds d\bold{x}_2\ldots d\bold{x}_N
\end{equation}
where $\bold{x}_{i}=\bold{r}_{i}s_{i}$ gathers the spatial $\bold{r}_{i}$ and spin $s_{i}$ coordinates of the $i-$th electron \footnote{For the shake of simplicity, the parametric dependence on the nuclear coordinates has been dropped.}.
In 1965, Walter Kohn and Lu Jeu Sham provided a systematical approach to evaluate the ground state electron density of a many-body system by introducing the so-called Kohn-Sham equation \cite{KohnPRA1965}.

\subsubsection*{The Kohn-Sham equation}
The Kohn-Sham equation is the time-independent Schr\"odinger equation of a fictitious system of non-interacting particles that generates the same density as a given system of interacting particles.
It can be written as
\begin{equation}\label{KSeq}
\Big[ -\frac{\hbar^{2}}{2m_e}\nabla^{2} + V_{\text{eff}}(\bold{r}) \Big] \phi_{i}(\bold{x}) = \epsilon_{i} \phi_{i}(\bold{x})
\end{equation}
where $\phi_i$ are the so called Kohn-Sham orbitals, $\epsilon_{i}$ are the corresponding energies and $V_{\text{eff}}$ the fictitious effective potential in which the non-interacting particles move:
\begin{equation}
V_{\text{eff}}(\bold{r})=-\sum_{\alpha=1}^{M}\frac{e^2 Z_{\alpha}}{|\bold{r}-\bold{R}_{\alpha}|}+V_{H}[\rho(\bold{r})]+V_{XC}[\rho(\bold{r})]
\end{equation}
where $V_{H}$ is the Hartree (Coulomb) potential:
\begin{equation}
V_{H}[\rho(\bold{r}_{i})] = e^2 \int \frac{\rho(\bold{r}')}{|\bold{r}-\bold{r}'|} d\bold{r}'
\end{equation}
and $V_{XC}$ is the exchange-correlation potential:
\begin{equation}
V_{XC}[\rho(\bold{r})]=\frac{\delta E_{XC}}{\delta \rho}
\end{equation}
$E_{XC}$ is the exchange-correlation energy.
If the exact forms of $E_{XC}$ and $V_{XC}$ where known, the Kohn-Sham strategy would provide the exact ground state energy.
Unfortunately, this is not the case and the exchange-correlation energy (potential) needs to be approximated through empirical formulations.
The central goal of modern DFT is finding better approximations to these two quantities.
As the particles of the Kohn-Sham system are non-interacting fermions, the ground state wave function can be written as a Slater determinant of the lowest energy solutions of eq. \ref{KSeq}:
\begin{equation}
\Psi(\bold{x}_1,\bold{x}_2,...,\bold{x}_n)= \frac{1}{\sqrt{N!}} \begin{vmatrix}
  \phi_{1}(\bold{x}_1) & \phi_{2}(\bold{x}_1) & \cdots & \phi_{N}(\bold{x}_1) \\
  \phi_{1}(\bold{x}_2) & \phi_{2}(\bold{x}_2) & \cdots & \phi_{N}(\bold{x}_2) \\
  \vdots               &              \vdots  & \ddots & \vdots               \\
  \phi_{1}(\bold{x}_N) & \phi_{2}(\bold{x}_N) & \cdots & \phi_{N}(\bold{x}_N) \\
 \end{vmatrix}
\end{equation}
By using eq. \ref{elecDens}, we can evaluate the ground state electron density:
\begin{equation}\label{GSdens}
\rho_{0}(\bold{r}) = \sum_{i=1}^{N} |\psi_{i}(\bold{r})|^2
\end{equation}
where $\psi_i(\bold{r})$ is the spatial part of the spin orbital $\phi_i(\bold{x})$, that is, $\phi_i(\bold{x})=\psi_i(\bold{r})\alpha(s)$ or $\phi_i(\bold{x})=\psi_i(\bold{r})\beta(s)$.
In practice, since the Kohn-Sham Hamiltonian depends on the Kohn-Sham orbitals (solutions of the eigenvalue problem) through the electron density, they are numerically found by performing a self-consistent field procedure.

\subsection{Static-exchange DFT}\label{section_staticExchangeDFT}

Standard DFT methods can accurately represent the electronic ground state of many-electron systems.
In order to describe photoionization processes one also needs to describe the electronic continuum.
In this work we have employed the static-exchange DFT method \cite{StenerIJQC1995,StenerTCA1999,StenerJPB2000,BachauRPP2001,ToffoliCP2002}, developed by Mauro Stener, Piero Decleva and collaborators, to evaluate transitions to continuum states.
The method makes use of the Kohn-Sham formalism to describe bound states and of the Galerkin approach to evaluate photoelectron wave functions in the field of the corresponding Kohn-Sham density.
Over the last decades, this methodology has provided accurate values of photoionization cross sections of small molecules as well as of medium and large size systems within the fixed-nuclei approximation \cite{StenerTCA1999,StenerJCP2006,ToffoliJCP2006,KoricaSC2010,CarrollJCP2013,KushawahaPNAS2013}, from small diatomic molecules such as N$_2$ to fullerenes.
More recently, the method was extended in collaboration with the group of Fernando Mart\'in to include the nuclear degrees of freedom, successfully evaluating vibrationally resolved cross sections of diatomic \cite{CantonPNAS2011,PlesiatPCCP2012,PlesiatJPB2012,KukkPRA2013,AyusoJESRP2014} and small polyatomic \cite{ArgentiNJP2012,PlesiatPRA2012,UedaJCP2013,PlesiatCEJP2013,PatanenJPB2014,AyusoJPCA2015} molecules, providing results which are in good agreement with experimental data.
In this section we explain the most relevant characteristics of the method. 

\subsubsection*{Electronic states within the static-exchange approximation}

The static-exchange DFT method makes use of single Slater determinants to define bound and excited (continuum) electronic states, ensuring that the Pauli exclusion principle is fulfilled.
The ground state wave function may be written as
\begin{equation}\label{elecGS}
\Psi_0(\bold{x}_1,\bold{x}_2,...,\bold{x}_n) = \Big| \phi_{1} \enskip \phi_{2} \ldots \phi_{N} \Big|
\end{equation}
where $N$ is the number of electrons, $\phi_i(\bold{x})=\psi_{i}(\bold{r})\alpha(s)$ if $i$ is odd and $\phi_i(\bold{x})=\psi_{i}(\bold{r})\beta(s)$ if $i$ is even.
For a closed-shell system, $\psi_{1}(\bold{r})=\psi_{2}(\bold{r})$, ... , $\psi_{N-1}(\bold{r})=\psi_{N}(\bold{r})$.
Continuum states are defined by promoting one electron from a bound spin orbital $\phi_{\alpha}$ to a continuum orbital $\phi_{\varepsilon lh}$ with kinetic energy $\varepsilon$ and angular quantum numbers $l$ and $h$, and can be written as
\begin{equation}\label{elecCont}
\Psi_{\alpha\varepsilon lh}(\bold{x}_1,\bold{x}_2,...,\bold{x}_n) = \Big| \phi_{1} \enskip \phi_{2} \ldots \phi_{\alpha -1} \enskip \phi_{\varepsilon lh} \enskip \phi_{\alpha +1} \ldots \phi_{N} \Big|
\end{equation}
Bound and continuum orbitals are expanded in a multicentric basis set of B-splines, as we explain as follows.

\subsubsection*{Multicentric B-spline basis set}

Traditional basis sets make use of Gaussian or Slater type orbital functions, which provide fast convergence for the lowest bound states with a reduced number of basis functions \cite{bookSzabo}.
However, these expansions are not adequate for the description of the rapidly oscillating continuum states, since numerical linear dependences rapidly come up as the basis set is increased due to the large overlap between functions with different centers.
In this context, basis sets of B-spline functions, which are piecewise polynomials, constitute a very powerful tool \cite{BachauRPP2001}.
B-spline functions are very flexible and due to its local nature they can describe accurately both bound and continuum orbitals without running into numerical dependencies \cite{BachauRPP2001}.
The present method evaluates bound and continuum orbitals in a multicentric basis set of B-splines, using symmetry-adapted \cite{BurkeJPB1972} linear combinations of real spherical harmonics with origin over different positions in the molecule:
\begin{itemize}
\item[$\circ$] A large one-center expansion (OCE) over the center of mass provides an accurate description of the long-range behavior of the continuum states.
\item[$\circ$] Small expansions, called off-centers (OC), located over the non-equivalent nuclei, complement the OCE.
They improve dramatically the convergence of the calculation, allowing to reduce the angular expansion in the OCE, since they can effectively describe the Kato cusps \cite{KatoCPAM1957} at the nuclear positions.
\end{itemize}

In the case of symmetric molecules, a large amount of computational effort can be saved by making use of point group symmetry and dividing the three-dimensional space into equivalent regions.
The basis set elements may be written as
\begin{equation}\label{basisExpansion}
\xi_{nlh\lambda\mu}^{p}=
\sum_{q\in \Lambda_{p}} \frac{1}{r_{q}} B^{\kappa}_{n}(r_{q})
\underbrace{\sum_{m} b_{mlh\lambda\mu}^{q} Y_{lm}(\theta_{q},\varphi_{q})}_{X_{lh\lambda\mu}^{p}(\theta_{q},\varphi_{q})}
\end{equation}
where $\Lambda_{p}$ represents a shell of equivalent centers ($p=0$ refers to the OCE), $q$ runs over the centers in the shell, $n$ is an index over the B-spline functions $B^{\kappa}_{n}$, whose order is $\kappa=10$,
$\lambda\mu$ are the indexes of the irreducible representation (see \cite{ToffoliCP2002}),
$h$ runs over the linearly independent angular functions, which are constructed as linear combinations of real spherical harmonics associated to a fixed angular quantum number $l$,
and the coefficients $b_{mlh\lambda\mu}^{q}$ are determined by symmetry \cite{BurkeJPB1972}, defining the so called symmetry-adapted spherical harmonics $X_{lh\lambda\mu}^{p}$, which are invariant under the symmetry operations of a given point group.
For instance, in the case of BF$_3$ (in the ground state equilibrium geometry), $p=1$ would represent the shell of equivalent F atoms ($q=1,2,3$ since there are 3 equivalent F atoms) and no more OC expansions would be required since the OCE would be located at the B atom (center of mass).\vspace{2 mm}

In each center $q$, the B-spline expansion reaches a maximum value $R^{p}_{\text{max}}$, which can be different for non-equivalent centers (different value of $p$, see eq. \ref{basisExpansion}).
A large vale of $R^{0}_{\text{max}}$ is required in the OCE in order to provide a good description of the oscillatory behavior of the continuum states.
One can control the overlap between the basis elements, avoiding running into linear dependences, by keeping small OC expansions ($R_{\text{max}}^{p>0} \simeq 1$ a.u.) since the Kato cusps are usually well localized at the atomic positions.
Angular expansions are truncated so $l$ takes values up to a maximum $l^{p}_\text{max}$, which can also be different for the non-equivalent centers.
In general, one can keep small values of $l_\text{max}$ in the OCs to complement the OCE in the description of the bound states, but a large angular expansion is usually required in the OCE, especially in the case of complex molecules and for the evaluation of continuum states with high kinetic energy.

\subsubsection*{Evaluation of bound orbitals}

There are several quantum chemistry packages available which can efficiently perform DFT calculations.
In this work, we have employed the Amsterdam Density Functional (ADF) program \cite{FonsecaTCA1998,VeldeJCC2001,ADF2013} to evaluate the ground state electron density of the molecules we have investigated using a double or a triple $\zeta$-polarization plus basis set (taken from the ADF library). 
Electronic exchange and correlation effects have been accounted for with the VWN \cite{VoskoCJP1980} local density approximation functional in some cases, or with the LB94 \cite{LeeuwenPRA1994}, depending on the characteristics of particular problem.
Besides providing a reliable description of bound states, these two functionals have been found to be suitable for the description of the long range behavior of the continuum states.
The electron density provided by the ADF calculation is projected into a multicentric B-spline basis set like the one described in the previous section. 
Then, the corresponding Hamiltonian $\bold{H}$ and overlap $\bold{S}$ matrices are constructed: 
\begin{align}
H_{nn'll'hh'\lambda\mu}^{pp'} &= \langle \xi_{nlh\lambda\mu}^{p}| H | \xi_{n'l'h'\lambda\mu}^{p'} \rangle \label{Hmatrix} \\
S_{nn'll'hh'\lambda\mu}^{pp'} &= \langle \xi_{nlh\lambda\mu}^{p}| \xi_{n'l'h'\lambda\mu}^{p'} \rangle \label{Smatrix}
\end{align}
By definition, both $\bold{H}$ and $\bold{S}$ are symmetric matrices.
Since the OC expansions cannot overlap, $H_{nn'll'hh'\lambda\mu}^{pp'}$ and $S_{nn'll'hh'\lambda\mu}^{pp'}$ are zero if $p\ne p'$, unless $p=0$ or $p'=0$ (elements of the OCE).
By solving the eigenvalue problem given by the secular equation
\begin{equation}\label{secularEq}
\bold{Hc}=\varepsilon\bold{Sc}
\end{equation}
we obtain the set coefficients $\bold{c}$ that define the bound orbitals in the B-spline basis.
Of course, both the basis set employed in the ADF calculation and the B-spline basis set need to be dense enough so the two calculations provide the same sets of orbitals.

\subsubsection*{Evaluation of continuum orbitals}\label{section_continuumOrbitals}

The continuum spectrum of an operator constitutes a family of eigenfunctions whose eigenvalues are a continuum variable.
The set of discrete solutions of eq. \ref{secularEq} whose energy is higher than the ionization threshold can be interpreted as a representation of the continuum, but with a different (arbitrary) normalization condition: $\varphi(R_{\text{max}}^{0})=0$, and normalized at the same level as the bound states: to a Kronecker delta.
Of course, the characteristics of these solutions depend on the numerical expansion and one needs to use a dense basis set and a large value of $R_{\text{max}}^{0}$ so their asymptotic behavior is properly represented.
In order to compute measurable quantities such as photoionization cross sections, one has to set the proper normalization of the continuum states: to a Dirac delta, and impose the adequate scattering boundary conditions, in the case of a molecule, of a multichannel problem.
Here we have employed the Galerkin \cite{BachauRPP2001} approach to evaluate photoelectron states at different energies and the correct boundary conditions have been imposed to the solutions, as we explain in this section.

\paragraph*{The Galerkin approach.}

The present method can yield the continuum wave function at any photoelectron energy using a fixed basis set.
The traditional eigenvalue problem given by eq. \ref{secularEq} does not admit non-trivial solutions ($\bold{c} \ne \bold{0}$) for an arbitrary value of energy $\varepsilon$.
However, one can obtain approximate solutions by finding the coefficients that minimize the residual vector $(\bold{Hc}-\varepsilon \bold{Sc})$, with $\bold{c} \ne \bold{0}$ by solving the eigenvalue problem
\begin{equation}\label{EVP_galerkin}
\bold{A}(\varepsilon)\bold{c}=a\bold{c}
\end{equation}
where $\bold{A}(\varepsilon)=\bold{H}-\varepsilon \bold{S}$.
The eigenfunctions corresponding to the lowest eigenvalues $a_{0}, a_{1},...$ can be taken as approximate solutions with energy $\varepsilon$ if their eigenvalues are close to zero.
It has been observed that, for $n$ partial waves, one can always find a set of $n$ eigenvalues $\left\{a_{i}\right\}_{i=1}^{n}$ whose moduli are sufficiently small and well separated from the others, provided that the basis set is dense and flexible enough.
Due to the lack of boundary conditions, $\bold{A}(\varepsilon)$ is not a Hermitian matrix and therefore its eigenvalues ${a_{i}}$ and eigenvectors ${\bold{c}_{i}}$ are, in general, complex.
Nonetheless since $\bold{A}(\varepsilon)$ is real they appear in conjugate pairs, i.e., for each pair $(a,\bold{c})$ that satisfies \ref{EVP_galerkin}, so does $(a^{*},\bold{c}^{*})$.
This later property makes possible to avoid complex representations just by taking $\mathfrak{R}(\bold{c})$ and $\mathfrak{I}(\bold{c})$ as independent solutions. 
Although these solutions do not satisfy the adequate boundary conditions of a multichannel scattering problem, they constitute a complete set and therefore can be combined to provide linear combinations that do.
In the next lines we explain how the Galerkin solutions can be renormalized so they accurately describe an electron being scattered from the molecular potential.

\paragraph*{Renormalization of continuum sates.}

Photoelectron wave functions describe a particle being ejected from an atom or a molecule.
Therefore, they must be solutions of the scattering Schr\"odinger equation
\begin{equation}\label{eq:scattering_hamiltonian}
\hat{H}_{sc}\varphi^{-}(\bold{r})=\varepsilon\varphi^{-}(\bold{r})
\end{equation}
where $\hat{H}_{sc}$ is the scattering Hamiltonian:
\begin{equation}
\hat{H}_{sc}=-\frac{1}{2}\nabla^{2}-U_{\alpha}(\bold{r})
\end{equation}
and $U_{\alpha}(\bold{r})$ is the potential generated by the residual ion.
In our case, electronic exchange and correlation effects are included in the potential through the use of a functional.
At long distances, the ionic potential can be approximated by that of a positive charge, i.e.:
\begin{equation}
\lim_{r \to \infty} U_{\alpha}(\bold{r})=\frac{1}{r}
\end{equation}
where $r=|\bold{r}|$.
The scattering Hamiltonian (eq. \ref{eq:scattering_hamiltonian}) does not admit analytical eigenfunctions in the case of complex potentials.
However, as $r$ increases, it tends to the Coulomb Hamiltonian, $\hat{H}_{c}$:
\begin{equation}
\lim_{r \to \infty} H_{sc} = H_{c} = -\frac{1}{2}\nabla^{2}-\frac{1}{r}
\end{equation}
which does have analytical solutions: the regular $F_{\varepsilon l}(r)$ and the irregular $G_{\varepsilon l}(r)$ Coulomb functions.
At long distances, they can be written as
\begin{align}
F_{\varepsilon l}(r)&=\sqrt{\frac{2}{\pi k}}\frac{1}{r}\sin{(kr-l\frac{\pi}{2}-\eta\log{2kr}+\sigma_{l})}\\
G_{\varepsilon l}(r)&=\sqrt{\frac{2}{\pi k}}\frac{1}{r}\cos{(kr-l\frac{\pi}{2}-\eta\log{2kr}+\sigma_{l})}
\end{align}
where $k$ is the momentum and $\sigma_{l}$ is the Coulomb phase shift:
\begin{equation}
\sigma_{l}=\arg{\Gamma(l+1+i\eta)}
\end{equation}
and $\Gamma$ is the Euler's Gamma function.
The asymptotic boundary conditions of the photoelectron scattering wave functions can be written in terms of the Coulomb functions:
\begin{equation}
\varphi^{-}_{\varepsilon lh} = F_{\varepsilon l}(r) X_{lh} + \sum_{l'}\pi K_{ll'} G_{\varepsilon l}(r) X_{l'h}
\end{equation}
where the $\bold{K}$ matrix is related to the usual scattering matrix $\bold{S}$ \cite{bookTaylor} by
\begin{equation}
I-i\pi \bold{K}=(I+i\pi \bold{K})\bold{S}
\end{equation}
The set of continuum states provided by the Galerkin approach in the basis set of B-splines, however, satisfy arbitrary boundary conditions of the form:
\begin{equation}
\varphi_{\varepsilon lh} = \sum_{l'} a_{ll'} F_{\varepsilon l'}(r) X_{l'h} + \sum_{l'}\pi K_{ll'} G_{\varepsilon l}(r) X_{l'h}
\end{equation}
where the sets of coefficients $\{a_{ll'}\}$ and $\{b_{ll'}\}$ can be obtained by comparing the radial part of the wave functions $\varphi_{\varepsilon lh}$ and its first derivatives at $r=R_{\text{max}}$ with those of the Coulomb functions.
They define two matrices $\bold{A}$ and $\bold{B}$ that can be used to obtain the correct wave functions:
\begin{equation}
\varphi^{-}_{\varepsilon lh} = \bold{A}^{-1} \varphi_{\varepsilon lh}
\end{equation}

The resulting wave functions $\varphi^{-}_{\varepsilon lh}$ have the proper $K$-matrix normalization, with $\pi \bold{K}=-\bold{A}^{-1}\bold{B}$.

\subsubsection*{Dipole-transition matrix elements}

As explained in the previous chapter, in order to evaluate the electronic wave packet generated in a molecule upon ionization by ultrashort laser pulses (eq. \ref{coefFOPT}) or to compute photoionization cross sections (eq. \ref{CS}), the dipole-transition matrix elements are required.
Here we indicate how to evaluate the element corresponding to a transition from the electronic ground state $\Psi_0$ (eq. \ref{elecGS}) to a continuum state $\Psi_{\alpha \varepsilon lh}$ (eq. \ref{elecCont}) upon interaction with linearly polarized light.
In our description of the wave function, the residual ion remains frozen (static-exchange approximation), which reduces the problem to the calculation of the coupling between the bound orbital where the electron is taken from $\phi_{\alpha}$ and the continuum orbital where is promoted to $\phi_{\varepsilon lh}$, that is,
\begin{equation}\label{elecDME}
\mu_{\boldsymbol\epsilon}^{\alpha \varepsilon lh} = \langle\Psi_{\alpha \varepsilon lh}(\bold{\bar{x}})| \boldsymbol\mu_{\boldsymbol\epsilon}^\text{e}(\bold{\bar{r}})|\Psi_0(\bold{\bar{x}})\rangle = \langle \phi_{\varepsilon lh}(\bold{r}) | \boldsymbol\epsilon \bold{r} |\phi_\alpha(\bold{r}) \rangle
\end{equation}
where $\boldsymbol\epsilon$ is the polarization vector of the electric field.
Dipole transition elements can be used to evaluate cross sections.

\subsubsection*{Photoionization cross section within the fixed-nuclei approximation}

Making use of eqs. \ref{CSphotoionization} and \ref{elecDME} and summing incoherently over all photoelectron symmetries (all possible values of $l$ and $h$), we can evaluate total photoionization cross sections in the framework first-order perturbation theory within the fixed-nuclei approximation:
\begin{equation}\label{CS_photo}
\sigma_{\boldsymbol\epsilon \alpha}(\varepsilon)=\frac{4\pi^{2}\omega}{\hbar c} \sum_{lh} \big| \mu_{\boldsymbol\epsilon}^{\alpha \varepsilon lh} \big| ^2 
\end{equation}
As indicated in section \ref{section_CS}, for the case of randomly oriented molecules, one needs to compute $\sigma_{\boldsymbol\epsilon \alpha}$ for three orthogonal directions of the polarization vector of the field $\boldsymbol\epsilon$ and then average the results incoherently.

\subsection{Time-dependent DFT}\label{section_TDDFT}

An improvement of the static-exchange DFT method described in the previous section in order to describe the coupling between different photoionization channels is the time-dependent DFT approach.
The method uses many concepts from the static-exchange version.
Here we present the most relevant concepts, for a more complete description, see \cite{StenerJCP2005}. \vspace{2 mm}

The linear response of the electron density to an external field can be evaluated using the scheme proposed by Zangwill and Soven by defining an effective self-consistent field potential:
\begin{equation}\label{TDDFT_SCFPotential}
V_{\text{SCF}}(\bold{r},\omega)=V_{\text{EXT}}(\bold{r},\omega)+\delta V(\bold{r},\omega)
\end{equation}
where $\omega$ is the frequency of the radiation, $V_{\text{EXT}}(\bold{r},\omega)$ is the external dipole potential and $\delta V(\bold{r},\omega)$ is the induced potential, which is given by the sum of the Hartree and exchange-correlation screening due to the redistribution of the electrons:
\begin{equation}\label{TDDFT_inducedPotential}
\delta V_{\text{SCF}}(\bold{r},\omega) = \int \frac{\delta(\bold{r},\omega)}{|\bold{r}-\bold{r}'|}d\bold{r}' +\frac{\partial V_{XC}}{\partial \rho} \Big|_{\rho=\rho(\bold{r})} \delta \rho(\bold{r},\omega)
\end{equation}
$\rho(\bold{r})$ is the unperturbed electron density and $\delta(\bold{r},\omega)$ denotes the induced density in the adiabatic local density approximation \cite{GrossAQC1990}, which can be expressed in terms of the dielectric susceptibility $\chi$ and the self-consistent field potential:
\begin{equation}\label{TDDFT_suscep}
\delta \rho(\bold{r},\omega)= \int \chi(\bold{r},\bold{r}',\omega) V_{\text{SCF}}(\bold{r}',\omega) d\bold{r}'
\end{equation}
By inserting eq. \ref{TDDFT_suscep} into eq. \ref{TDDFT_inducedPotential}, we obtain:
\begin{equation}\label{TDDFT_inducedPotential2}
\delta V(\bold{r},\omega) = \iint K(\bold{r},\bold{r'}) \chi(\bold{r},\bold{r}',\omega) V_{\text{SCF}}(\bold{r},\omega) d\bold{r}d\bold{r}'
\end{equation}
where $K(\bold{r},\bold{r'})$ represents the Hartree and exchange-correlation kernel:
\begin{equation}
K(\bold{r},\bold{r'}) = \frac{1}{|\bold{r}-\bold{r'}|} + \delta(\bold{r}-\bold{r}') \frac{V_{\text{XC}}}{d\rho'} \Big|_{\rho'=\rho(\bold{r})}
\end{equation}
Eq. \ref{TDDFT_inducedPotential2} is solved with respect to $\delta V(\bold{r},\omega)$ in a basis set of B-splines basis as the one employed in the static-exchange DFT approach and the adequate boundary conditions for the photoelectron wave functions are imposed.
Then, the electronic dipole-transition matrix elements are evaluated using $V_{\text{SCF}}$ instead of the dipole operator.
The time-dependent DFT method can accurately describe interchannel coupling effects and autoionization resonances at the linear-response level.
For this reason, it is more suitable than the static-exchange version for the description of correlation effects due to the coupling between different ionization channels.

\section{Inclusion of the nuclear motion}\label{section_nuclear}

The nuclear motion may play an important role in molecular photoionization since the energy of the incident photon is usually distributed between electrons and nuclei.
In general, molecules undergo vertical transitions upon photoionization because the electronic emission occurs suddenly and the nuclei have no time to rearrange.
This can lead to (several) vibrational excitations in the parent ion, thus generating superpositions of vibronic states $\sum_{\nu} c_{n\nu}\chi_{n\nu}(\bold{\bar{R}})$, 
where $\chi_{n\nu}(\bold{\bar{R}})$ are the (final) vibrational wave functions, $\nu$ and $n$ being the vibrational and electronic quantum numbers, respectively, 
and the expansion coefficients $c_{n\nu}$ are approximately given the overlaps with the initial wave function $\chi_{0}(\bold{\bar{R}})$, i.e., the Franck-Condon factors $\langle \chi_{n\nu}(\bold{\bar{R}}) |\chi_{0}(\bold{\bar{R}})\rangle$.
Vibrational excitations in the parent ion are experimentally observable even in the case of inner-shell photoionization, thanks to the advent of the third generation of synchrotron radiation sources and high-energy-resolution detection techniques \cite{BilderbackJPB2005}.
Of course, in order to describe these situations, the nuclear degrees of freedom must be taken into account.
Here we present a method for including the nuclear motion at the Born-Oppenheimer level applicable to diatomic molecules and to small polyatomics in situations in which only one vibrational mode is active.

\subsection{The nuclear Hamiltonian}

Within the Born-Oppenheimer approximation, the nuclear Hamiltonian (see eq. \ref{Nevp}) of a \textbf{diatomic AB molecule} in a given electronic state $n$ can be written as
\begin{equation}\label{HnucBO}
\hat{H}_{N} = -\frac{1}{2M} \nabla^{2}_{\bold{R}_{\text{CM}}} \underbrace{ -\frac{1}{2\mu}\nabla^{2}_{R}+E_n(R) }_{\hat{H}_{\text{int}}}
\end{equation}
where $M$ and $\mu$ are the total and the reduced mass of the system, respectively, $R$ is the relative distance between the two nuclei, $\bold{R}_{\text{CM}}$ is the position of the center of mass and $E_n$ is the potential energy curve associated to the electronic state $n$ in the Born-Oppenheimer approximation. 
In this formulation, it is clear that the total kinetic energy is composed by a translational motion of the center of mass (first term in the left side of eq. \ref{HnucBO}) and an internal motion (second term).
The translational motion is not quantized and just contributes to the total energy by adding a constant energy to the eigenvalues.
For this reason, here we will focus on finding the eigenstates of the internal Hamiltonian $\hat{H}_{\text{int}}$, which in polar coordinates is given by
\begin{equation}\label{HpolarCoord}
\hat{H}_{\text{int}} = -\frac{\hbar}{2\mu} \Big( \frac{\partial^2}{\partial R^2}+\frac{2}{R} \frac{\partial}{\partial R}\Big) + \frac{\hat L^2}{2 \mu R^2} + E_n(R)
\end{equation}
where $\hat{L}^2$ is the square of the total angular momentum operator.
The eigenfunctions of $\hat{H}_{\text{int}}$ can be written as a product of a radial part $\chi_{n\nu}(R)$, describing the vibrational motion, and a spherical harmonic $Y_l^m(\theta,\phi)$ that accounts for the molecular rotation.
Since the rotational energy is significantly smaller than the vibrational quantum, we can neglect this term and describe the nuclear motion in terms of vibrational states $\chi_{\nu}(R)$, which are the eigenfunctions of the vibrational Hamiltonian:
\begin{equation}\label{Hvib}
\hat{H}_{\text{vib}}(R) = -\frac{\hbar}{2\mu} \Big( \frac{\partial^2}{\partial R^2}+\frac{2}{R} \frac{\partial}{\partial R}\Big) + E_n(R)
\end{equation}
Vibrational eigenfunctions can be written as
\begin{equation}
\chi_{n\nu}(R) = \frac{\zeta_{n\nu}(R)}{R}
\end{equation}
which simplifies the eigenvalue problem to
\begin{equation}\label{Vevp}
\bigg[ -\frac{\hbar}{2\mu} \frac{\partial^2}{\partial R^2} + E_n(R) \bigg]  \zeta_{n\nu}(R) = E_{n\nu} \zeta_{n\nu}(R)
\end{equation}
Vibrational eigenstates are evaluated in a basis set of B-spline functions, as we explain in section \ref{evaluationVibrationalEigenstates}.

\paragraph*{Polytomic molecules.} The motion of a set of particles can be decomposed into translation and rotation of the center of mass and vibration of its particles.
A $n-$particle system has $3n-5$ vibrational degrees of freedom (vibrational modes) if it is linear and $3n-6$ otherwise \cite{bookLevine}.
A complete description of molecular vibration within the Born-Oppenheimer approximation would require including in the total wave function the corresponding vibrational eigenfunctions depending on all vibrational degrees of freedom.
In this work we have investigated inner shell photoionization of polyatomic AB$_{n}$ molecules, A being the central atom and B the surrounding atomic centers, symmetrically displaced around A.
Recent experimental data \cite{ThomasJCP2007,ThomasJCP2008} has shown that, in this scenario, the totally symmetric stretching mode (TSSM), that in which the B atoms move symmetrically towards A, is the most affected by the structural rearrangement that accompanies core ionization.
For this reason, in this work we have restricted nuclear motion to the TSSM coordinate.
The dynamics of the B atoms along the TSSM can be understood in terms of a virtual particle moving in a monodimensional well $V(R)$, defined by the $n$ chemical bonds, with a reduced mass\footnote{The mass of the central atom ($m_{A}$) does not contribute to $\mu$ since it remains frozen along the TSSM coordinate.} $\mu=3M_{B}$.
Then, eq. \ref{HpolarCoord} remains valid with $R=|\bold{R}_{B}-\bold{R}_{A}|$, the TSSM coordinate.

\subsection{Evaluation of vibrational eigenstates} \label{evaluationVibrationalEigenstates}

Vibrational eigenstates are evaluated by solving the eigenvalue problem given by eq. \ref{Vevp} in a basis set of B-splines
\begin{equation}
\zeta(R)_{n\nu} = \sum_{i=1}^{N_{\text{max}}} c_{n\nu i} B_{i}(R)
\end{equation}
where $n$ and $\nu$ are the electronic and vibrational quantum numbers, respectively, and $i$ is an index going over the $N_{\text{max}}$ B-spline functions $B_{i}$, which are defined up to a value of $R_{\text{max}}$.
The corresponding secular equation in its matrix form, $\bold{Hc}=E\bold{Sc}$, where $\bold{H}$ and $\bold{S}$ are the Hamiltonian and the overlap matrices in the B-spline basis set, is solved using a standard diagonalization procedure. 
Since B-spline functions are piecewise polynomials, the elements of $\bold{H}$ and  $\bold{S}$ are computed exactly using a Gauss-Legendre integration method.

\subsubsection*{Bound states}
The resolution of the secular equation provides an orthonormal set of stationary states.
Those whose energy is lower than the molecular dissociation limit constitute the bound part of the spectrum.
Of course, the number of bound states and the energy spacing depends on the shape of the potential well (depth and width) and on the reduced mass of the system, but not on the parameters of the B-spline expansion, provided the basis set has been wisely chosen.

\subsubsection*{Continuum states} \label{electronic_continuum_states}
The stationary solutions of the secular equation with energy higher than the dissociation limit constitute a discretized representation of the vibrational continuum with the arbitrary boundary condition $\chi(R_{\text{max}})=0$.
As in the case of the electronic states (section \ref{section_continuumOrbitals}), the number of continuum states and their energy spacing depend on the parameters of the numerical expansion and one needs to employ a large value of $R_{\text{max}}$ and a dense grid of B-splines so the asymptotic behavior of the continuum wave functions can be properly described.
However, the situation here is more simple since this is a mono-channel scattering problem and the adequate normalization of the true continuum states $\chi_{E_{\nu}}$ can be set by multiplying the solutions coming from the diagonalization procedure by a factor:
\begin{equation}
 \chi_{E_{\nu}}=\sqrt{\rho(E_{\nu})}\chi_{\nu}
\end{equation}
where $\rho_{E_{\nu}}$ is the density of states \cite{BachauRPP2001}.
Working with discretized states in a box, $\rho(E_\nu) $ can be approximated by
\begin{gather}\label{DOS}
\rho(E_\nu) = \Big|\frac{\partial E(\nu')}{\partial \nu'}\Big|_{\nu'=\nu} \simeq \frac{2}{E_{\nu+1}-E_{\nu-1}}
\end{gather}
which has been proved to be a good approximation \cite{BachauRPP2001}.

\subsection{Vibrationally resolved cross sections}\label{section_VRCS}

Let us consider a transition from the ground state $\Phi_{0\nu}(\bold{x},R)$, with $\nu=0$, to a state $\Phi_{\alpha\varepsilon lh\nu'}(\bold{x},R)$ in which an electron has been emitted from the $\alpha$ molecular orbital with $\varepsilon$ kinetic energy and $lh$ symmetry, leaving the residual ion in the $\alpha\nu$ vibronic state.
These are Born-Oppenheimer states and can be written as
\begin{align}
 \Phi_{0\nu}(\bold{x},R)&=\Psi_{0}(\bold{x},R)\chi_{0\nu}(R) \label{BOgs} \\
 \Phi_{\alpha\varepsilon lh\nu'}(\bold{x},R)&=\Psi_{\alpha\varepsilon lh}(\bold{x},R)\chi_{\alpha\nu'}(R) \label{BOcs}
\end{align}
The corresponding dipole transition matrix element upon interaction with linearly polarized light is given by:
\begin{equation}\label{vibronicDME}
\mu_{\boldsymbol\epsilon}^{\alpha\varepsilon lh\nu' \leftarrow 0\nu} = \int \langle \Phi_{f}(\bold{x},R) | \boldsymbol\mu_{\boldsymbol\epsilon}| \Phi_{i}(\bold{x},R) \rangle dR
\end{equation}
where $\boldsymbol\mu_{\boldsymbol\epsilon}$ is the total dipole operator ($\hat{\boldsymbol\epsilon}$ is the polarization vector of the field), which is the sum of two contributions:
\begin{equation}\label{dipoleOp}
\boldsymbol\mu_{\boldsymbol\epsilon} = \underbrace{ \hat{\boldsymbol\epsilon} \sum_n \bold{r}_n }_{\boldsymbol\mu_{\boldsymbol\epsilon}^{\text{elec}}} + \underbrace{ \hat{\boldsymbol\epsilon} \sum_\alpha Z_{\alpha} \bold{R}_n }_{\boldsymbol\mu_{\boldsymbol\epsilon}^{\text{nuc}}} 
\end{equation}
Inserting eqs. \ref{BOgs}, \ref{BOcs} and \ref{dipoleOp} into eq. \ref{vibronicDME}, we have 
\begin{align*}
\mu_{\boldsymbol\epsilon}^{\alpha\varepsilon lh\nu' \leftarrow 0\nu} &= \int \underbrace{ \langle \Psi_{\alpha\varepsilon lh}(\bold{x},R) | \boldsymbol\mu_{\boldsymbol\epsilon}^{\text{elec}} | \Psi_{0}(\bold{x},R) \rangle }_{\mu_{\boldsymbol\epsilon}^{\alpha \varepsilon lh\leftarrow 0}} \chi_{\alpha\nu'}(R)\chi_{0\nu}(R) dR \\
& + \int \langle \Psi_{\alpha\varepsilon lh}(\bold{x},R) | \Psi_{0}(\bold{x},R) \rangle  \chi_{\alpha\nu'}(R) \boldsymbol\mu_{\boldsymbol\epsilon}^{\text{nuc}} \chi_{0\nu}(R) dR
\end{align*}
where the second term is zero due to the orthogonality of electronic states.
Then, the vibronic dipole-transition matrix element can be written as
\begin{equation}\label{vibronicDTME}
\mu_{\boldsymbol\epsilon}^{\alpha\varepsilon\nu' lh \leftarrow 0\nu} = \int \mu_{\boldsymbol\epsilon}^{\alpha \varepsilon lh\leftarrow 0} \chi_{\alpha\nu'}(R)\chi_{0\nu}(R) dR
\end{equation}
An expression for the photoionization cross section upon interaction with monochromatic light is obtained inserting eq. \ref{vibronicDTME} into eq. \ref{CSphotoionization} and summing incoherently over all photoelectron symmetries (all values of $l$ and $h$):
\begin{equation}
\sigma_{\boldsymbol\epsilon}^{\alpha\nu'}(\varepsilon)=\frac{4\pi^{2}\omega}{\hbar c} \sum_{lh} \bigg| \int \mu_{\boldsymbol\epsilon}^{\alpha \varepsilon lh\leftarrow 0} \chi_{\alpha\nu'}(R)\chi_{0\nu}(R) dR \bigg|^{2}
\end{equation}
where $\omega$ is the photon energy, which is related to the photoelectron energy $\varepsilon$ through the equation $\varepsilon = \hbar\omega -E_{\alpha}^{\nu\nu'}$, where $E_{\alpha}^{\nu\nu'}=E_{\alpha\nu'}-E_{0\nu}$ is the energy required to produce the ion in the $\alpha\nu'$ vibronic state.
For the case of randomly oriented molecules, one can compute $\sigma_{\boldsymbol\epsilon}^{\alpha\nu'}$ for three orthogonal directions of $\hat{\boldsymbol\epsilon}$ and then the results incoherently (eq. \ref{CSrandom}).

%% file: Chapters/Chapter3.tex
\chapter{Electron dynamics initiated by attosecond pulses}
\label{chapter3}
\fancyhead[LE]{\fontsize{11pt}{11pt}\selectfont Chapter 3. Electron dynamics initiated by attosecond pulses}
\fancyhead[RO]{\fontsize{11pt}{11pt}\selectfont \nouppercase{\rightmark}}

The development of attosecond technology has enabled the real-time observation of electron motion in atoms, molecules and solids \cite{KrauszRMP2009}.
Experimentally, it is now possible to generate laser pulses of durations of a few tens of attoseconds.
These durations are of the order of the period of revolution of the first Bohr orbit, which is 150 attoseconds, thus opening the way to track and to manipulate electron dynamics at its natural time scale.
Due to their wide spectral bandwidth, attosecond pulses create coherent superpositions of electronic states, inducing an ultrafast response in the target.
In this chapter we present a method to evaluate the electronic wave packet generated in a molecule upon attosecond ionization and the subsequent charge redistribution, applying the concepts that have been previously introduced in chapters \ref{chapter1} and \ref{chapter2}.

\section{Wave packet dynamics}

Attosecond XUV pulses can efficiently ionize molecules from several shells, creating coherent superpositions of electronic states, i.e., electronic wave packets.
In general, the ultrafast electronic response to prompt ionization can be described in the framework of the fixed-nuclei approximation since it usually precedes the onset of the nuclear motion.
In this work we have employed the static-exchange Density Functional Theory method, explained in section \ref{section_staticExchangeDFT}, to evaluate the electronic structure of molecules.
The electronic wave packet generated by attosecond ionization can be written as:
\begin{equation}\label{WFexpansion}
\Phi(\bold{\bar{x}},t) = c_{0}(t)\Psi_{0}(\bold{\bar{x}}) + \sum_{\alpha lh}\int c_{\alpha\varepsilon lh}(t) \Psi_{\alpha\varepsilon lh}(\bold{\bar{x}}) d\varepsilon
\end{equation}
where $\bold{\bar{x}}=(\bold{x}_{1},...,\bold{x}_{N})$ stands for the spatial and spin coordinates of the $N$ electrons in the molecule, $\Psi_{0}(\bold{\bar{x}})$ is the electronic ground state, $\Psi_{\alpha\varepsilon lh}(\bold{\bar{x}})$ represents a continuum state in which an electron has been promoted from the $\alpha$ orbital to a continuum orbital with kinetic energy $\varepsilon$ and angular quantum numbers $l$ and $m$ and the time-dependence of the wave function is included in the spectral coefficients $c_{0}$ and $c_{\alpha lh}$, which satisfy the normalization condition:
\begin{equation}
|c_{0}(t)|^{2} + \sum_{\alpha lh} \int |c_{\alpha\varepsilon lh}(t)|^{2} d\varepsilon = 1
\end{equation}
At $t=0$ the system is assumed to be in the ground state, i.e., $|c_{0}(0)|^2=1$ and $c_{\alpha\varepsilon lh}(0)=0$.
If the attosecond pulse is weak, most of the population will remain in the ground state, i.e., $|c_{0}(t)|^2\simeq 1$, and the time-dependent coefficients can be evaluated using first-order perturbation theory, as explained in section \ref{section_TDPT}.
We can thus make use of eq. \ref{coefFOPT} to evaluate the continuum spectral coefficients which, for our particular case, reads:
\begin{equation}\label{coeff1}
c_{\alpha\varepsilon lh}(t) = -\frac{i}{\hbar} \langle \Psi_{\alpha\varepsilon lh}(\bold{\bar{x}}) | \boldsymbol\mu_{\boldsymbol\epsilon} | \Psi_{0}(\bold{\bar{x}}) \rangle e^{-\frac{i}{\hbar}(E_{\alpha}+\varepsilon ) t} \int_{0}^{t} \text{E}(\tau) e^{\frac{i}{\hbar}(E_{\alpha}+\varepsilon-E_{0})\tau} d\tau
\end{equation}
where $\hat{\boldsymbol\epsilon}$ is the polarization direction of the electric field $\text{E}$ and $E_{\alpha}$ is the energy of an ion with a hole in the $\alpha$ molecular orbital.
After the interaction with the pulse ($t>T$), the integral in eq. \ref{coeff1} can be substituted by the Fourier transform of the electric field $\mathcal{F}_{\{\text{E}\}}$:
\begin{equation}\label{coeff2}
c_{\alpha\varepsilon lh}(t>T) = -\frac{i}{\hbar} \langle \Psi_{\alpha \varepsilon lh}(\bold{\bar{x}}) | \boldsymbol\mu_{\boldsymbol\epsilon} | \Psi_{0}(\bold{\bar{x}}) \rangle e^{-\frac{i}{\hbar}(E_{\alpha}+\varepsilon ) t} \mathcal{F}_{\{\text{E}\}}\bigg(\frac{E_{\alpha}+\varepsilon-E_{0}}{\hbar}\bigg)
\end{equation}
where the dependence on time is that of the stationary phases $e^{-\frac{i}{\hbar}(E_{\alpha}+\varepsilon ) t}$ as the wave packet evolves freely.
The interferences between the spectral components of the wave packet can be imprinted in different observables.

\subsubsection*{Electron density}

As explained in section \ref{section_continuumOrbitals}, we employ a discretization technique to describe the electronic continuum.
The perturbed part of the time-dependent wave function, that is, the part of the wave function that does not contain the ground state (see eq. \ref{WFexpansion}), can be approximated by:
\begin{equation}\label{WFexpansion_pert}
\Phi^{\text{pert}}(\bold{\bar{x}},t) \simeq \sum_{\alpha nlh} \tilde{c}_{\alpha nlh}(t) \Psi_{\alpha\varepsilon lh}(\bold{\bar{x}})
\end{equation}
where the integral has been replaced by a discrete sum, $n$ is an index on the discretized photoelectron energies $\varepsilon_{n}$ and the coefficients of the discretized expansion $\tilde{c}_{\alpha nlh}$ are related those in eq. \ref{WFexpansion} by:
\begin{equation}
\tilde{c}_{\alpha nlh}=\frac{c_{\alpha lh}(\varepsilon_{n})}{\sqrt{D_{n}}}
\end{equation}
where $D_{n}$ is the density of states \cite{BachauRPP2001}, which, as discussed in section \ref{electronic_continuum_states}, can be approximated as $D_{n}\simeq\frac{2}{\varepsilon_{n+1}-\varepsilon_{n-1}}$.
The time-dependent electron density can be evaluated by inserting eq. \ref{WFexpansion_pert} into eq. \ref{elecDens}:
\begin{equation}\label{elecDens_Nelec}
 \rho(\bold{r})=N\int  \ldots \int  |  \sum_{\alpha nlh} \tilde{c}_{\alpha nlh}(t) \Psi_{\alpha\varepsilon_n lh}(\bold{\bar{x}})|^2 ds d\bold{x}_2\ldots d\bold{x}_N
\end{equation}
where $N$ is the number of electrons in the neutral molecule.
After some rearrangement, eq. \ref{elecDens_Nelec} can be written as:
\begin{align}\label{elecDens_Nelec2}
\begin{split}
\rho(\bold{r})&=N \sum_{\alpha nlh} |\tilde{c}_{\alpha nlh}(t)|^{2} \int\ldots\int |\Psi_{\alpha\varepsilon_n lh}(\bold{\bar{x}})|^2 ds d\bold{x}_2\ldots d\bold{x}_N\\
&+N  \sum_{\substack{nn'll'hh' \\ (nlh)\neq (n'l'h') \\ \text{same spin}}} \sum_{\alpha} \Big[ \tilde{c}_{\alpha nlh}^{*}(t)\tilde{c}_{\alpha n'l'h'}(t) \Big] \int\ldots\int \Psi_{\alpha\varepsilon_n lh}(\bold{\bar{x}})^* \Psi_{\alpha\varepsilon_{n'}l'h'}(\bold{\bar{x}}) ds d\bold{x}_2\ldots d\bold{x}_N \\
&+N \sum_{\substack{\alpha\alpha' \\ \alpha\neq\alpha'\\ \text{same} \\ \text{spin}}} \sum_{nlh} \Big[ \tilde{c}_{\alpha nlh}^{*}(t)\tilde{c}_{\alpha'nlh}(t) \Big] \int\ldots\int \Psi_{\alpha\varepsilon_n lh}(\bold{\bar{x}})^* \Psi_{\alpha'\varepsilon_n lh}(\bold{\bar{x}}) ds d\bold{x}_2\ldots d\bold{x}_N
\end{split}
\end{align}
where we have taken into account that for $\alpha\neq\alpha'$ and $(n,l,h)\neq(n',l',h)$
\begin{equation}
\int\ldots\int  \Psi_{\alpha\varepsilon_n lh}(\bold{\bar{x}})^* \Psi_{\alpha'\varepsilon_{n'} l'h'}(\bold{\bar{x}}) ds d\bold{x}_2\ldots d\bold{x}_N =0
\end{equation}
Eq. \ref{elecDens_Nelec2} can be simplified by making use of the following relations:
\begin{itemize}
\item[$\circ$]  The integral of the absolute square of a stationary state is given by
\begin{equation}\label{SlaterProp1}
\int\ldots\int |\Psi_{\alpha\varepsilon_n lh}(\bold{\bar{x}})|^2 ds d\bold{x}_2\ldots d\bold{x}_N =\frac{1}{N} \Big( \sum_{\substack{\alpha' \\ \alpha'\neq\alpha}} |\varphi_{\alpha'}(\bold{r})|^2 + |\varphi_{\varepsilon_n lh}(\bold{r})|^2 \Big)
\end{equation}
\item[$\circ$]  If $(n,l,h)\neq(n',l',h')$, then
\begin{equation}\label{SlaterProp2}
\int\ldots\int \Psi_{\alpha\varepsilon_n lh}(\bold{\bar{x}})^* \Psi_{\alpha \varepsilon_{n'} l'h'}(\bold{\bar{x}}) ds d\bold{x}_2\ldots d\bold{x}_N =\frac{1}{N} \varphi_{\varepsilon_n lh}(\bold{r}) \varphi_{\varepsilon_{n'} l'h'}(\bold{r})
\end{equation}
\item[$\circ$]  If $\alpha\neq\alpha'$, then
\begin{equation}\label{SlaterProp3}
\int\ldots\int \Psi_{\alpha\varepsilon_n lh}(\bold{\bar{x}})^*\Psi_{\alpha'\varepsilon_n lh}(\bold{\bar{x}})^*ds d\bold{x}_2\ldots d\bold{x}_N=-\frac{1}{N} \varphi_{\alpha}(\bold{r}) \varphi_{\alpha'}(\bold{r})
\end{equation}
provided the ionic substates $\alpha$ and $\alpha'$ have the same spin, otherwise the integral given in \ref{SlaterProp3} is zero.
\end{itemize}
Making use of these properties (eqs. \ref{SlaterProp1}, \ref{SlaterProp2} and \ref{SlaterProp3}), eq. \ref{elecDens_Nelec2} can be simplified to:
\begin{align}
\label{elecDens_NelecFinal}
\begin{split}
\rho(\bold{r},t)&=\sum_{\alpha nlh} |\tilde{c}_{\alpha nlh}(t)|^{2} \Big( \sum_{\alpha'} |\varphi_{\alpha'}(\bold{r})|^2 + |\varphi_{\varepsilon_n lh}(\bold{r})|^2 \Big) \\
& + \sum_{\substack{nn'll'hh' \\ \text{same spin}}} \sum_{\alpha} \Big[ \tilde{c}_{\alpha nlh}^{*}(t)\tilde{c}_{\alpha n'l'h'}(t) \Big] \varphi_{\varepsilon_n lh}(\bold{r}) \varphi_{\varepsilon_{n'} l'h'}(\bold{r}) \\
& - \sum_{\substack{\alpha\alpha' \\ \text{same} \\ \text{spin}}} \sum_{nlh} \Big[ \tilde{c}_{\alpha nlh}^{*}(t)\tilde{c}_{\alpha'nlh}(t) \Big] \varphi_{\alpha}(\bold{r})\varphi_{\alpha'}(\bold{r})
\end{split}
\end{align}
We can see that the electron density is the sum of three contributions.
The first term is a stationary term that does not depend on time.
The second is constructed using continuum orbitals and thus describes the emission of the photoelectron wave.
The third term, built from bound (Kohn-Sham) orbitals, accounts for the charge redistribution along the molecular skeleton.
The evaluation of the full electron density might become tedious in some situations due to the large number of continuum orbitals one may need to employ to describe the photoelectron wave (term 2 in eq. \ref{elecDens_Nelec}).
Al alternative approach to evaluate the ultrafast charge redistribution occurring in the residual ion upon attosecond ionization is to employ the reduced density matrix of the ionic subsystem, as we explain in the next section.

\section{Evolution of the ionic subsystem}

We seek to analyze the evolution of the hole generated in the molecular target upon attosecond ionization.
The residual ion is an open system that remains coupled to the emitted electron.
Therefore, it can be fully characterized in terms of its reduced density matrix, whose elements can be constructed from the spectral coefficients (eqs. \ref{coeff1} and \ref{coeff2}):
\begin{equation}\label{redDensMat}
\gamma_{\alpha\alpha'}^{\text{(ion)}} (t)= \sum_{lh}\int c_{\alpha\varepsilon lh}(t) c_{\alpha'\varepsilon lh}^{*}(t) d\varepsilon
\end{equation}
where the double sum runs over the ionic states from which the electron has been emitted with the same spin.
The trace of the reduced density matrix contains the population of each ionic state and the off-diagonal terms provide the coherence between pairs of states.
In the case of ionization with monochromatic light, all off-diagonal terms would be zero (except those involving degenerate states) since the parent ion would be in an incoherent superposition of states.
This is the situation one would expect to find in experiment with synchrotron radiation where the energy of the incident photons is well defined \cite{BilderbackJPB2005}.
Due to their wide spectral bandwidths, attosecond XUV pulses can generate coherent superpositions of electronic states, allowing to investigate ultrafast dynamics with the required time resolution \cite{KrauszRMP2009}.

\subsubsection*{Electron density of the residual ion}

The reduced density matrix (eq. \ref{redDensMat}) contains all the information about the ionic subsystem and therefore can retrieve any observable depending on its coordinates.
In particular, the time-dependent electron density is given by:
\begin{equation}\label{elecDens_ion}
\rho^{(\text{ion})}(\bold{r},t) = \sum_{\alpha} \gamma_{\alpha\alpha}^{\text{(ion)}}(t) \sum_{\alpha'} |\varphi_{\alpha'}(\bold{r})|^2 - \sum_{\substack{\alpha\alpha' \\ \text{same}\\ \text{spin}}} \gamma_{\alpha\alpha'}^{\text{(ion)}}(t) \varphi_{\alpha}(\bold{r})\varphi_{\alpha'}(\bold{r})
\end{equation}
Where the first term is time-independent since the trace of the reduced density matrix is constant.
By comparing eq. \ref{elecDens_ion} with the electron density of the full system including the photoelectron (eq. \ref{elecDens_Nelec}) we can see that the former contains all the terms describing the ultrafast dynamics occurring in the parent ion upon ultrafast ionization.
Only the terms describing the photoelectron emission, i.e., those containing continuum orbitals $\varphi_{\varepsilon_{n}lh}(\bold{r})$, are not included in eq. \ref{elecDens_ion}.

\subsubsection*{Hole density}

An interesting observable is the density of the hole generated upon ionization, defined by Lorenz S. Cederbaum and coworkers \cite{CederbaumCPL1999} as the difference between the electron density of the ion, $\rho^{(\text{ion})}(\bold{r},t)$, and that of the (initial) neutral molecule, $\rho_0(\bold{r})$, which does not depend on time:
\begin{equation}\label{hole_density_2}
Q(\bold{r},t) = \rho_0(\bold{r})-\rho^{(\text{ion})}(\bold{r},t) =\sum_{\substack{\alpha\alpha' \\ \text{same}\\ \text{spin}}} \gamma_{\alpha\alpha'}^{\text{(ion)}}(t) \varphi_{\alpha}(\bold{r})\varphi_{\alpha'}(\bold{r})
\end{equation}
where the ground state density is given by
\begin{equation}
\rho_{0}(\bold{r})=\sum_{\alpha} \varphi_{\alpha}^{2}(\bold{r})
\end{equation}
and we have assumed that the reduced density matrix of the ionic subsystem is normalized to unity, i.e.,
\begin{equation*}
\sum_{\alpha} \gamma_{\alpha\alpha}^{\text{(ion)}}=1
\end{equation*}
Fluctuations in the hole density might arise if several ionic states are populated coherently, that is, if off-diagonal elements of the reduced density matrix are not zero.
Although, up to now, no experiment has been able to measure the hole density of an isolated molecule directly, the ultrafast charge redistribution accompanying sudden ionization can be imprinted in observables that are experimentally accessible. 
For instance, some fragmentation channels may be sensitive to the localization of the hole created in the molecule upon attosecond ionization \cite{CalegariScience2014,CalegariIEEE2015}.

%% file: Chapters/Chapter4.tex
\chapter{Interferences in molecular photoionization}
\label{chapter4}
\fancyhead[LE]{\fontsize{11pt}{11pt}\selectfont Chapter 4. Interferences in molecular photoionization}
\fancyhead[RO]{\fontsize{11pt}{11pt}\selectfont \nouppercase{\rightmark}}

X-rays can ionize matter from their inner and valence shells, producing short\hyp{}wavelength electrons that can be diffracted by the surrounding atomic centers.
Consequently, scattered photoelectrons convey structural information about the system, which can be extracted by fitting experimental photoelectron spectra to analytical formulas.
Based on this principle, the NEXAFS (near-edge X-ray absorption fine structure) and the EXAFS (extended X-ray absorption fine structure) techniques \cite{SayersPRL1971,PettiferNat2005} can retrieve structural information of crystals and of bulk amorphous materials \cite{ChambersAP1991,HofmannNat1994,BresslerCR2004,GlatzelCCR2005,WoodruffSSR2007,ArcovitoPNAS2007,HaumannPNAS2008}, where the relatively low intensity of standard X-rays is compensated by the large number of particles in the sample.
However, obtaining structural information of isolated molecules is more difficult because of the low densities of the gas phase.
To overcome this difficulty, in addition to the development of the gas-phase EXAFS technique \cite{PangherPRL1993,PreserenJSR2001,SoderstromPRL2012}, other methods are being explored.
One of them takes advantage of the high brightness of the X-ray free electron lasers (XFEL), which can take time-resolved ``pictures'' through the so-called photoelectron holography \cite{KrasniqiPRA2010}.
However, its practical applications are still very limited due to the complexity and large dimensions of the recently operating XFEL facilities.
A more traditional method consists of measuring molecular-frame photoelectron angular distributions in photoionization with synchrotron radiation \cite{FukuzawaCPL2008,AdachiJPB2012,WilliamsPRL2012,WilliamsJPB2012}.
Also, recent work on diatomic molecules \cite{LiuJPB2006,CantonPNAS2011,PlesiatPRA2012,PlesiatPCCP2012} has shown that even the angle-integrated photoelectron spectra might be a valuable tool for structural determination.
Here we present an overview of our most significant results on photoionization of diatomic (CO, F$_2$) and small polyatomic (BF$_3$, CF$_4$) molecules with synchrotron radiation.
We show that the interferences arising between different ionization paths may encode structural information that can be extracted by analyzing the role of the nuclear motion.
This chapter constitutes only a summary of the work attached in appendices \ref{appendix1}, \ref{appendix2}, \ref{appendix3}, \ref{appendix4} and \ref{appendix5}, performed in collaboration with Piero Decleva (Universit\`a degli Studi di Trieste) and with the experimental groups of Catalin Miron (Synchrotron SOLEIL), Kiyosi Ueda (Synchrotron SPring-8), Edwin Kukk (University of Turku) and Thomas Darrah Tomas (University of Oregon).

\section{Intramolecular scattering in inner-shell photoionization} 
\label{section_innerShell}

\begin{figure}[h!]
\centering
\includegraphics[width=\textwidth]{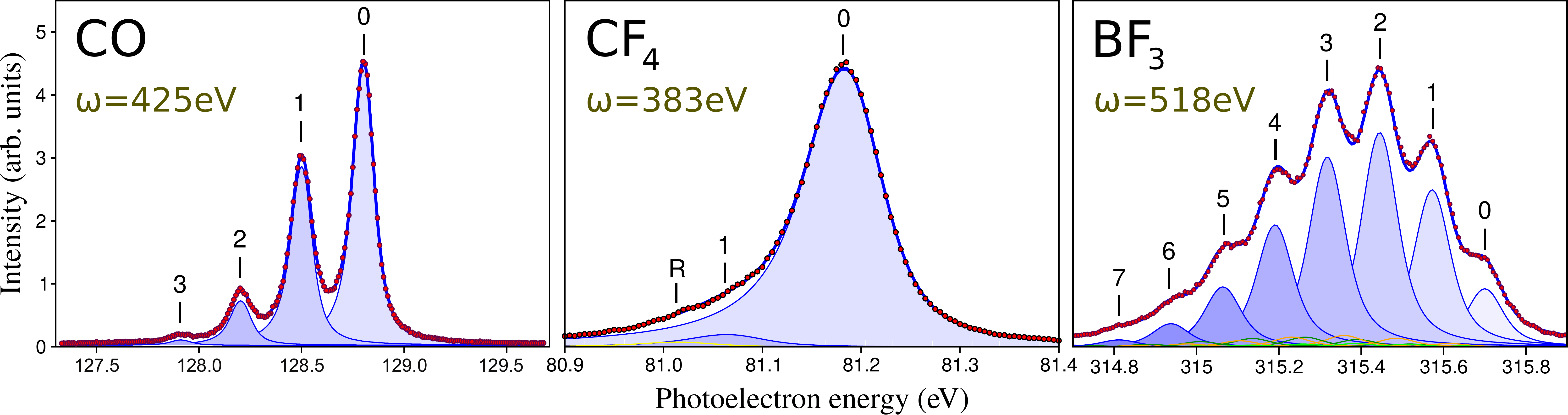}
\caption{Photoelectron spectra of CO (left), CF$_4$ (center) and BF$_3$ (right) taken SOLEIL at $h\nu = 425$, $383$ and $518$ eV, respectively.
Experimental results: red circles.
Thick blue line: fit of the experimental data.
Thin blue lines enclosing shaded areas: vibrational progression associated with the symmetric stretching mode.
Other thin lines: contribution of other modes resulting from recoil.
Peak labels indicate the vibrational quantum numbers of the core-hole species.
\newline}
\label{fig_expSpectra}
\includegraphics[width=\textwidth]{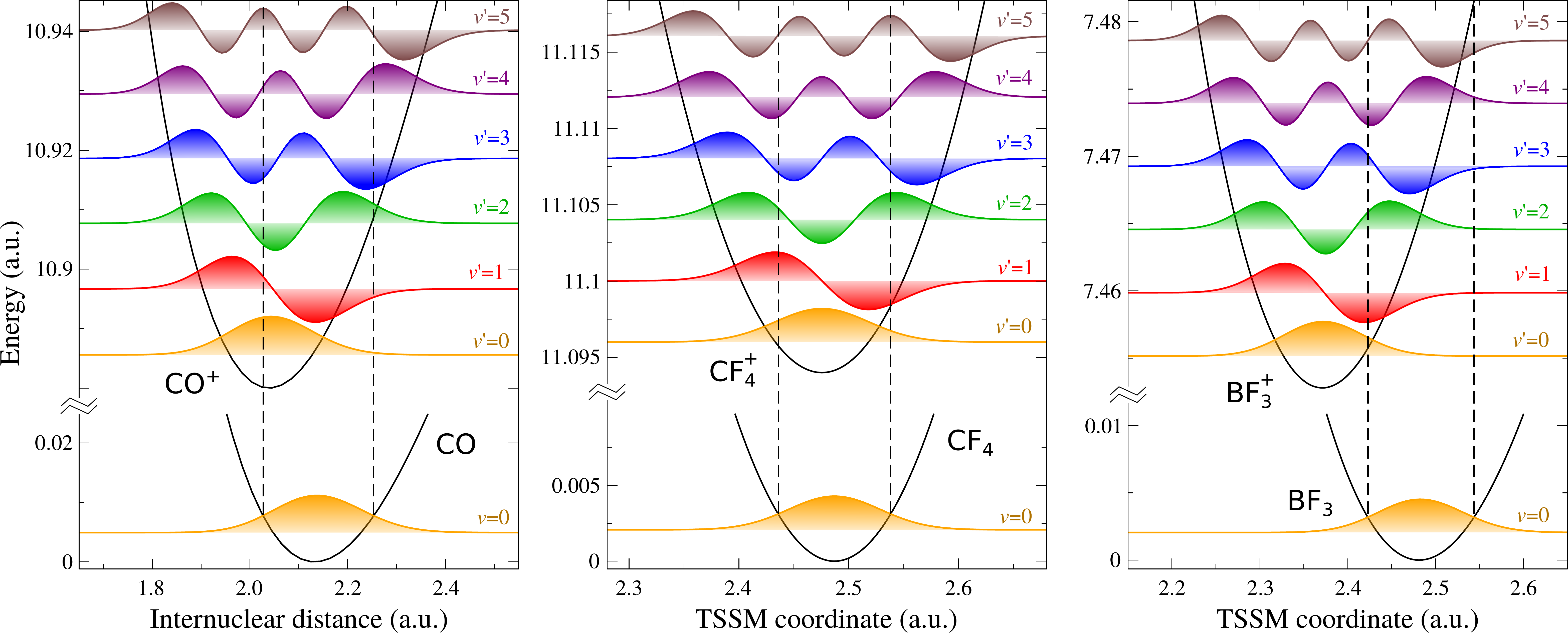}
\caption{Potential energy curves of the electronic ground state of CO (left), CF$_4$ (center) and BF$_3$ (right) and of the core-hole species generated upon C 1s (CO and CF$_4$) and B 1s (BF$_3$) ionization along the internuclear distance of CO and the TSSM coordinate of CF$_4$ and BF$_3$.
They have been constructed using reliable spectroscopic parameters available in the literature (CO: \cite{KempgensJPB1997,HergenhahnJPB2004}, CF$_4$: \cite{ThomasJCP2002,ThomasJCP2008}, BF$_3$: \cite{UedaPRA1992,KuchitsuJCP1996,NIST_chemistrybook,KirkpatrickJME2006,ThomasJCP2007}).
The relevant vibrational eigenstates are shown: the ground state of the neutral molecules (orange) and the low-lying states of the core-hole species (different colors), as well as the corresponding FC regions (dashed black lines).}
\label{fig_PECs}
\end{figure}

We have investigated inner-shell photoionization of small molecules, where an electron is ejected from a 1s orbital of a first-row atom.
Fig. \ref{fig_expSpectra} shows the photoelectron spectra of CO, BF$_3$ and CF$_4$ taken at photon energies of 425, 383 and 518 eV, respectively, at PLEIADES beamline \cite{PLEIADESurl} at SOLEIL Synchrotron.
The experimental spectrum of BF$_3$ and CF$_4$ (fig. \ref{fig_expSpectra}) shows several vibrational excitation peaks in the the totally-symmetric stretching mode (TSSM), which is the most affected by the electronic rearrangement accompanying core ionization \cite{ThomasJCP2007,ThomasJCP2008}.
Potential energy curves of the electronic ground state and of the core-hole species generated upon C 1s and B 1s ionization of CO, CF$_4$ and BF$_3$, repectively, are shown in fig. \ref{fig_PECs}, as well as the relevant vibrational eigenfunctions.
They have been evaluated using the harmonic (CF$_4$) and the Morse (CO, BF$_3$) approximations using reliable spectroscopic parameters available in the literature (CO: \cite{KempgensJPB1997,HergenhahnJPB2004}, CF$_4$: \cite{ThomasJCP2002,ThomasJCP2008}, BF$_3$: \cite{UedaPRA1992,KuchitsuJCP1996,NIST_chemistrybook,KirkpatrickJME2006,ThomasJCP2007}).
Photoionization of BF$_3$ leads to a large progression of vibrational levels $\nu'$, reaching up to $\nu'=7$, as a consequence of the favorable Franck\hyp{}Condon (FC) overlap between the initial and several final-state vibrational wave functions due to the large bond contraction accompanying core-ionization ($\Delta R_{\text{BF}}=-0.110 \text{ a.u.}$ \cite{ThomasJCP2007}).
The progression is limited to only two vibrational levels in the spectrum of CF$_4$ because the potential energy curves of the neutral and the ionic species (fig. \ref{fig_PECs}) are very similar in the region close to the equilibrium geometry ($\Delta R_{\text{CF}}=-0.0115 \text{ a.u.}$ \cite{ThomasJCP2008}).
The spectrum of CO shows an intermediate situation: 4 vibrational excitations in the only vibrational mode of the parent ion ($\Delta R_{\text{CO}}=-0.0932 \text{ a.u.}$ \cite{KempgensJPB1997}, see fig. \ref{fig_PECs}).\vspace{2 mm}

\begin{figure}[h!]
\centering
\includegraphics[width=\textwidth]{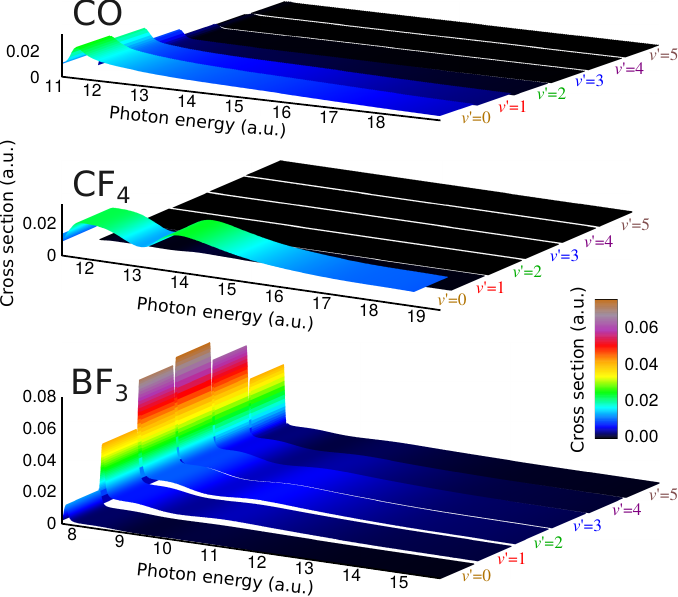}
\caption{Vibrationally resolved C 1s photoionization cross section of CO (upper figure) and CF$_4$ (center) and B 1s photoionization cross section of BF$_3$ (lower figure).
The nuclear motion in the polyatomic molecules BF$_3$ and CF$_4$ has been restricted to the TSSM.}
\label{fig_absoluteCS}
\end{figure}

Fig. \ref{fig_absoluteCS} shows the vibrationally resolved C 1s and B 1s photoionization cross sections of CO, CF$_4$ and BF$_3$, respectively, as a function of the photon energy.
They have been calculated as explained in section \ref{section_VRCS}, using the static-exchange Density Functional Theory (DFT) method within the Born-Oppenheimer approximation, and in the case of BF$_3$ and CF$_4$ the nuclear motion has been restricted to the TSSM coordinate.
In good agreement with the spectrum shown in fig. \ref{fig_expSpectra}, we observe a large progression of vibrational excitations upon B 1s ionization of BF$_3$, $\nu'=2$ and $\nu'=3$ being the dominant contributions in the entire energy range.
Only the low-lying vibrational eigenstates or the parent ion are excited in C 1s ionization of CO and CF$_4$, as experimentally found.
In all cases, we can distinguish that the photoionization cross sections (fig. \ref{fig_absoluteCS}) exhibit sharp increases near the ionization threshold due to the presence of shape resonances \cite{DillPRL1975,PiancastelliJESRP1999,ShimizuJCP1997}.
The origin of these structures can be understood in terms of a quasi-bound state embedded in the electronic continuum as a consequence to the existence of a small barrier in the molecular potential.
More subtle structures arise at higher energies due to photoelectron diffraction by the surrounding atomic centers (O in CO, F in BF$_3$ and CF$_4$).
However, the rapid decrease of the cross sections with the photon energy usually washes out scattering effects in the high\hyp{}energy region.
A better analysis can be performed by taking ratios between vibrationally resolved cross sections ($\nu-$ratios), since the decay is the same for each vibrational component.
Experimentally, presenting the cross sections as $\nu-$ratios is advantageous since certain calibration problems that one would face in case of absolute cross sections can be avoided.

\begin{figure}[H]
\centering
\includegraphics[width=\textwidth]{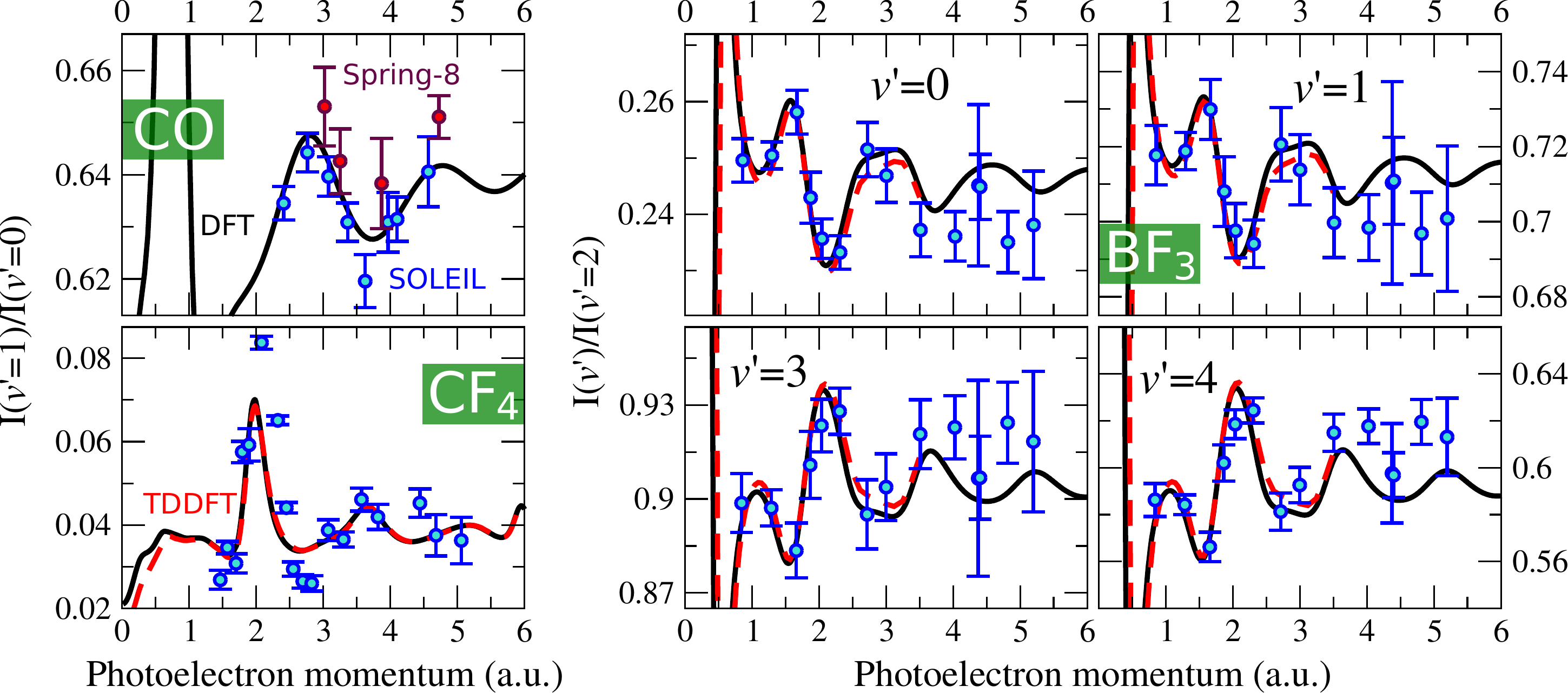}
\caption{Ratios between inner-shell vibrationally resolved photoionization cross sections of CO (upper left), CF$_4$ (lower left) and BF$_3$ (right), shown in fig. \ref{fig_absoluteCS}
Circles with error bars: experimental data including statistical errors taken at SOLEIL and Spring-8.
Black dashed lines: results from the static-exchange DFT calculations.
Red full lines: results of the TDDFT calculations (BF$_3$ and CF$_4$).
Horizontal dashed-dotted lines: ratios predicted by the FC approximation.}
\label{fig_vratios}
\end{figure}

Fig. \ref{fig_vratios} shows the experimental and theoretical $\nu-$ratios a function of the photoelectron momentum.
For the three systems, the $\nu-$ratios are calculated taking the largest contribution as a reference, which is $\nu'=0$ for CO and CF$_4$ and $\nu'=2$ for BF$_3$.
The shape resonances appear now as even sharper structures close to the ionization threshold.
At higher energies, the $\nu-$ratios exhibit pronounced oscillations superimposed to a nearly flat background which are a consequence of intramolecular scattering.
The periodicity of the oscillations is $2k_eR$, where $k_e$ is the photoelectron momentum and $R$ the distance between emitting and diffracting atoms, as in the well known EXAFS equation \cite{sternPRB1974}.
Our interpretation in terms of intramolecular scattering is supported by the good agreement with a first Born model \cite{PlesiatPRA2012} (see appendices \ref{appendix1} and \ref{appendix2}).

The agreement between theory and experiment is very good in all cases.
The excellent agreement between the results provided by the static-exchange and the time-dependent DFT methods indicates that interchannel coupling does not play an important role in core-ionization and that therefore one can rely on the former to interpret the experimental findings.
In the case of BF$_3$, the overall shape of the $\nu-$ratios is very similar for all $\nu \ge 3$; for $\nu'=0$ and $\nu'=1$, the oscillations are essentially identical but appear inverted.
The reason is that all vibrational contributions are referred to $\nu'=2$.
If we choose $\nu'=4$ instead, then the first four $\nu-$ratios would be inverted.
The same behavior is observed in CO and CF$_4$.
These results suggest that, for a given molecule, all $\nu-$ratios carry the same structural information.
In fact, as discuss in appendices \ref{appendix1} and \ref{appendix2}, the information contained in each individual $\nu-$ratio can be gathered in a generalized $\nu-$ratio.
In the case of BF$_3$, where the vibrational progression reaches up to $\nu'=7$, the use of a generalized $\nu-$ratio is very useful because it improves dramatically the statistical significance of the experimental data.

\subsubsection*{Extracting structural information}

\begin{figure}[h]
\centering
\includegraphics[width=\textwidth]{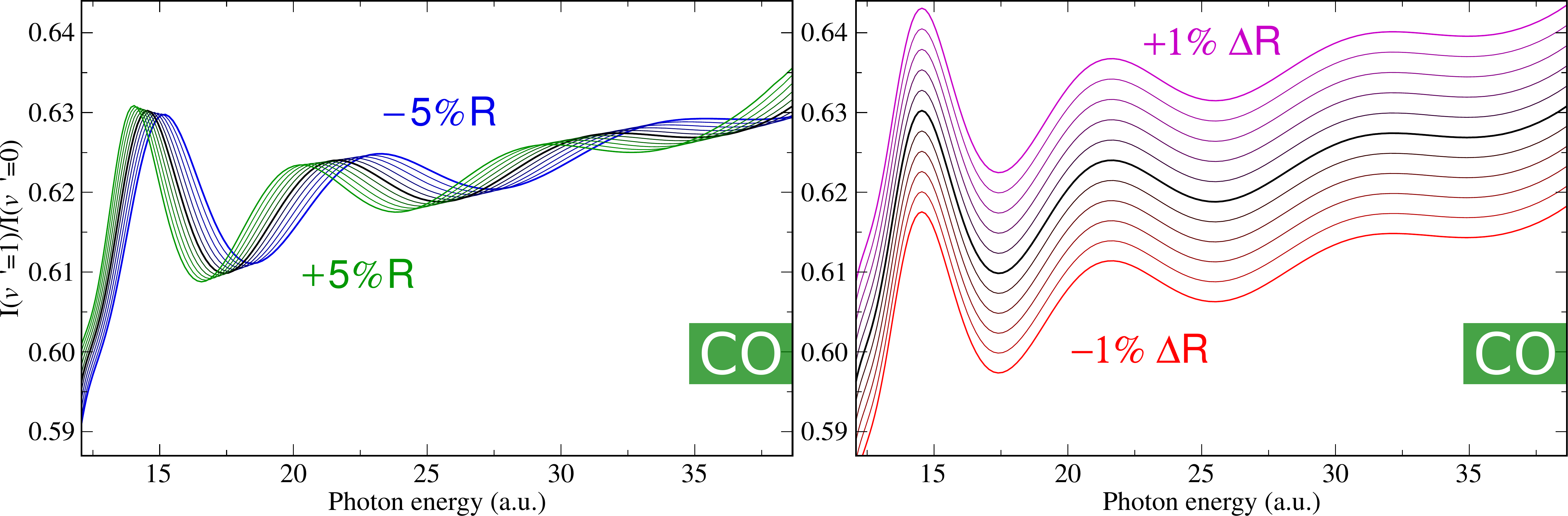}
\caption{Vibrational branching ratio $I(\nu'=1)/I(\nu'=0)$ corresponding to C 1s photoionization of CO calculated using the bibliographic values of $R$ and $\Delta R$ \cite{KempgensJPB1997,HergenhahnJPB2004} (both panels: black lines), 
increasing/decreasing $R$ in steps of $1\%$ (left pannel: green and blue lines)
and increasing/decreasing $\Delta R$ in steps of $0.2\%$ (right pannel: pink and red lines).
\newline\newline\newline}
\label{fig_vratios-CO}
\centering
\includegraphics[width=\textwidth]{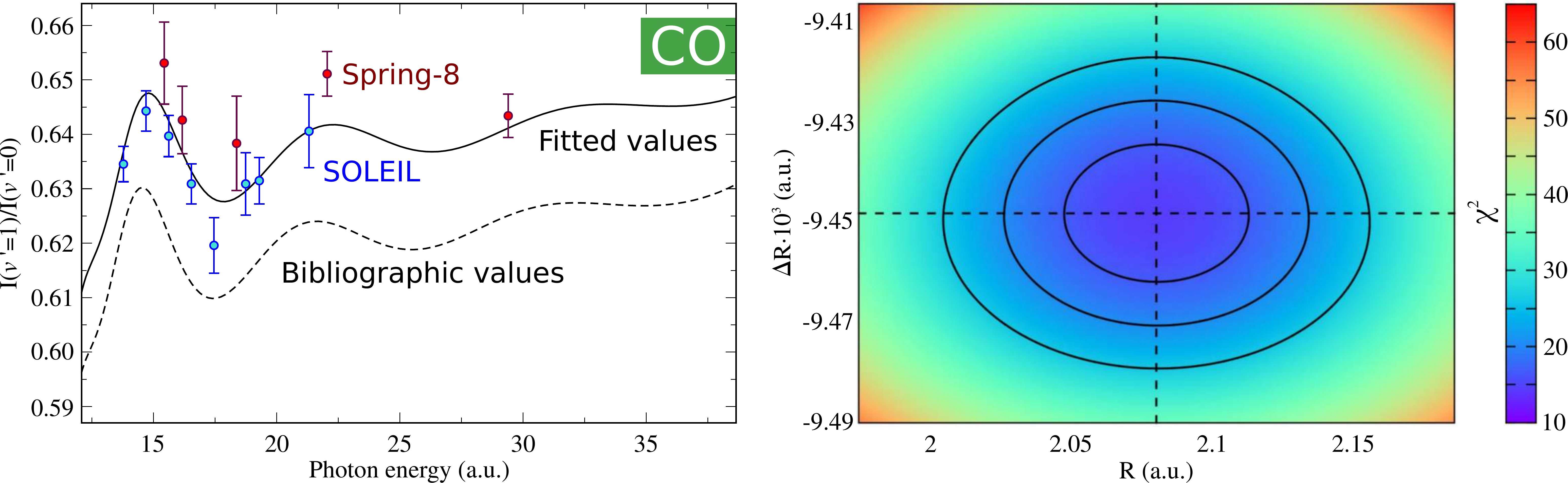}
\caption{Left panel: vibrational branching ratio $I(\nu'=1)/I(\nu'=0)$ corresponding to C 1s photoionization of CO measured at Spring-8 (red points) and SOLEIL (blue points)
and calculated using the bibliographic values of $R$ and $\Delta R$ \cite{KempgensJPB1997,HergenhahnJPB2004} (dashed black line) and those that provide the best agreement with the experimental points (full black line).
The optimum values of $R$ and $\Delta R$ have been obtained by performing a $\chi^2$ minimization procedure.
Right panel: $\chi^2$ function as a function of $R$ and $\Delta R$.
The black lines correspond to the limits of the confidence regions for confidence levels: $68.3\%$ ($1 \sigma$), $95.4\%$ ($2 \sigma$) and $99.7\%$ ($3 \sigma$).}
\label{vratios-fitCO}
\end{figure}

As the oscillatory patterns found in the $\nu-$ratios are due to intramolecular scattering, they convey structural information about the system.
In appendix \ref{appendix4}, we present a systematical approach for extracting this information.
As a proof of principle, we have applied it to the simultaneous determination of the internuclear distance of CO ($R=R_{\text{CO}}$) and the bond contraction ($\Delta R = R_{\text{CO}^+}-R_{\text{CO}}$) upon C 1s ionization.
Fig. \ref{fig_vratios-CO} illustrates of how the $\nu-$ratio $I(\nu'=1)/I(\nu'=0)$ changes when $R$ and $\Delta R$ are modified independently.
Since the bond contraction sets the overlap between the initial and the final vibrational wave functions, the $\nu-$ratios are shifted vertically when varying $\Delta R$ (left panel in fig. \ref{fig_vratios-CO}).
This effect is very sensitive because the FC overlap is strongly affected by small modifications of $\Delta R$, as can be seen in fig. \ref{fig_PECs}.
When $R$ is modified (right panel in fig. \ref{fig_vratios-CO}), the periodicity of the high-energy oscillations changes because they are due to intramolecular scattering and therefore depend on the distance between the emitting (C) and diffracting (O) centers.\vspace{2 mm}

We have performed a $\chi$-square minimization procedure in order to find the values of $R$ and $\Delta R$ that provide the theoretical $\nu$-ratio that is in best agreement with the experimental points measured at SOLEIL and Spring-8 (fig. \ref{vratios-fitCO}).
These values are $R =2.09 \pm 0.03$ a.u and $\Delta R = -0.0945 \pm 0.00014$ a.u.
Details of the fitting procedure can be found in appendix \ref{appendix4}.
The $\nu-$ratios computed using these values and those taken from the literature \cite{KempgensJPB1997,HergenhahnJPB2004}: $R _{\text{lit}}=2.1322$ a.u. and $\Delta R_{\text{lit}} = -0.0932$ a.u. are shown in fig. \ref{vratios-fitCO} as well as the experimental points.
We note that the literature values lay inside the confidence intervals provided by the method.
The $\chi$-square function of the fit is shown in \ref{vratios-fitCO} as a function of $R$ and $\Delta R$, as well as the and the isocurves that determine the bidimensional confidence regions for different confidence levels.
Due to the lack of local minima in the $\chi$-square function, we note that the fitting procedure always converges to the same values, regardless the initial guess.

\section{Photoionization of F$_2$: multicenter emission}
\label{section_F2}

Young's double slit interferences are expected to arise when an electronic wave is coherently emitted from two (several) atomic centers. 
As found by Cohen and Fano in the sixties \cite{CohenPR1966}, these interferences are imprinted in the angle-integrated photoelectron spectra of homonuclear diatomic molecules.
This phenomenon has attracted the interest of various authors \cite{SemenovJPB2006,LiuJPB2006,EharaJCP2006,FernandezPRL2007,CantonPNAS2011,ArgentiNJP2012,PlesiatPCCP2012} in the last few years.
We have investigated multicenter emission effects in the fluorine molecule by analyzing dissociative and non-dissociative ionization from inner and valence shells.
Here we present a brief summary of our work, which is explained in detail in appendix \ref{appendix5}.

\begin{figure}[h]
\centering
\includegraphics[scale=0.5]{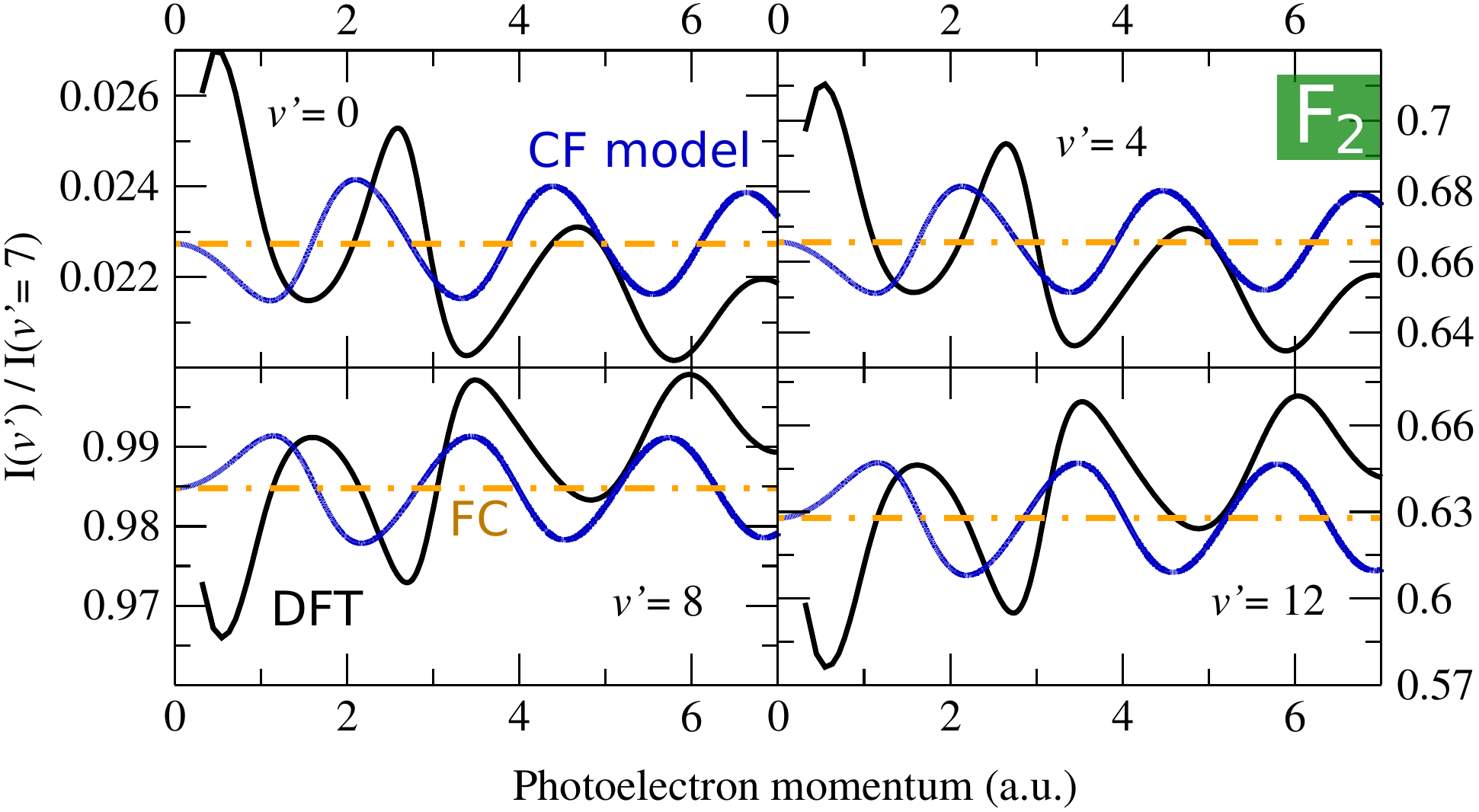}
\caption{Ratios between vibrationally resolved $3\sigma_{g}$ photoionization cross sections photoionization of F$_{2}$.
Black lines: DFT theory; blue dashed lines: Cohen-Fano model; orange dashed lines: FC values.}
\label{fig_vratios-F2}
\end{figure}

Fig. \ref{fig_vratios-F2} shows some $\nu-$ratios corresponding to ionization from the $3\sigma_{g}$ orbital as a function of the photoelectron momentum.
We have chosen $\nu'=7$ as the common denominator because it constitutes the largest contribution to the total cross section.
As in fig. \ref{fig_vratios}, the $\nu-$ratios show pronounced oscillations around the FC value as a function of the photoelectron momentum.
However, in this case they are not due to photoelectron diffraction but to multicenter emission as the electron is ejected from an orbital that is delocalized between the two fluorine atoms.
In order to confirm this assumption, we have extended formula developed by Cohen and Fano in the sixties \cite{CohenPR1966} to account for the vibrational motion, as is \cite{CantonPNAS2011}.
The results of the model are included in fig. \ref{fig_vratios-F2}, where we can see that the agreement with the static-exchange theory is very good.
As discussed in appendix \ref{appendix5}, the agreement is not so good in the case of $1\pi_{g}$ and $1\pi_{u}$ ionization because $\pi$ orbitals concentrate most of the electron density outside the molecular axis. 
However, even in those situations, the amplitudes and periodicities of the oscillations predicted by the Cohen-Fano formula are close to the ones provided by the static-exchange DFT theory.
Only the phases are not properly described.

\section{Conclusions}

In summary, we have found measurable evidence of intramolecular scattering and multicenter emission occurring in the photoionization of small molecules at high photoelectron energies.
The details of the work presented here can be found in appendices \ref{appendix1}, \ref{appendix2}, \ref{appendix3}, \ref{appendix4} and \ref{appendix5}.
When an electron is emitted from a very localized region of a molecule, such as the C 1s (B 1s) orbital in CO or CF$_4$ (BF$_3$), features due to photoelectron diffraction are expected to arise, 
whereas in the case of ionization from a delocalized orbital, like those in F$_2$, one should expect to observe Cohen-Fano-like interferences.
Vibrationally resolved photoelectron spectroscopy allows to detect these high\hyp{}energy interferences in an elegant and consistent way, because
(i) the problem of the rapid decrease of the ionization probability with the photon energy can be avoided by monitoring the ratios between vibrationally resolved cross sections, and
(ii) the effect of the interferences manifest differently in different final vibrational states.
The combination of state-of-the-art DFT-like calculations and high-resolution third-generation synchrotron facilities has enabled to explore these non\hyp{}Franck\hyp{}Condon effects both theoretically and experimentally,
demonstrating that the nuclear response to intramolecular electron diffraction / multicenter emission is observable and can provide structural information of both the neutral molecule and the ionized species.

%% file: Chapters/Chapter5.tex
\chapter{Ultrafast electron dynamics in aminoacids}
\label{chapter5}
\fancyhead[LE]{\fontsize{11pt}{11pt}\selectfont Chapter 5. Ultrafast electron dynamics in aminoacids}
\fancyhead[RO]{\fontsize{11pt}{11pt}\selectfont \nouppercase{\rightmark}}

Intramolecular charge transfer is the trigger of important chemical and biological processes, such as photosynthesis \cite{EberhardARG2008}, cellular respiration \cite{CordesCSR2009} or DNA damage \cite{bookChapterBecker2007}.
The study of charge transfer within isolated complex molecules was pioneered by R. Weinkauf and coworkers in the 90s \cite{WeinkaufJPC1994,WeinkaufJPC1995,WeinkaufJPC1996}.
They were able to track the motion of a positive hole generated upon ionization through up to 12 sigma bonds of a tetrapeptide by analyzing light absorption shifts.
However, the time resolution in their experiments was limited by the durations of the pulses they employed, that were around 200 ns.
One decade later, by using femtosecond pulses, they could measure how long it took for a positive charge to move from the phenyl to the amino group in the PENNA molecule: $(80\pm20)$ fs \cite{LehrJPCA2005}.
In their paper, they suggested that the charge transfer was probably mediated by the nuclear motion through a conical intersection.
Motivated by the pioneering work of R. Weinkauf and collaborators \cite{WeinkaufJPC1994,WeinkaufJPC1995,WeinkaufJPC1996}, Lorenz S. Cederbaum and coworkers demonstrated that electron correlation can drive ultrafast charge dynamics in a time scale that is faster than the onset of the nuclear motion \cite{CederbaumCPL1999}.
This phenomenon has been referred to as charge migration to distinguish it from charge transfer mediated by nuclear motion and, 
over the last two decades, it has been widely investigated in a large number of molecules of biological interest \cite{RemaclePNAS2006,CederbaumCPL1999,BreidbachJCP2003,BreidbachPRL2005,HennigJPCA2005,KuleffJCP2005,KuleffCP2007}.\vspace{2 mm}

\begin{figure}[h!]
\centering
\includegraphics[scale=0.5]{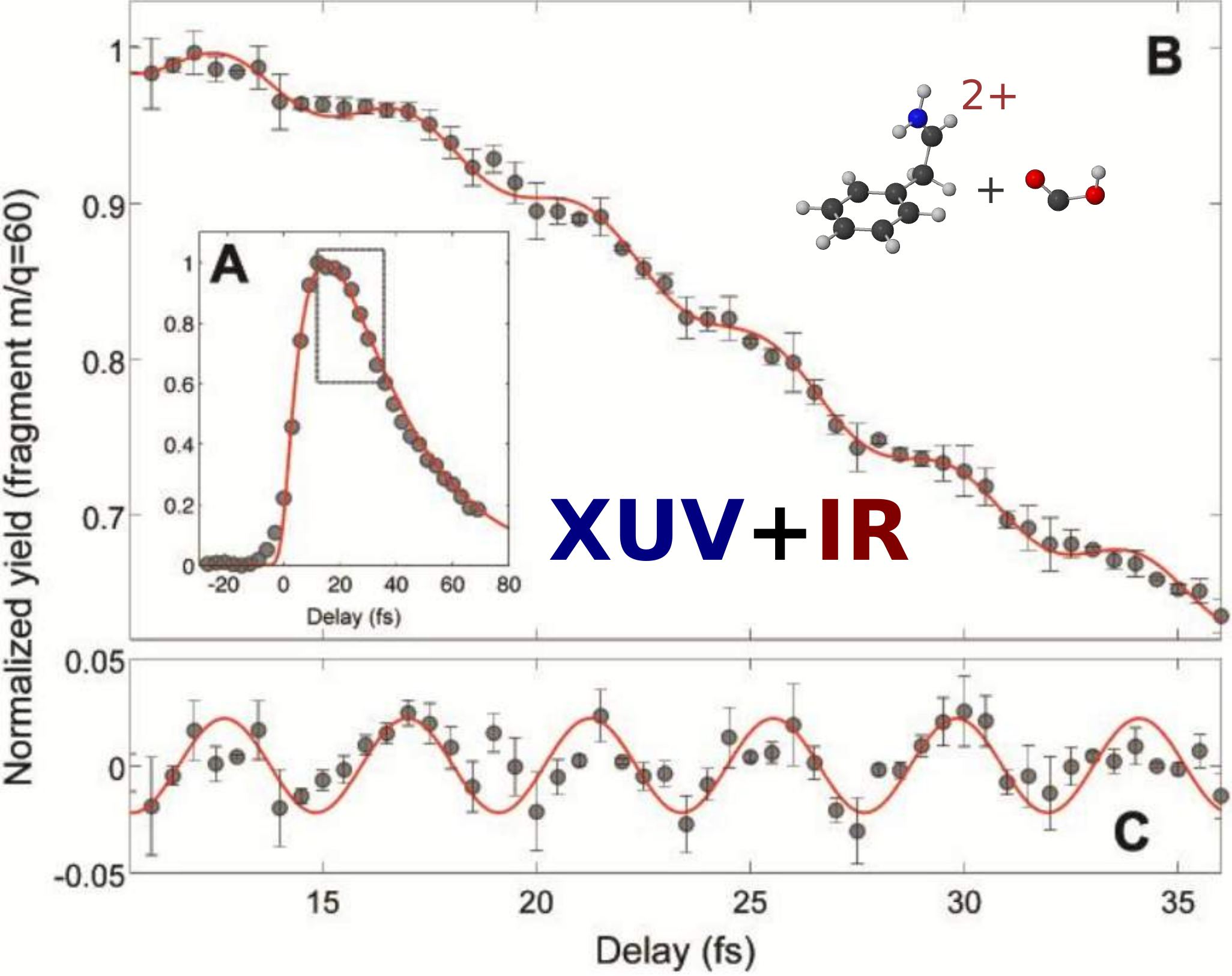}
\caption{Experimental data from the group of Mauro Nisoli.
(A) Yield of doubly charged immonium ion (mass/charge = 60) as a function of pump-probe delay, measured with 3-fs temporal steps.
The red line is a fitting curve with an exponential rise time of 10 fs and an exponential relaxation time of 25 fs.
(B) Yield of doubly charged immonium ion versus pump-probe delay measured with 0.5-fs temporal steps, within the temporal window shown as dotted box in (A).
Error bars show the standard error of the results of four measurements.
The red line is the fitting curve given by the sum of the fitting curve shown in (A) and a sinusoidal function of frequency 0.234 PHz (4.3-fs period).
(C) Difference between the experimental data and the exponential fitting curve displayed in (A). Red curve is a sinusoidal function of frequency 0.234 PHz}
\label{fig_fragmentationYield_phe}
\end{figure}

The application of attosecond technology to the study of complex molecules has lead to the experimental demonstration of charge migration in a biomolecule: the amino acid phenylalanine \cite{CalegariScience2014,CalegariIEEE2015}.
The $\alpha-$amino acids consist of a central carbon atom ($\alpha$ carbon) linked to an amino group ($-$NH$_2$), a carboxyl group ($-$COOH), a hydrogen atom and a side chain (R), which in the case of phenylalanine is a benzyl group.
A two-color pump-probe technique was used in the experiment.
Charge dynamics were initiated by isolated XUV sub-300-as pulses, with photon energy in the spectral range between 15 and 35 eV and probed by 4-fs, waveform-controlled visible/near infrared (VIS/NIR, central photon energy of 1.77 eV) pulses (see appendices \ref{appendix6} and \ref{appendix7}).
Ionization induced by the attosecond pulse occurred in a sufficiently short time interval to exclude substantial electron rearrangement during the excitation process.
The yield for the production of doubly charged immonium ions was measured as a function of the time delay between the attosecond pump pulse and the VIS/NIR probe pulse.
Fig. \ref{fig_fragmentationYield_phe}a shows the results on a 100-fs time scale.
The experimental data display a rise time of $(10\pm2)$ fs and an exponential decay with time constant of $(25\pm2)$ fs (this longer relaxation time constant is in agreement with earlier experimental results reported in \cite{BelshawJPCL2012}).
Fig. \ref{fig_fragmentationYield_phe}b shows a 25-fs-wide zoom of the pump-probe dynamics, obtained by reducing the delay step between pump and probe pulses from 3 to 0.5 fs.
An oscillation of the dication yield is clearly visible.
For a better visualization, fig. \ref{fig_fragmentationYield_phe}c shows the same yield after subtraction of an exponential fitting curve.
The data have been fitted with a sinusoidal function of frequency 0.234 PHz (corresponding to an oscillation period of 4.3 fs), with lower and upper confidence bounds of 0.229 and 0.238 PHz, respectively (see appendices \ref{appendix6} and \ref{appendix7}).
The ultrafast oscillations in the temporal evolution of the dication yield cannot be related to nuclear dynamics, which usually come into play on a longer temporal scale, ultimately leading to charge localization in a particular molecular fragment.
Therefore, these measurements constitute the first experimental observation of purely electron dynamics in a biomolecule. \vspace{2 mm}

In order to verify that the observed oscillations are not related to any nuclear dynamics, we have calculated the vibrational frequencies and the corresponding periods of phenylalanine by means of Density Functional Theory (DFT) using the B3LYP functional \cite{LeePRB1988,BeckeJCP1993} and a 6$-$311+g(3df,2p) basis set, implemented in the quantum chemistry package Gaussian 09 \cite{Gaussian09}.
The results are given in table \ref{table_vibrFreq_phe}.
Our calculations show that the highest vibrational frequency is 0.11 PHz, which corresponds to a period of 8.9 fs, associated with X-H stretching modes, whereas skeleton vibrations are even slower, so that one can rule out that the observed beatings are due to vibrational motion.
In any case, some influence of the nuclear motion cannot be completely excluded, because, for example, stretching of the order of a few picometers of carbon bonds can occur in a few femtoseconds, and this could modify the charge dynamics \cite{LunnemannCPL2008,MendiveTapiaJCP2013}.\vspace{2 mm}

\begin{table}
\small
\begin{center}
\begin{tabular}{|ccc|c|ccc|}
\cline{1-3}\cline{5-7}
Mode & Freq. (PHz)  & Period (fs) & & Mode & Freq. (PHz)  &  Period (fs)   \\ \cline{1-3}\cline{5-7}
 1  &       0.0011   &    894.8 & &   33  &    0.0335      &  29.8 \\ 
 2  &       0.0013   &    740.7 & &   34  &    0.0342      &  29.2 \\ 
 3  &       0.0020   &    499.2 & &   35  &    0.0353      &  28.3 \\ 
 4  &      0.0031    &   319.9  & &   36  &    0.0355      &  28.2 \\ 
 5  &      0.0057    &   174.2  & &   37  &    0.0362      &  27.6 \\ 
 6  &      0.0069    &   145.3  & &   38  &    0.0365      &  27.4 \\ 
 7  &      0.0074    &   134.5  & &   39  &    0.0373      &  26.8 \\ 
 8  &      0.0088    &   113.9  & &   40  &    0.0388      &  25.8 \\ 
 9  &      0.0114    &    88.0   & &   41  &    0.0399      &  25.1 \\
10  &     0.0123     &   81.6   & &   42  &    0.0403      &  24.8 \\ 
11  &     0.0125     &   79.8   & &   43  &    0.0410      &  24.4 \\ 
12  &     0.0146     &   68.4   & &   44  &    0.0413      &  24.2 \\ 
13  &     0.0148     &   67.6   & &   45  &    0.0421      &  23.8 \\ 
14  &     0.0170     &   59.0   & &   46  &    0.0446      &  22.4 \\ 
15  &     0.0181     &   55.2   & &   47  &    0.0446      &  22.4 \\ 
16  &     0.0191     &   52.5   & &   48  &    0.0459      &  21.8 \\ 
17  &     0.0193     &   51.7   & &   49  &    0.0486      &  20.6 \\ 
18  &     0.0214     &   46.7   & &   50  &    0.0493      &  20.3 \\ 
19  &     0.0220     &   45.5   & &   51  &    0.0500      &  20.0 \\ 
20  &     0.0232     &   43.1   & &   52  &    0.0544      &  18.4 \\ 
21  &     0.0234     &   42.8   & &   53  &    0.0907      &  11.0 \\ 
22  &     0.0256     &   39.0   & &   54  &    0.0916      &  10.9 \\ 
23  &     0.0260     &   38.5   & &   55  &    0.0929      &  10.8 \\ 
24  &     0.0266     &   37.7   & &   56  &    0.0946      &  10.6 \\ 
25  &     0.0276     &   36.2   & &   57  &    0.0947      &  10.6 \\ 
26  &     0.0283     &   35.3   & &   58  &    0.0950      &  10.5 \\ 
27  &     0.0299     &   33.5   & &   59  &    0.0952      &  10.5 \\ 
28  &    0.0300      &  33.3    & &   60  &    0.0956      &  10.5 \\ 
29  &    0.0302      &  33.1    & &   61  &    0.1050      &  9.5 \\ 
30  &    0.0304      &  32.9    & &   62  &    0.1070      &  9.3 \\ 
31  &    0.0315      &  31.7    & &   63  &    0.1125      &  8.9 \\ 
32  &    0.0331      &  30.2    & \multicolumn{1}{c}{} & & & \multicolumn{1}{c}{} \\ 
\end{tabular}
\end{center}
\caption{Frequencies (PHz) and periods (fs) of the vibrational modes of phenylalanine.}
\label{table_vibrFreq_phe}
\end{table}

\begin{figure}
\centering
\includegraphics[width=\textwidth]{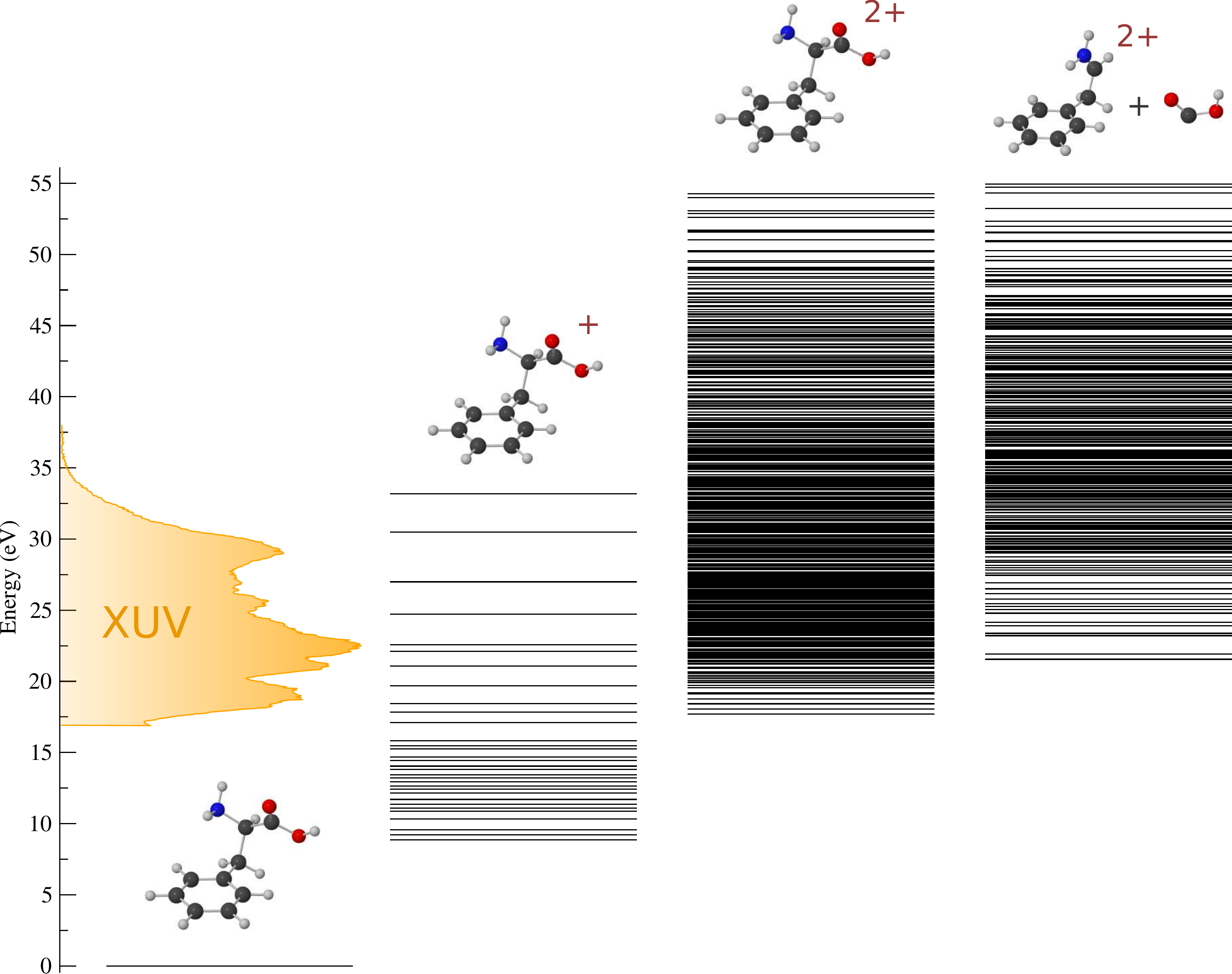}
\caption{Energy level diagram containing the states of singly charged phenylalanine populated by the XUV pulse, whose energy distribution is included as a shadowed area in the axis bar, the states of doubly-charged phenylalanine and those of the system doubly-charged immonium + neutral carboxyl.}
\label{fig_energyDiagram_phe}
\end{figure}

Fig. \ref{fig_energyDiagram_phe} shows an energy-level diagram of the electronic states of singly charged phenylalanine (phe$^{+}$) accessible by the XUV pulse, 
all the states of doubly-charged phenylalanine (phe$^{++}$) and those of the system doubly-charged immonium + carboxyl.
The energies of the latter two systems have been evaluated within the static-exchange approximation, which is accurate enough to provide a qualitative picture of the full process.
As can be seen, one can go from a highly excited state of phe$^{+}$ to the lowest states of the phe$^{++}$ or the dissociated system by absorbing just a few VIS/NIR photons (photon energy around 1.77 eV).
Of course, one cannot know how likely this transition will be, but one can unambiguously say that the process only requires absorption of very few VIS/NIR photons.
Even if these transitions were unlikely, e.g., due to unfavorable overlap between initial and final orbitals, the transition should be much more likely than others involving many photons even with favorable overlap.
Since the HOMO orbital of phenylalanine is substantially localized on the amino group, ionization by the VIS/NIR pulse is expected to occur from this part of the molecule.
Therefore, the removal of the second electron is sensitive to charge localization on the amino group. \vspace{2 mm}

\begin{figure}
\centering
\includegraphics[scale=0.25]{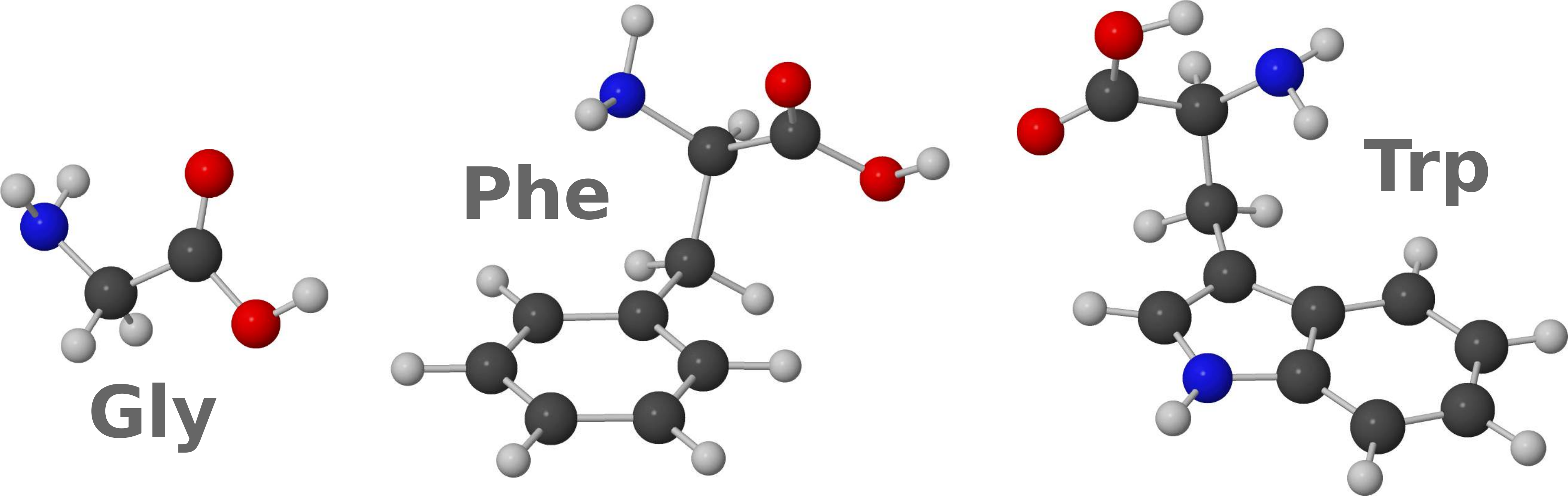}
\caption{Geometry of the most abundant conformers of glycine \cite{CsaszarJACS1992}, phenylalanine \cite{HuangTHEOCHEM2006} and tryptophan \cite{SnoekPCCP2001} at $430$ K, the temperature of the pump-probe experiment presented in fig. \ref{fig_fragmentationYield_phe}.}
\label{fig_geometries}
\end{figure}

In order to understand the origin of the ultrafast oscillations shown in fig. \ref{fig_fragmentationYield_phe}, we have evaluated the electronic wave packet generated in phenylalanine upon ionization by an attosecond pulse similar to that used in the experiment and the subsequent evolution of the electron density.
We aim to understand the influence of different radicals, so we have performed a systematical study including the amino acids glycine and tryptophan.
For the evaluation of the ionization amplitudes and the wave packet dynamics, we have employed the static-exchange DFT method within the formalism of time-dependent first-order perturbation theory, as explained in chapter \ref{chapter3}.
The most stable conformers of glycine \cite{CsaszarJACS1992}, phenylalanine \cite{HuangTHEOCHEM2006} and tryptophan \cite{SnoekPCCP2001} are depicted in fig. \ref{fig_geometries}.
It is well known that, even at room temperature, amino acids present several conformations due to their structural flexibility.
In the case of phenylalanine, 37 conformers have been theoretically found \cite{HuangTHEOCHEM2006}.
This chapter is organized as follows: first, we give the details about the evaluation of the electronic stationary states and show the relevant molecular orbitals of the three amino acids (section \ref{section_aa_MOs}) and the corresponding photoionization cross sections (section \ref{section_aa_CS}), restricting the analysis to the most stable conformers at the temperature of the experiment presented in fig. \ref{fig_energyDiagram_phe}.
Then, we analyze the wave packet dynamics initiated by an attosecond pulse similar to that used in the experiment (section \ref{section_aa_holeDynamics}), performing a Fourier analysis of the electron density on different portions of the molecules and comparing our results with experimental data and with previous theoretical work, when available.
We conclude by exploring the effect of molecular conformation (section \ref{section_conformation}) and the role of the photoelectron emission dynamics (section \ref{section_fullSystem}) in the charge migration mechanism.

\section{Evaluation of electronic states}
\label{section_aa_MOs}

We employ Slater determinants to represent electronic stationary states, as explained in section \ref{section_electronic}.
The corresponding bound (Kohn-Sham) and continuum orbitals have been evaluated in a basis set of B-splines and spherical harmonics.
In particular, we used a large one-center expansion (OCE) of a variable number of B-spline functions, from $118$ in glycine to $135$ in tryptophan, enclosed in a sphere of $30$ a.u. with origin in the center of mass, using spherical harmonics up to an angular momentum of $l=20$ (see eq. \ref{basisExpansion}).
The OCE was complemented with small off centers, located at the atomic positions, with sizes varying from $0.2$ to $1.6$ a.u., larger for the heavier nuclei since they accumulate more electron density. 
The angular expansion in each off-center was limited to $l=2$.
The LB94 \cite{LeeuwenPRA1994} functional was employed to account for electronic exchange and correlation effects.
An initial guess for the electronic density of the three amino acids was generated with the Amsterdam Density Functional (ADF) package \cite{FonsecaTCA1998,VeldeJCC2001,ADF2013} using a double $\zeta$-polarization plus (DZP) basis set in the case of glycine and a triple $\zeta$-polarization plus (TZP) \cite{bookLevine} for the more complex amino acids phenylalanine and tryptophan.
Since it is well known that the LB94 functional overestimates the molecular orbital eigenvalues, the first ionization potential of glycine was calculated using the outer-valence Green's function (OVGF) \cite{NiessenCPR1984} method implemented in Gaussian09 \cite{Gaussian09}, which can provide accurate values of the ionization potentials of the outer-valence shells. 
Then, the DFT/LB94 eigenvalues were shifted according to the energy difference between the first IP provided by the OVGF method and the DFT/LB94 calculation.
In order to obtain reliable values of the ionization energies of phenylalanine and tryptophan, we have employed the VWN \cite{VoskoCJP1980} local density approximation functional within the Slater transition state procedure \cite{SlaterAQC1972} using ADF with a TZP basis set.
The molecular geometries where previously optimized at the DFT/B3LYP \cite{LeePRB1988,BeckeJCP1993} level in a 6–311+g(3df,2p) basis set (6-31+g(d) in the case of tryptophan) using Gaussian09 \cite{Gaussian09}, starting from the approximate optimized geometries reported in \cite{CsaszarJACS1992,HuangTHEOCHEM2006,SnoekPCCP2001}.\vspace{2 mm}

\begin{figure}
\centering
\includegraphics[width=\textwidth]{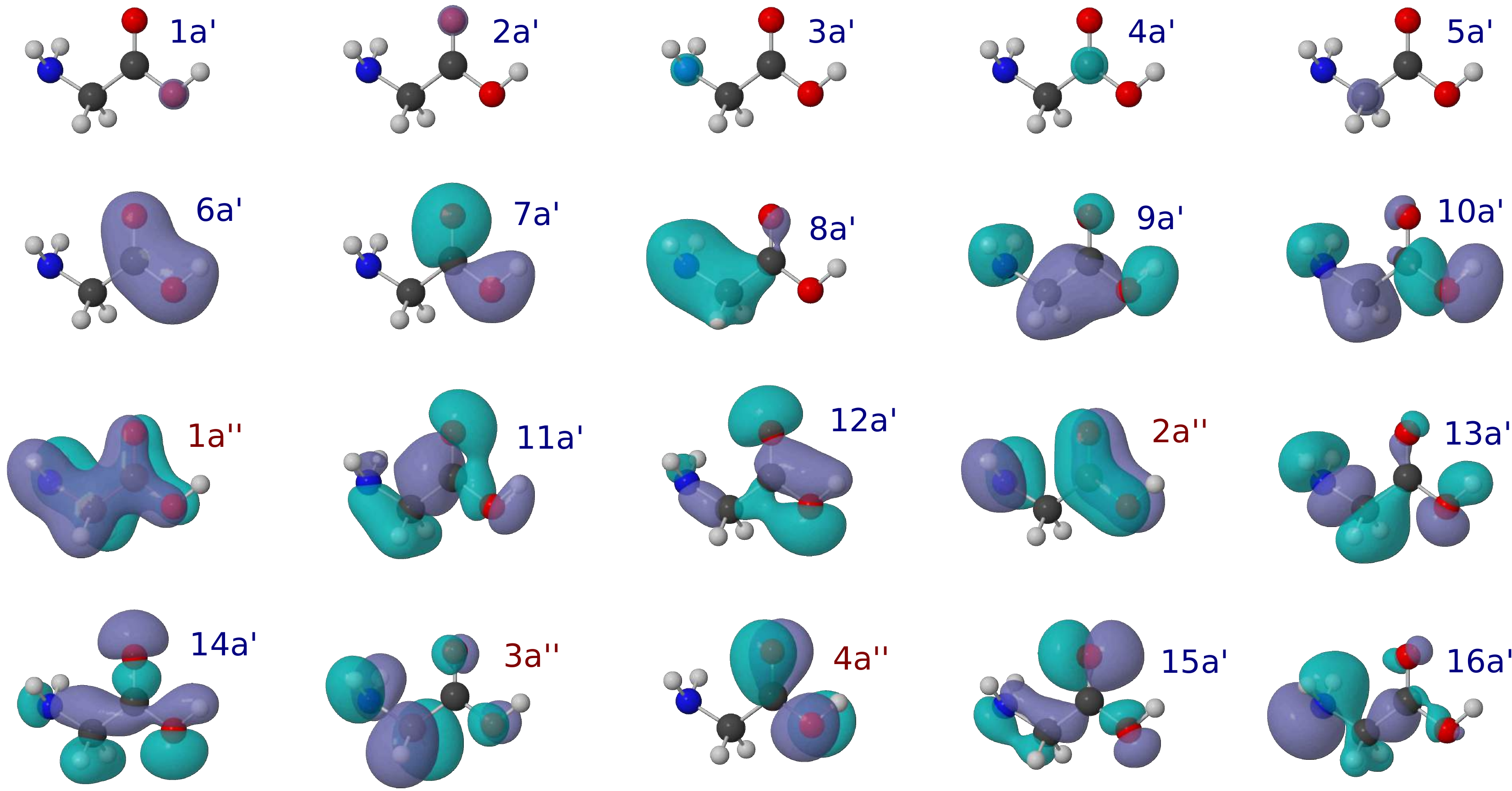}
\caption{Occupied Kohn-Sham orbitals of neutral glycine.
They have been calculated using the LB94 \cite{LeeuwenPRA1994} functional in a basis set of B-spline functions, as explained in the text.}
\label{fig_MOs_gly}
\end{figure}

\begin{figure}
\centering
\includegraphics[width=\textwidth]{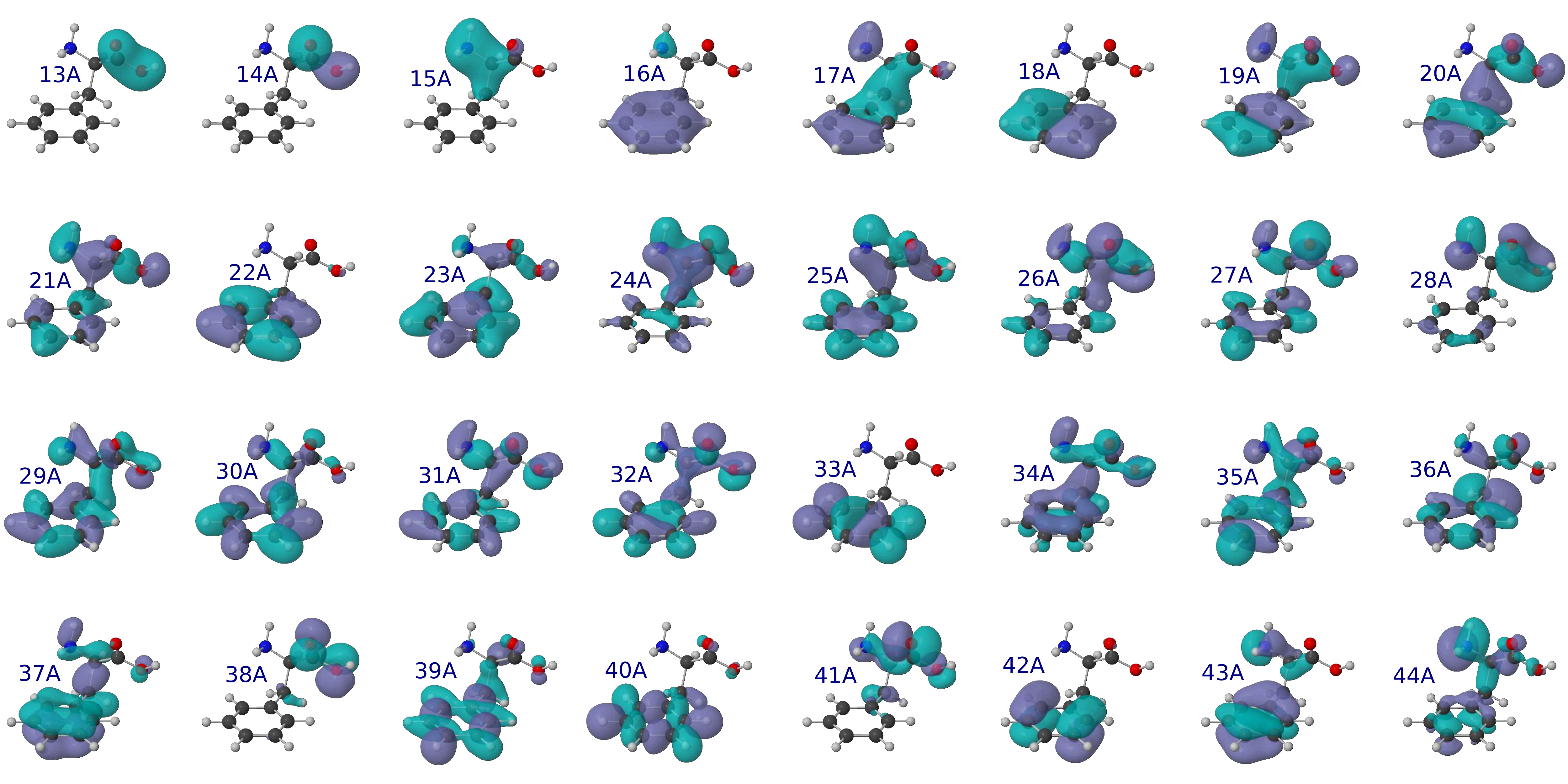}
\caption{Same as fig. \ref{fig_MOs_gly} for phenylalanine (core orbitals have been omitted).}
\label{fig_MOs_phe}
\end{figure}

\begin{figure}
\centering
\includegraphics[width=\textwidth]{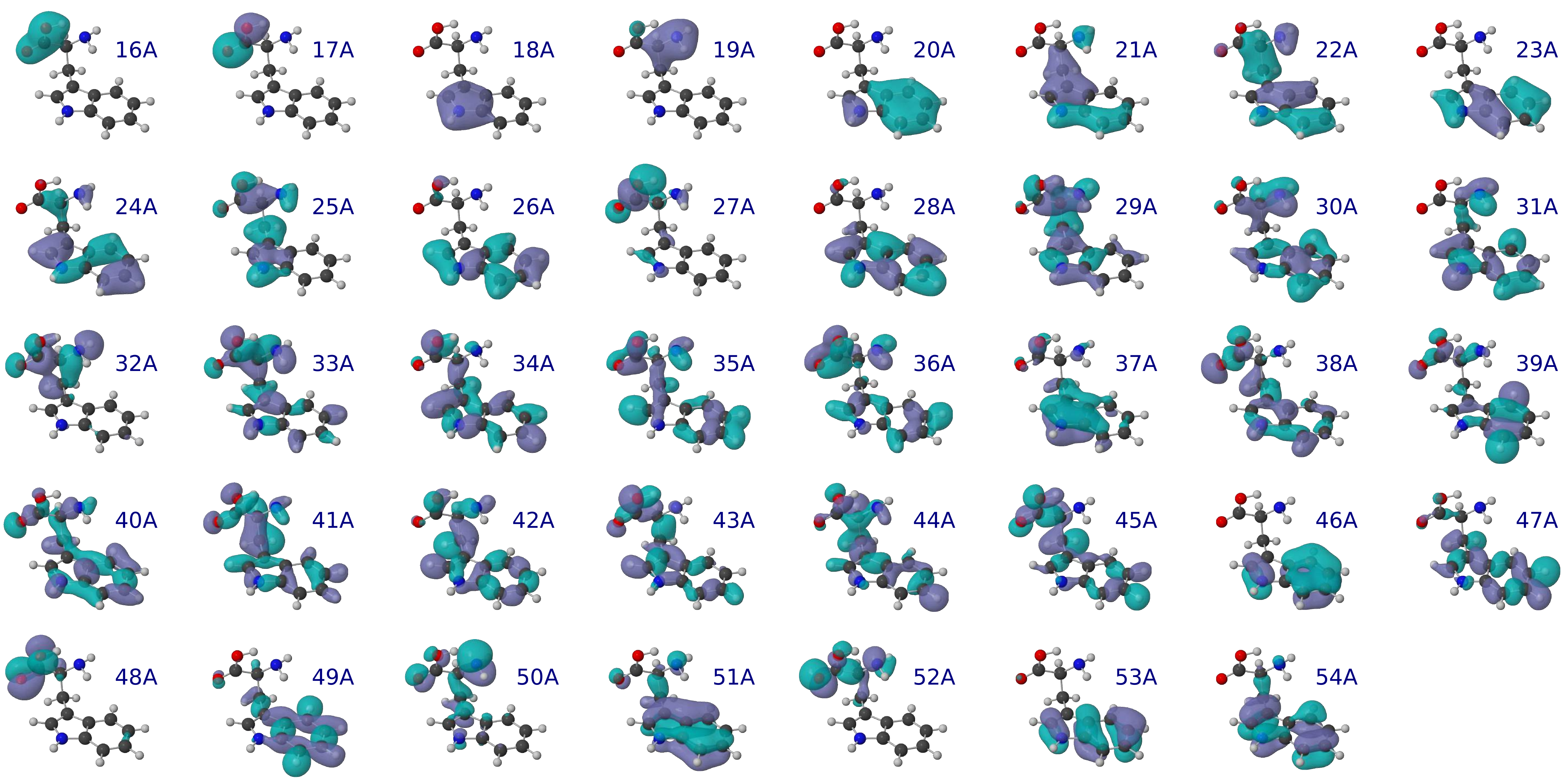}
\caption{Same as fig. \ref{fig_MOs_gly} for tryptophan (core orbitals have been omitted).}
\label{fig_MOs_trp}
\end{figure}


Figs. \ref{fig_MOs_gly}, \ref{fig_MOs_phe} and \ref{fig_MOs_trp} show the occupied Kohn-Sham orbitals of the ground state of glycine, phenylalanine and tryptophan, respectively, which were used to evaluate the corresponding wave functions according to equations \ref{elecGS} and \ref{elecCont}.
Glycine has planar symmetry and therefore belongs to the $C_s$ point group and its orbitals have either a$'$ or a$''$ symmetry.
As can be seen in fig. \ref{fig_MOs_gly}, a$'$ orbitals are symmetric with respect to reflection thought the mirror plane and a$''$ orbitals are antisymmetric and thus contain a nodal plane.
Phenylalanine and tryptophan belong to the $C_1$ point group because they are not invariant under any symmetry transformation except for the identity operation and therefore all their orbitals have A symmetry. \vspace{2 mm}

Neutral glycine, phenylalanine and tryptophan have $40$, $88$ and $108$ electrons, respectively.
Therefore, the corresponding electronic ground states constitute closed-shell systems that can be accurately described using $20$, $44$ and $54$ molecular orbitals.
Core orbitals are those constituted by the 1s orbitals of the ``heavy'' atoms (C, N and O).
The energy required to remove an electron from a core orbital in these molecules ranges from around $290$ eV in the case of C 1s to $400$ eV for N 1s and $535$ eV for O 1s.
They are thus not accessible with usual attosecond pulses generated via HHG. 
Valence orbitals, with ionization potentials ranging from around $10$ to $20$ eV, are the easiest to ionize.
They are highly delocalized, especially in phenylalanine and tryptophan since they contain an aromatic ring.
Inner-valence orbitals present an intermediate situation, with ionization energies from around $20$ to $35$ eV.
In order to describe the interaction with an XUV pulse, capable of ionizing from all valence and inner-valence shells, the corresponding ionization amplitudes need to be evaluated.

\section{Photoionization cross sections}
\label{section_aa_CS}

\begin{figure}
\centering
\includegraphics[width=12cm]{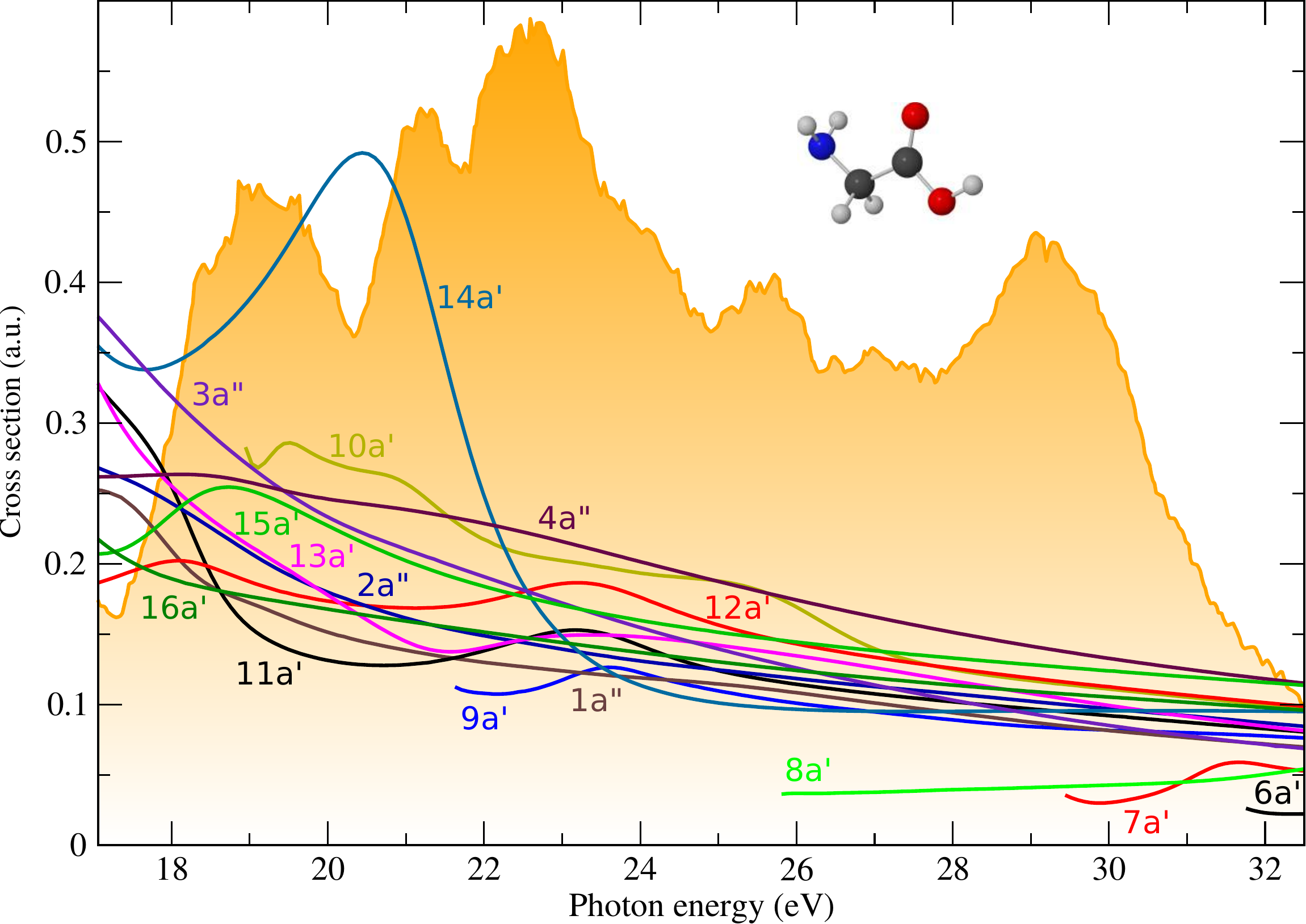}
\caption{Photoionization cross sections of glycine from different molecular orbitals calculated using the static-exchange DFT method.
Numbers and colors denote the molecular orbitals from where the electron is emitted in each case (fig. \ref{fig_MOs_gly}).
The energy spectrum of the attosecond pulse employed in the experiment presented in fig. \ref{fig_fragmentationYield_phe} is represented by a thick orange curve lying over a shaded area.}
\label{fig_CS_gly}
\end{figure}

\begin{figure}
\centering
\includegraphics[width=12cm]{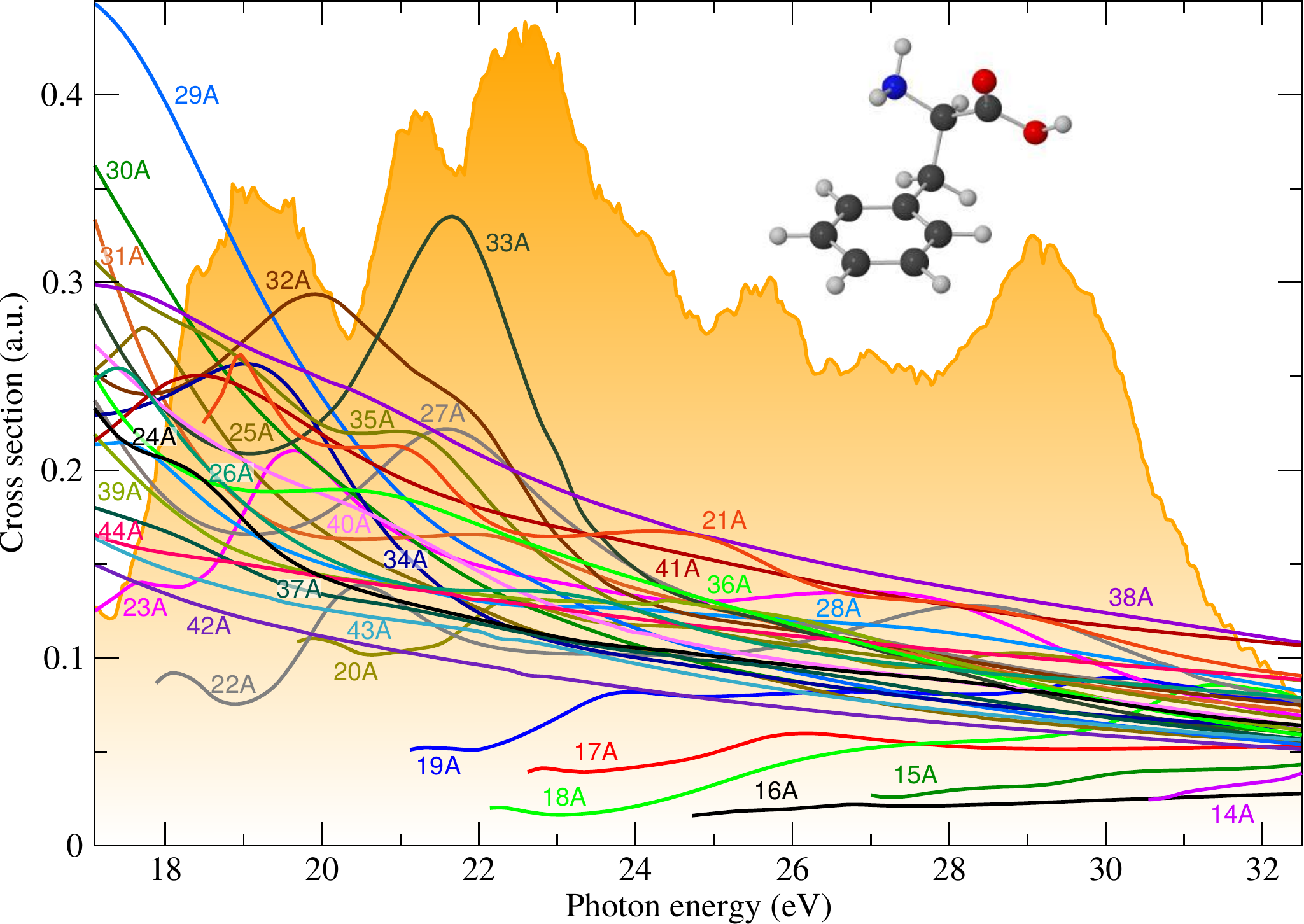}
\caption{Same as fig. \ref{fig_CS_gly} for phenylalanine (orbitals are shown in fig. \ref{fig_MOs_phe}).\vspace{4 mm}}
\label{fig_CS_phe}
\end{figure}

\begin{figure}
\centering
\includegraphics[width=12cm]{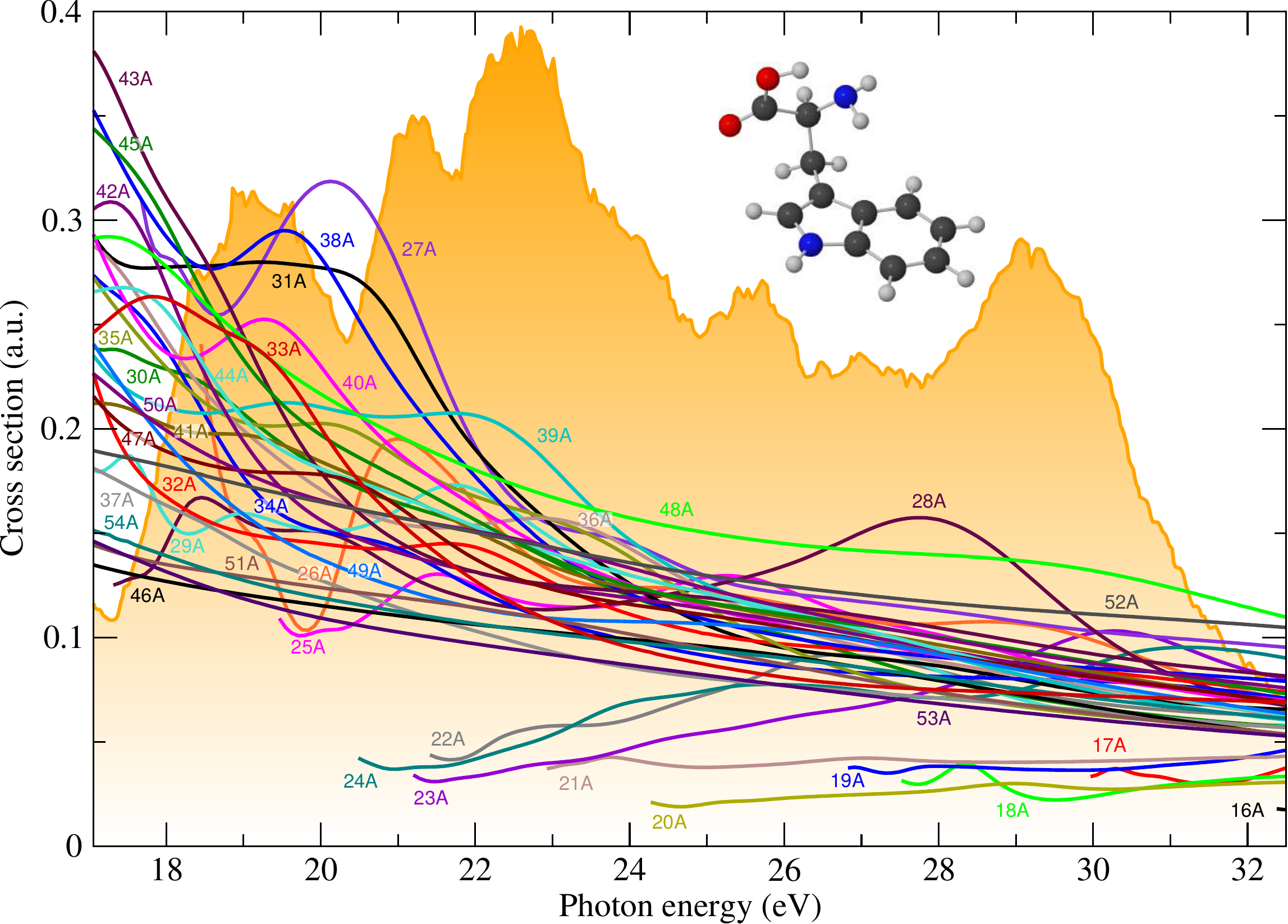}
\caption{Same as fig. \ref{fig_CS_gly} for tryptophan (orbitals are shown in fig. \ref{fig_MOs_trp}).}
\label{fig_CS_trp}
\end{figure}

We have evaluated photoionization cross sections of glycine, phenylalanine and tryptophan from all valence and inner-valence shells in the framework of the fixed-nuclei approximation, as explained in section \ref{section_CS}.
The results are shown in figs. \ref{fig_CS_gly}, \ref{fig_CS_phe} and \ref{fig_CS_trp} in the energy range accessible by the attosecond XUV pulse used in the experiment presented in fig. \ref{fig_fragmentationYield_phe}.
The energy spectrum of the pulse is depicted in the figs.by a thick orange curve lying over a shaded area.
As expected, the cross sections decay with the photon energy.
In some ionic channels, we can see sharp structures near the threshold that can be understood in terms of shape resonances \cite{DillPRL1975,PiancastelliJESRP1999,ShimizuJCP1997} due to the existence of small barriers in the complex molecular potentials.
Unfortunately, the use of a single excitation approach prevents us from observing any possible signature coming from multiple (doubly, triply) excited electronic states of the molecule embedded in the ionization continuum. \vspace{2 mm}

From the figures, it is clear that the three molecules will be efficiently ionized from most valence and inner-valence shells upon interaction with the attosecond pulse.
Only core electrons will remain unaffected.
In fact, for any energy within the bandwidth, we find similar contribution from different ionization thresholds.
For instance, using (monochromatic) synchrotron radiation of $22$ eV in phenylalanine, electrons would be ejected from orbitals $19$A...$44$A and $33$A would represent the largest contribution.
This would lead to a very delocalized hole in the parent ion. 
However, monochromatic light would generate an incoherent superposition of ionic states and the hole would not migrate.
Due to their broad energy bandwidths, attosecond pulses can generate coherent superpositions of electronic states, i.e., electronic wave packets, by emitting electrons with the same energy from different molecular orbitals and thus induce charge dynamics along the molecular skeleton.
Note that this scenario differs from that considered in most previous works on charge migration \cite{CederbaumCPL1999,KuleffJCP2005,RemaclePNAS2006,KuleffCP2007,MignoletJPB2014,KuleffJPB2014}, where the initial hole is created in a given molecular orbital.

\subsubsection*{Comparison with experimental photoelectron spectra}

\begin{figure}[!b]
\centering
\includegraphics[scale=0.42]{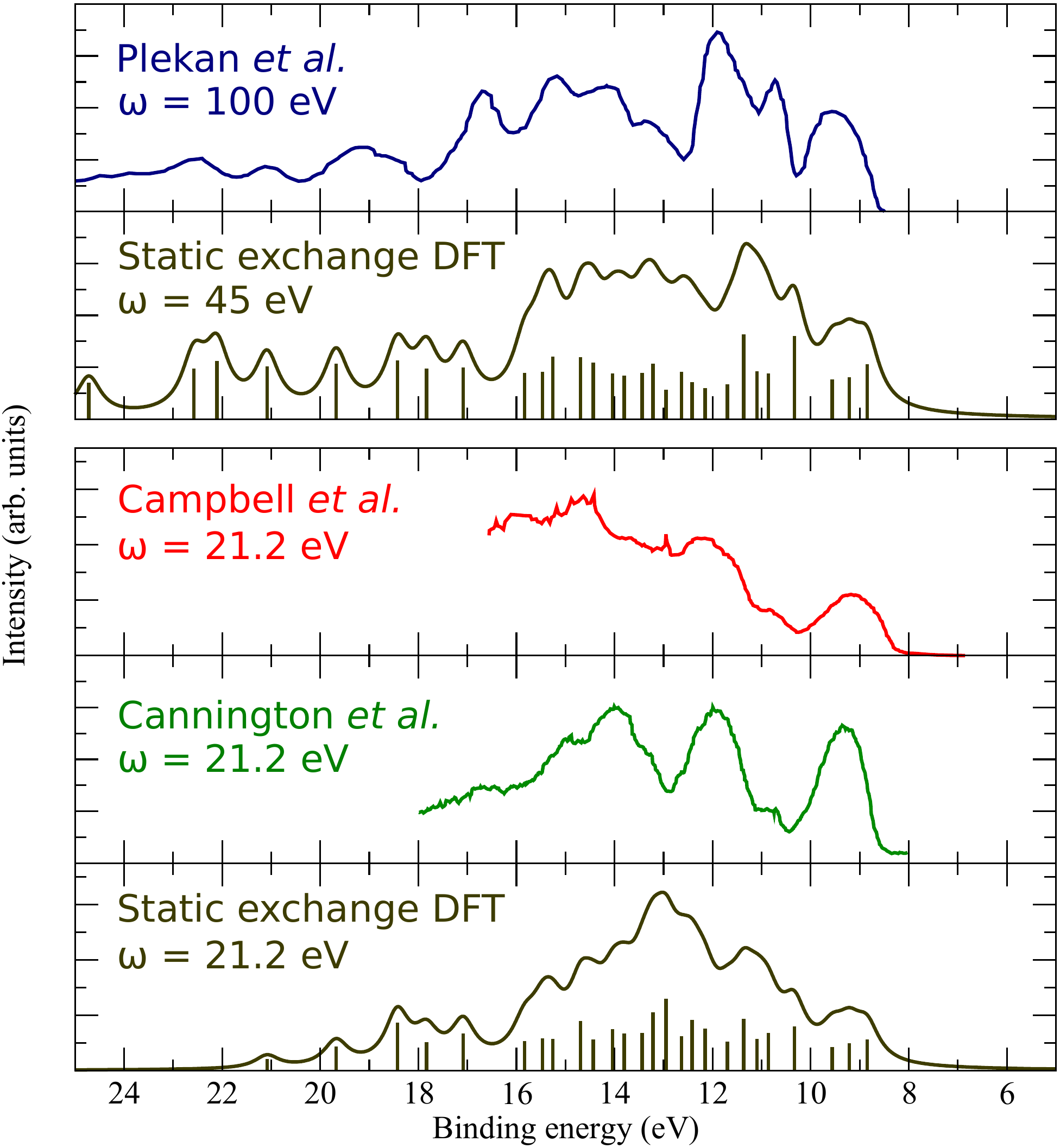}
\caption{Comparison between the calculated photoelectron spectra of phenyalanine at 45 eV and 21.2 eV and the experimental spectra reported by Plekan \emph{et al.} \cite{PlekanMP2008} at 100 eV and Campbell \emph{et al.} \cite{CampbellJACS1994} and Cannington \emph{et al.} \cite{CanningtonJESRP1983} at 21.2 eV}
\label{fig_photoelectronSpectra_phe}
\end{figure}

In order to test the validity of our description of the ionization process, we have calculated the photoelectron spectra of phenylalanine for photon energies of 21.2 and 45 eV using the cross section values at those energies (shown in fig. \ref{fig_CS_phe}) and the corresponding ionization potentials.
Then, we have compared our results with synchrotron \cite{PlekanMP2008} and He(I) \cite{CampbellJACS1994,CanningtonJESRP1983} radiation spectra available in the literature. 
For the comparison with the experiments, we have convoluted our infinitely resolved lines with a Lorentzian function of $0.3-$eV width at half maximum to account for the vibrational broadening and experimental energy resolution, which is rather limited in these and earlier experiments (the experiments cannot resolve the individual peaks).
The comparison between theory and experiment is shown in fig. \ref{fig_photoelectronSpectra_phe}.
As can be seen, the agreement is reasonably good.
We notice however that the experiment of Plekan \emph{et al.} \cite{PlekanMP2008} was performed at a photon energy of 100 eV, which is substantially higher than ours.
The other two earlier experiments \cite{CampbellJACS1994,CanningtonJESRP1983} were performed at a photon energy of 21.2 eV.
Our results are in better agreement with the most recent experiment, especially for binding energies below 15 eV.
This energy range includes the states that play an important role in the ultrafast charge dynamics initiated in the molecule by the attosecond pulse considered in this work (represented in fig. \ref{fig_CS_phe}), as we show in section \ref{section_aa_holeDynamics}.
The photoelectron spectrum is expected to be more sensitive to the choice of photon energy as we approach the threshold.
The reason is that the ionization amplitudes can strongly vary with photon energy for values below $25-30$ eV, while the variation becomes smother for larger values, as can be seen in fig. \ref{fig_CS_phe}.
This is most likely the reason we find a better agreement between the high-energy spectra.

\section{Ultrafast electron dynamics initiated by attosecond pulses}
\label{section_aa_holeDynamics}

The first theoretical predictions of the possibility of observing ultrafast charge migration upon prompt ionization of an organic molecule can be attributed to Lorenz S. Cederbaum and collaborators \cite{CederbaumCPL1999}.
In that work, an electronic wave packet is generated by sudden electron removal from a Hartree Fock (HF) molecular orbital of the difluoropropadienone molecule.
Then, the hole generated in the electronic structure moves though the molecular skeleton because the prepared state is not a stationary state of the ionic Hamiltonian but a linear combination of several.
Over the years, they have investigated charge migration in a large number of organic molecules (see, for instance \cite{CederbaumCPL1999,KuleffJCP2005,KuleffCP2007,KuleffJPB2014}).
Here we consider a different scenario:
(i) we are using an ultrashort pulse with a broad energy bandwidth to create an electronic wave packet in the parent ion, and
(ii) we compute the scattering states to obtain the actual photoionization amplitudes.\vspace{2 mm}

In order to verify the validity of our time-propagation method, we have compared our results with those obtained by Kuleff, Breidbach and Cederbaum for the case of glycine \cite{KuleffJCP2005}.
To perform a meaningful comparison, we have started from the same initial wave function as in \cite{KuleffJCP2005}.
The corresponding HF orbitals ha-

\begin{figure}[H]
\centering
\includegraphics[width=\textwidth]{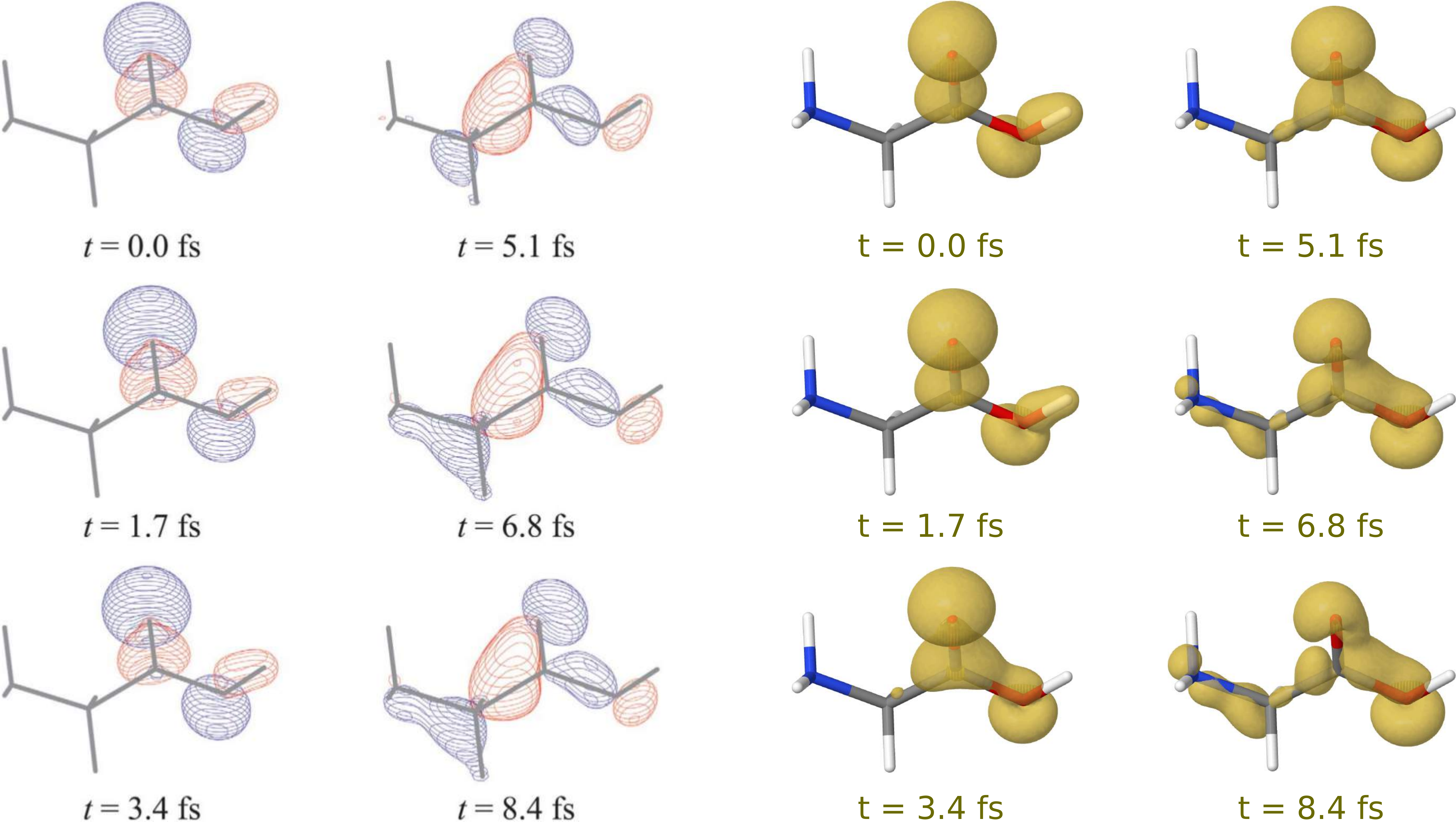}
\caption{Time evolution of the hole generated in glycine upon sudden ionization from the $11$a$'$ HF orbital.
Left figure: natural charge orbitals calculated by A. Kuleff \emph{et al.} \cite{KuleffJCP2005} (the square of the natural orbitals provides the hole density).
Reprinted with permission from \cite{KuleffJCP2005}. Copyright 2005, AIP Publishing LLC.
Right figure: hole density, evaluated using the present approach.\newline}
\label{fig_Kuleff_11a}
%
\centering
\includegraphics[width=\textwidth]{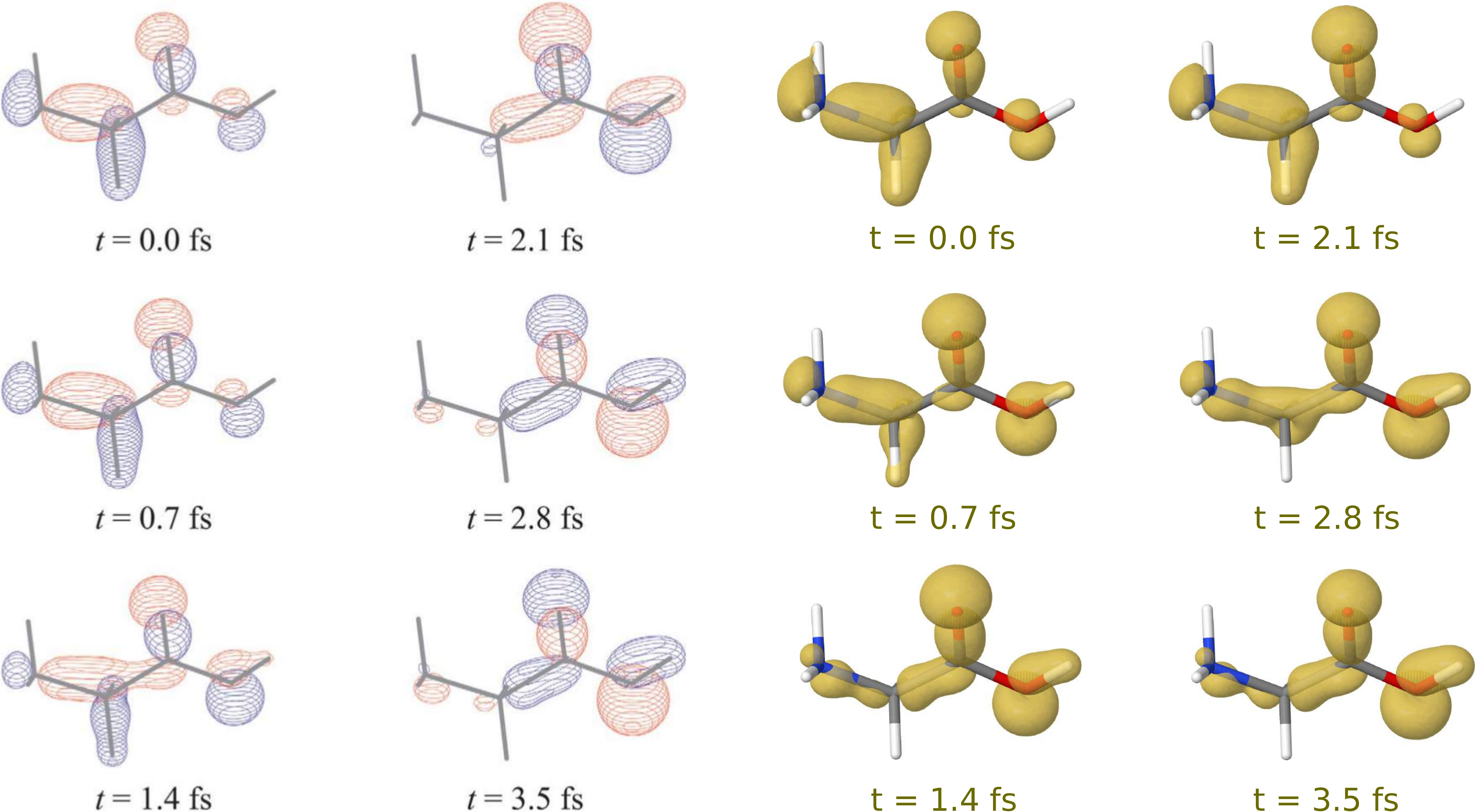}
\caption{Same as fig. \ref{fig_Kuleff_11a} for the case of sudden ionization from the $14$a$'$ HF orbital of glycine.
Left figure reprinted with permission from \cite{KuleffJCP2005}. Copyright 2005, AIP Publishing LLC.}
\label{fig_Kuleff_14a}
\end{figure}

\noindent ve been evaluated using Gaussian 09 \cite{Gaussian09} with a DZP basis set.
Then, to study the time evolution of the hole density, we have projected the initial state onto the Slater determinants built from the KS orbitals (shown in fig. \ref{fig_MOs_gly}) that we use to represent ionic stationary states.
The projection leads to a coherent superposition of ionic states that is let evolve freely as dictated by the relative phases resulting from the corresponding energy differences.
The evolution of the hole dynamics is shown in figs. \ref{fig_Kuleff_11a} and \ref{fig_Kuleff_14a} for the cases of sudden ionization from the $11$a$'$ and the $14$a$'$ HF orbitals, respectively.
As can be seen, the agreement between our results and those previously reported \cite{KuleffJCP2005} is quite satisfactory.
In the latter reference, two-holes-one-particle (2h1p) configurations were explicitly included in the time propagation, so a direct comparison with their results provides an answer about the role played by those configurations, which are not included in our time propagation scheme.
If we considered sudden ionization from the inner-valence HF orbitals of glycine instead, as in \cite{KuleffCP2007}, the agreement would not be so spectacular because in that case 2h1p configurations play a role.
However, these are not expected to be important in the dynamics studied here.
The main reason is that a transition from the ground state to doubly-excited (shake-up) state is a two-electron process and therefore is less likely to occur via 1 photon absorption than a direct transition to a one-hole (1h) state.\vspace{2 mm}

In the following, we present our results on the time-evolution of the hole generated in the electronic structure of the amino acids glycine, phenylalanine and tryptophan upon interaction with an attosecond XUV pulse similar to that used in the experiment illustrated in fig. \ref{fig_fragmentationYield_phe}.
As discussed in the previous section, attosecond pulses do not remove electrons from only one molecular orbital but from several of them, generating coherent superpositions of ionic states.
Ionization amplitudes have been evaluated for all open channels (15 for glycine, 32 for phenylalanine and 39 for tryptophan) using the static-exchange DFT method \cite{StenerIJQC1995,StenerTCA1999,StenerJPB2000,BachauRPP2001,ToffoliCP2002} (see section \ref{section_staticExchangeDFT}), which has been thoroughly tested in systems of similar complexity, and time-dependent first-order perturbation theory (see section \ref{section_TDPT}).
From the ionization amplitudes, we have evaluated the reduced density matrices of the ionic subsystems using eq. \ref{redDensMat}. 
Then, the hole densities were calculated as the difference between the electronic densities of the neutral molecules, which do not depend on time, and the electronic densities of the ions (eq. \ref{hole_density_2}).
Because in the experiments the molecules are not aligned, we performed calculations considering three orthogonal orientations with respect to the polarization vector of the attosecond pulse.
The results were then averaged assuming randomly oriented molecules.\vspace{2 mm}

\begin{figure}[H]
\centering
\includegraphics[width=\textwidth]{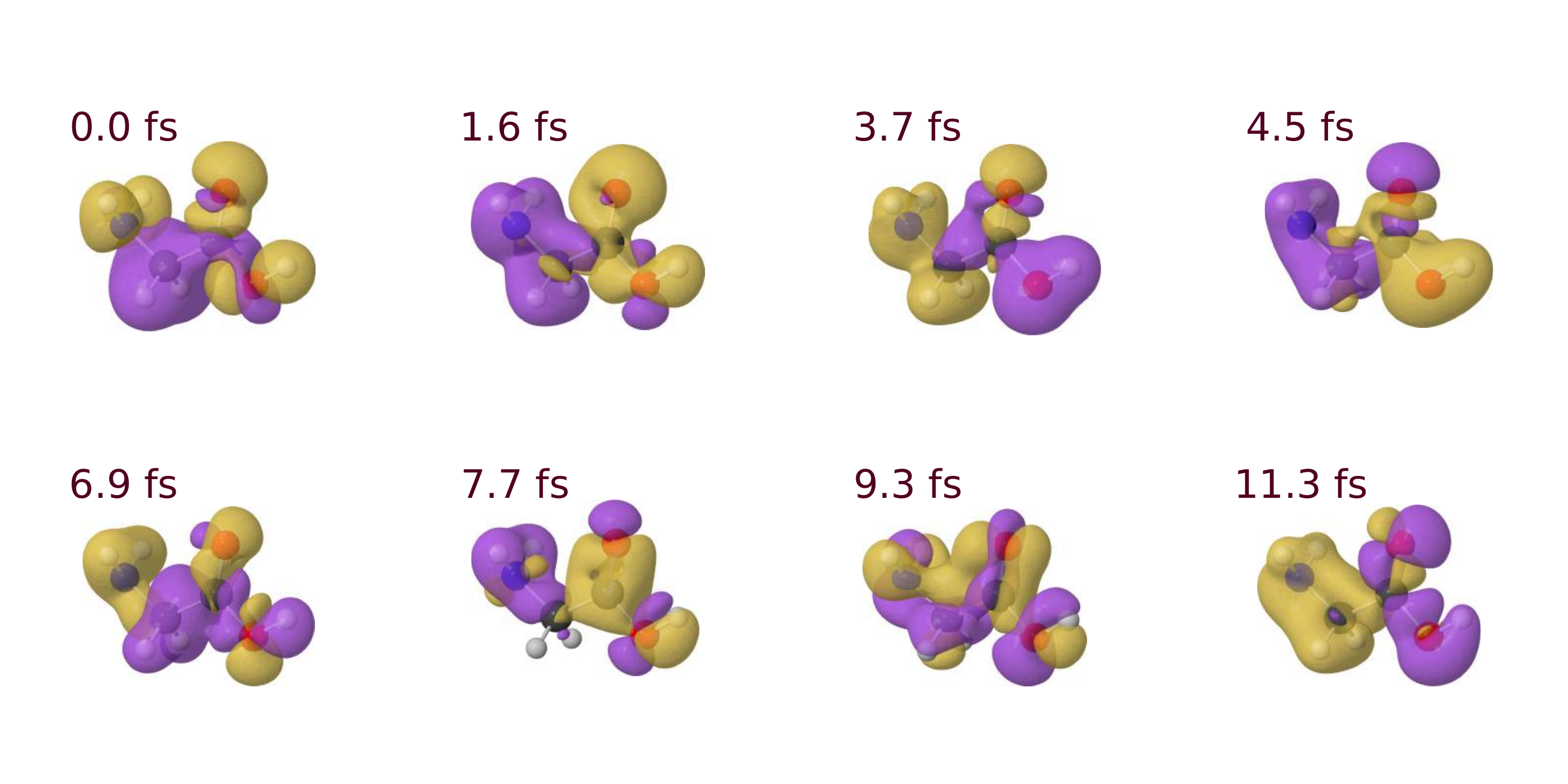}
\caption{Relative variation of the hole density on glycine with respect to its time-averaged value as a function of time.
Isosurfaces of the relative hole density are shown for cutoff values of $10^{-4}$ a.u. (yellow) and $-10^{-4}$ a.u. (purple).
Time is with reference to the end of the XUV pulse (first snapshot).}
\label{fig_snapshots_gly}
\end{figure}

\begin{figure}[H]
\includegraphics[width=\textwidth]{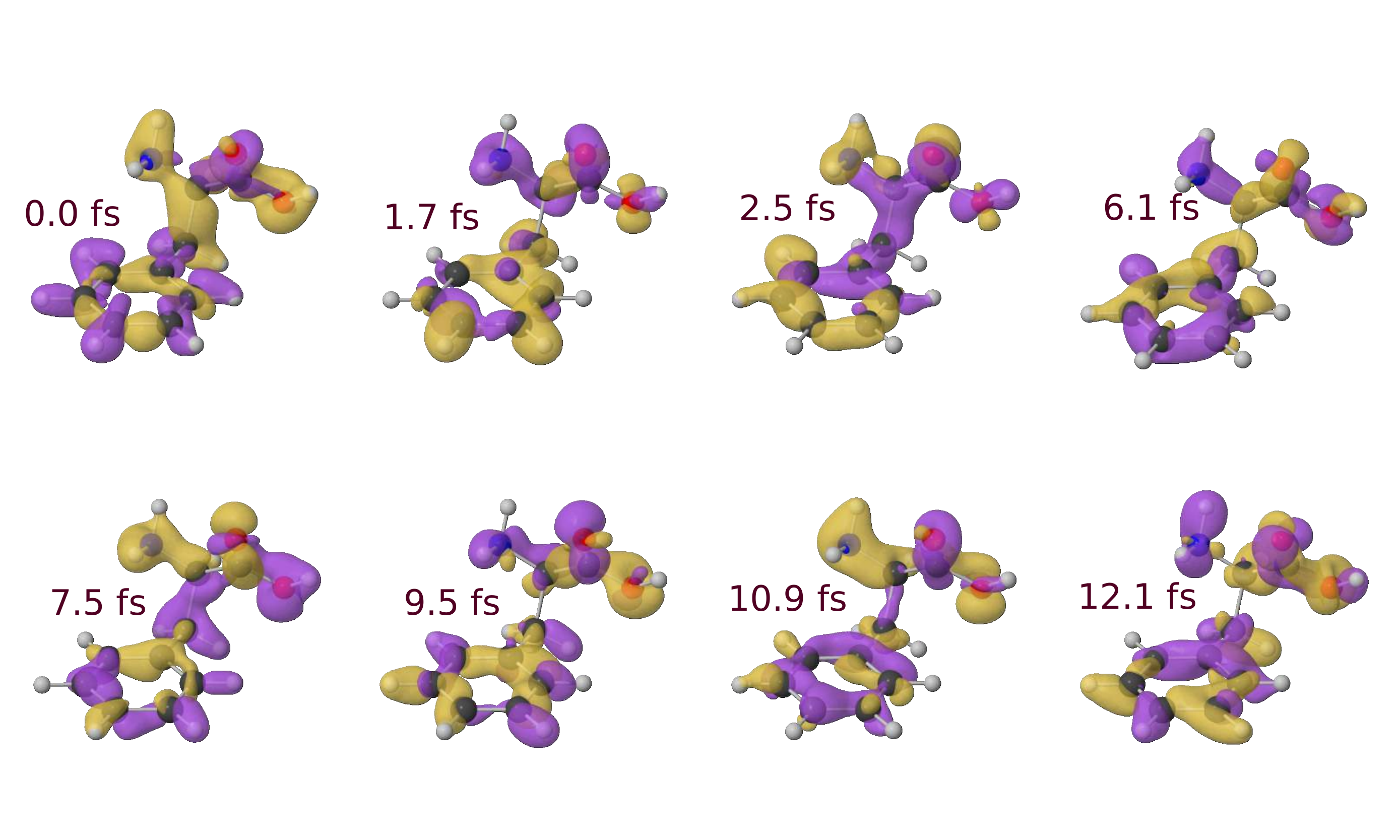}
\caption{Same as fig \ref{fig_snapshots_gly} for phenylalanine.}
\label{fig_snapshots_phe}
\end{figure}

\begin{figure}[H]
\includegraphics[width=\textwidth]{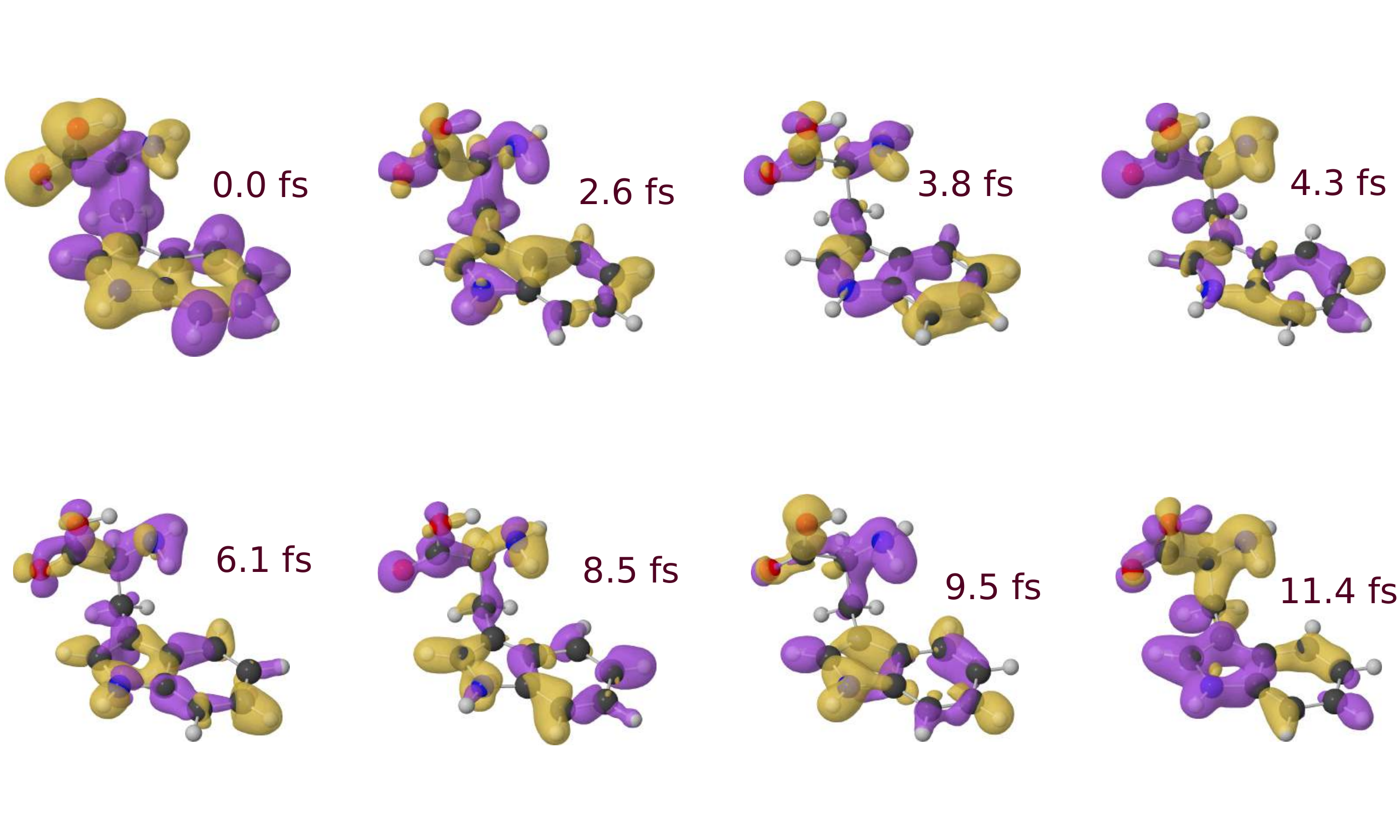}
\caption{Same as fig \ref{fig_snapshots_gly} for tryptophan.}
\label{fig_snapshots_trp}
\end{figure}

Figs. \ref{fig_snapshots_gly}, \ref{fig_snapshots_phe} and \ref{fig_snapshots_trp} display snapshots of the relative variation of the hole density with respect to the time-averaged values for glycine, phenylalanine and tryptophan, respectively.
In spite of the very delocalized nature of the hole densities resulting from the broadband XUV excitations, substantial redistributions take place on a sub-femtosecond scale.
These charge dynamics cannot be associated with simple migrations between two sites of the molecules, as found in most previous theoretical work \cite{CederbaumCPL1999,KuleffJCP2005,RemaclePNAS2006,KuleffCP2007,KuleffJPB2014}.
However, despite the complexity of the charge configuration calculated in a realistic (i.e., experimentally accessible) situation, the concept of charge migration is still valid.

\subsection{Fourier analysis on the amino group}

\begin{figure}
\centering
\includegraphics[width=\textwidth]{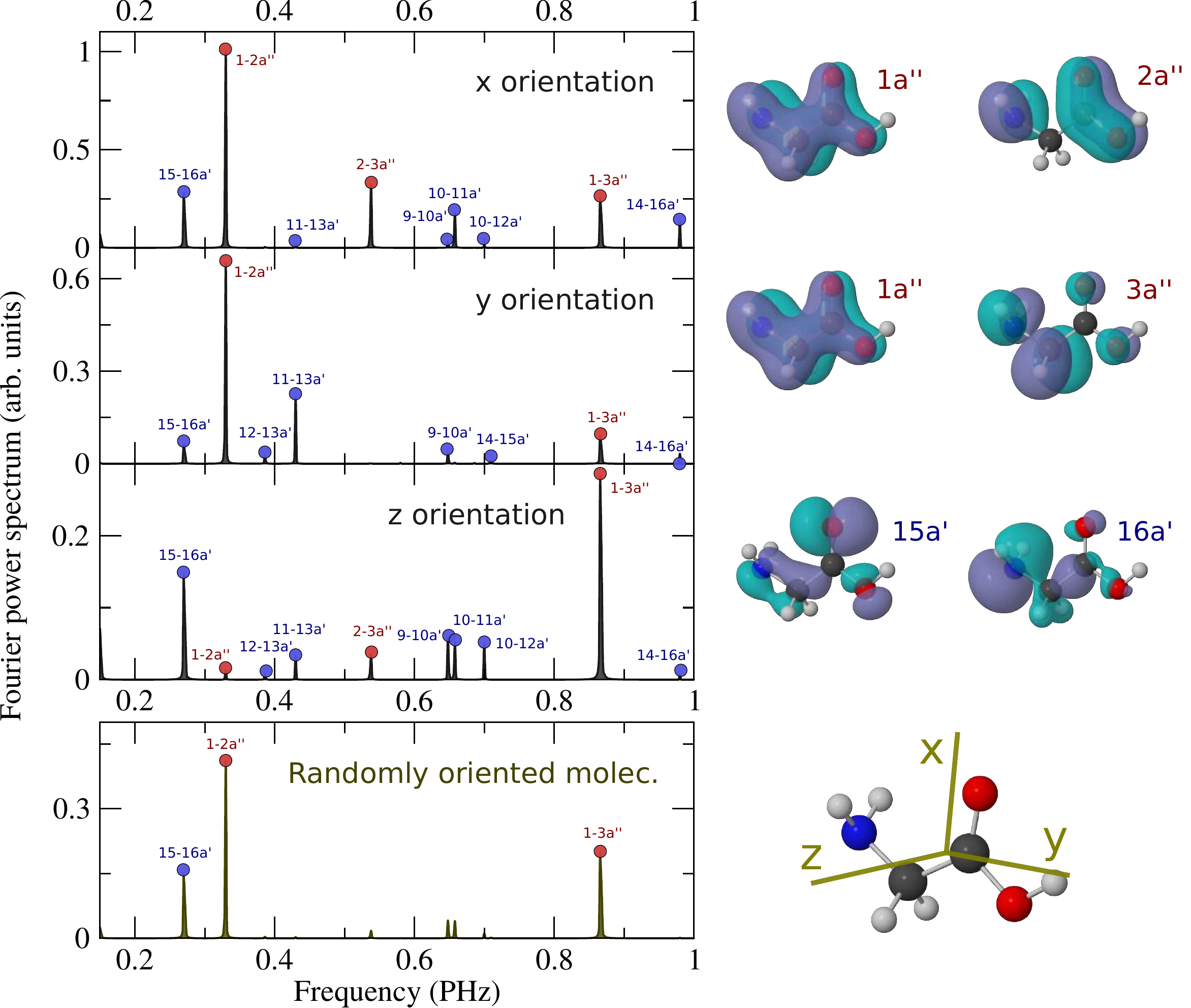}
\caption{Fourier power spectra of the hole density integrated over the amino group of glycine.
Results are shown for three orthogonal orientations of the molecule with respect to the polarization vector of the electric field associated to the attosecond XUV pulse and for the case of randomly oriented molecules.
In order to obtain well resolved peaks in frequency, the hole density has been evaluated up to 500 fs.
The states that give rise to the dominant peaks are indicated in the spectra by labels that denote the molecular orbitals where the holes have been created (molecular orbitals are shown in fig. \ref{fig_MOs_gly}).}
\label{fig_FPS_amino_gly}
\end{figure}

\begin{figure}
\centering
\includegraphics[width=\textwidth]{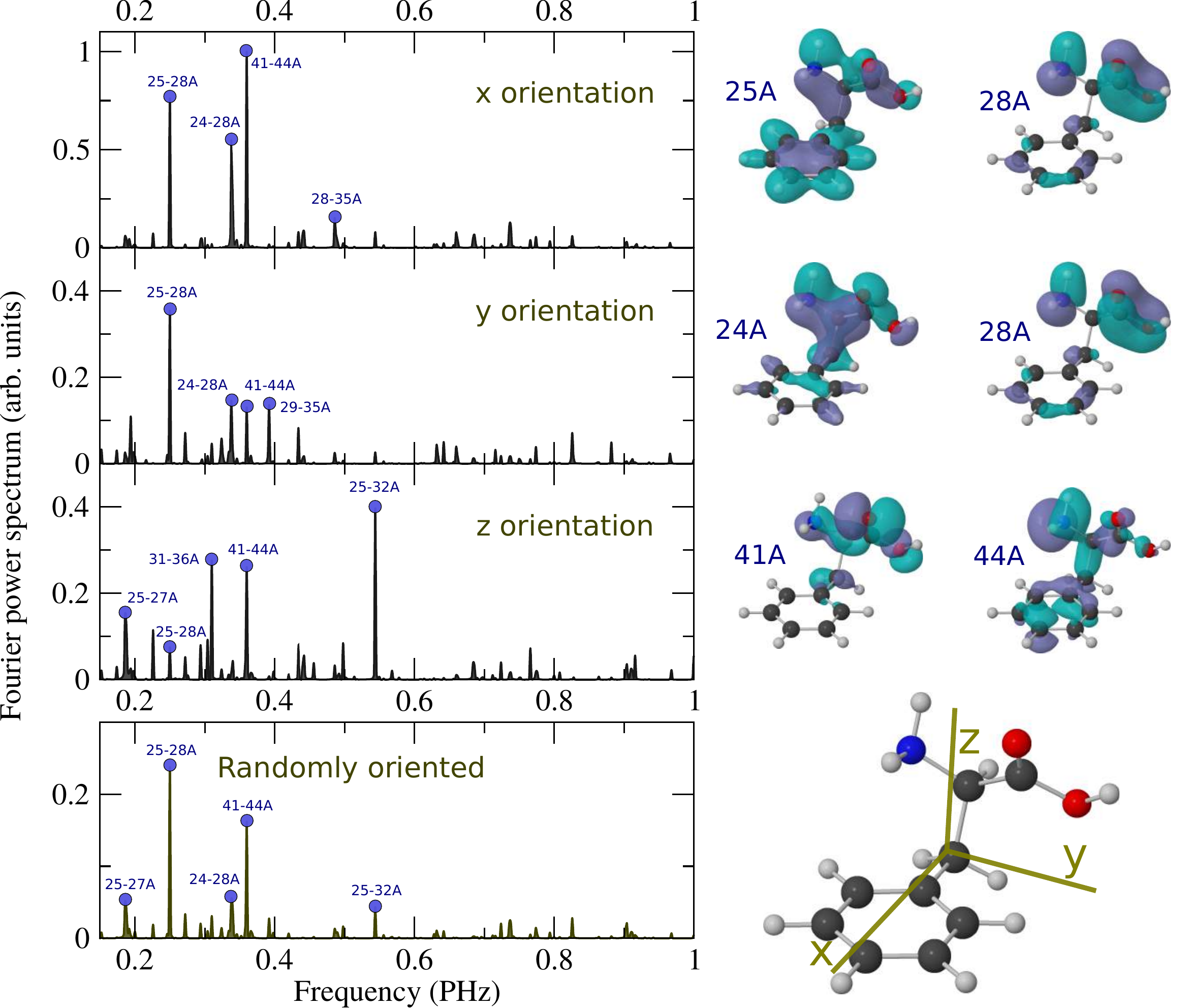}
\caption{Same as fig. \ref{fig_FPS_amino_gly} for phenylalanine (molecular orbitals are shown in fig. \ref{fig_MOs_phe}).}
\label{fig_FPS_amino_phe}
\end{figure}

\begin{figure}
\centering
\includegraphics[width=\textwidth]{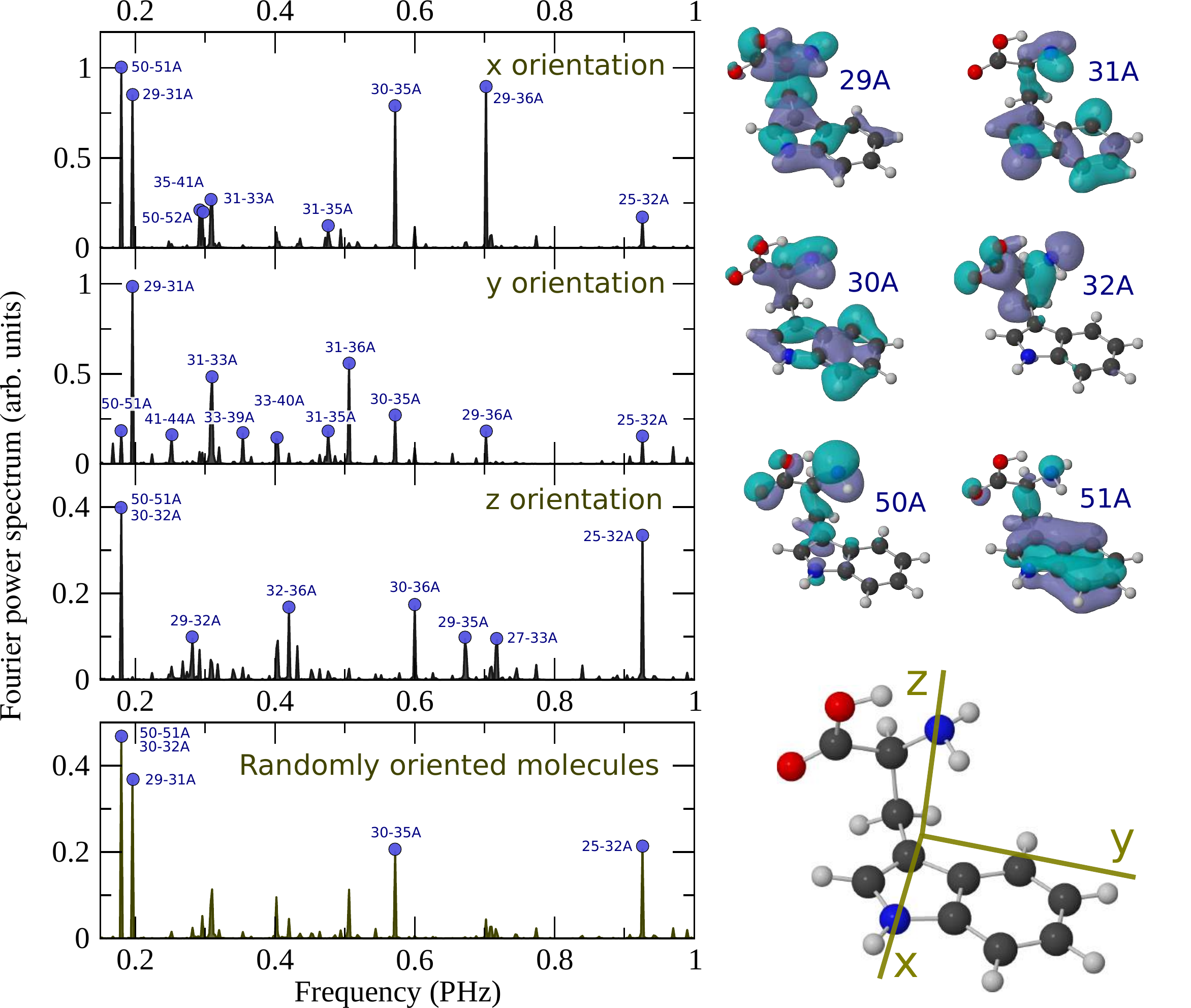}
\caption{Same as fig. \ref{fig_FPS_amino_gly} for tryptophan (molecular orbitals are shown in fig. \ref{fig_MOs_trp}).}
\label{fig_FPS_amino_trp}
\end{figure}

In order to perform a more quantitative analysis of the charge dynamics, we have integrated the hole density around different portions of the molecule.
Here we present our results on the amino group.
In the case of phenylalanine, these results can be compared with the fragmentation yield of double-charged immonium presented in fig. \ref{fig_fragmentationYield_phe} since, as already discussed, this signal is sensitive to hole localization on the amino group.
A full analysis including different atomic centers is presented in section \ref{section_fourier_atoms}.
Figs. \ref{fig_FPS_amino_gly}, \ref{fig_FPS_amino_phe} and \ref{fig_FPS_amino_trp} show the Fourier power spectra of the hole density on the amino group of glycine, phenylalanine and tryptophan, respectively, for three orthogonal orientations of the molecules with respect to the polarization vector of the field (indicated in the figures) and for the case of randomly oriented molecules.
The peaks appearing in the Fourier spectra provide information about the frequency and the intensity of the charge fluctuations observed in figs. \ref{fig_snapshots_gly}, \ref{fig_snapshots_phe} and \ref{fig_snapshots_trp}.
As expected, the complexity of the spectra increases with the number electrons in the molecule as so does the number of 1h states are accessible by the pulse.\vspace{2 mm}

To better understand the observed dynamics, we have identified the ionic states that are responsible for the most important beatings (indicated in figs. \ref{fig_FPS_amino_gly}, \ref{fig_FPS_amino_phe} and \ref{fig_FPS_amino_trp}).
Each beating frequency is given by the energy difference between the pair of states that originate it.
The main interferences involve states with holes in orbitals that are delocalized between the amino group and another common part of the molecule, which allows charge migration between the two sites.
Also, we can see that the dominant beatings involve states which are close in energy.
In fact, the greater the energy spacing between two ionic states, the weaker the interference between them is expected to be.
This can be easily understood in terms of the coherences between the ionic states populated by the attosecond pulse, which are given by the off-diagonal terms of reduced density matrix (see eq. \ref{redDensMat}). 
If the energy spacing $\Delta E_{\alpha\alpha'}=|E_{\alpha}-E_{\alpha'}|$ between two ionic states $\alpha$ and $\alpha'$ is large, then the kinetic energy ranges in which the photoelectron is emitted when the two states are populated are very different and therefore the corresponding element in the reduced density matrix $\gamma_{\alpha\alpha'}$ is small.
If $\Delta E_{\alpha\alpha'}$ were greater than the energy bandwidth of the pulse (and the two ionic states were accessible), they would be populated incoherently because the photoelectron kinetic energy ranges would not overlap at all and therefore $\gamma_{\alpha\alpha'}$ would be zero.
Because of the large energy bandwidth of the attosecond pulse considered here, all 1h states are populated coherently, but coherences between states with similar energy are, in general, higher.\vspace{2 mm}

In the spectrum of glycine (fig. \ref{fig_FPS_amino_gly}), we can only see interferences between states having the same symmetry because states with different symmetry are populated incoherently.
The reason is that, for a given orientation of the field, electrons emitted from two orbitals ($\alpha$, $\alpha'$) with different symmetry have, in general, different symmetry and therefore the corresponding coherence term ($\gamma_{\alpha\alpha'}$) is zero (see eq. \ref{redDensMat}).
In the case of phenylalanine and tryptophan (figs. \ref{fig_FPS_amino_phe} and \ref{fig_FPS_amino_trp}), although there are no strict selection rules due the lack of global symmetry elements, approximate selection rules will apply.
For instance, one of the most intense beatings in the amino group of phenylalanine occurs between the states with holes in the 41A and the 44A orbitals, both having nodal planes that contain the C$-$N and the C$=$O bonds with very similar orientations.\vspace{2 mm}

\begin{figure}[!b]
\centering
\includegraphics[width=\textwidth]{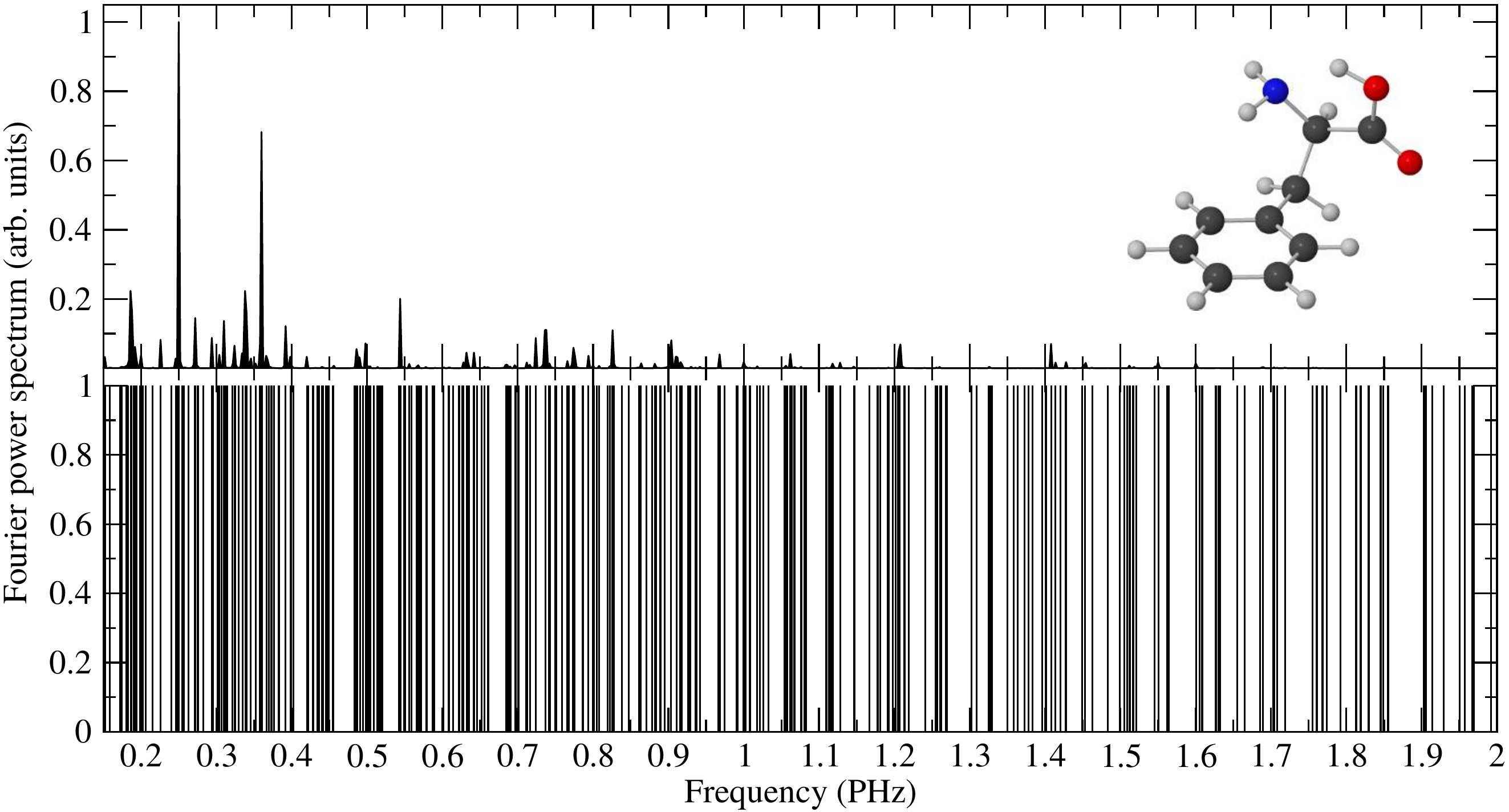}
\caption{Fourier power spectra of the calculated hole density integrated over the amino group of phenylalanine.
Upper panel: results from the actual calculation (as in the lower panel of fig. \ref{fig_FPS_amino_phe}).
Lower panel: results obtained by using an equal weight for all ionic states accessible by the XUV pulse.}
\label{fig_FPS_barcode}
\end{figure}

\begin{figure}[H]
\centering
\includegraphics[width=\textwidth]{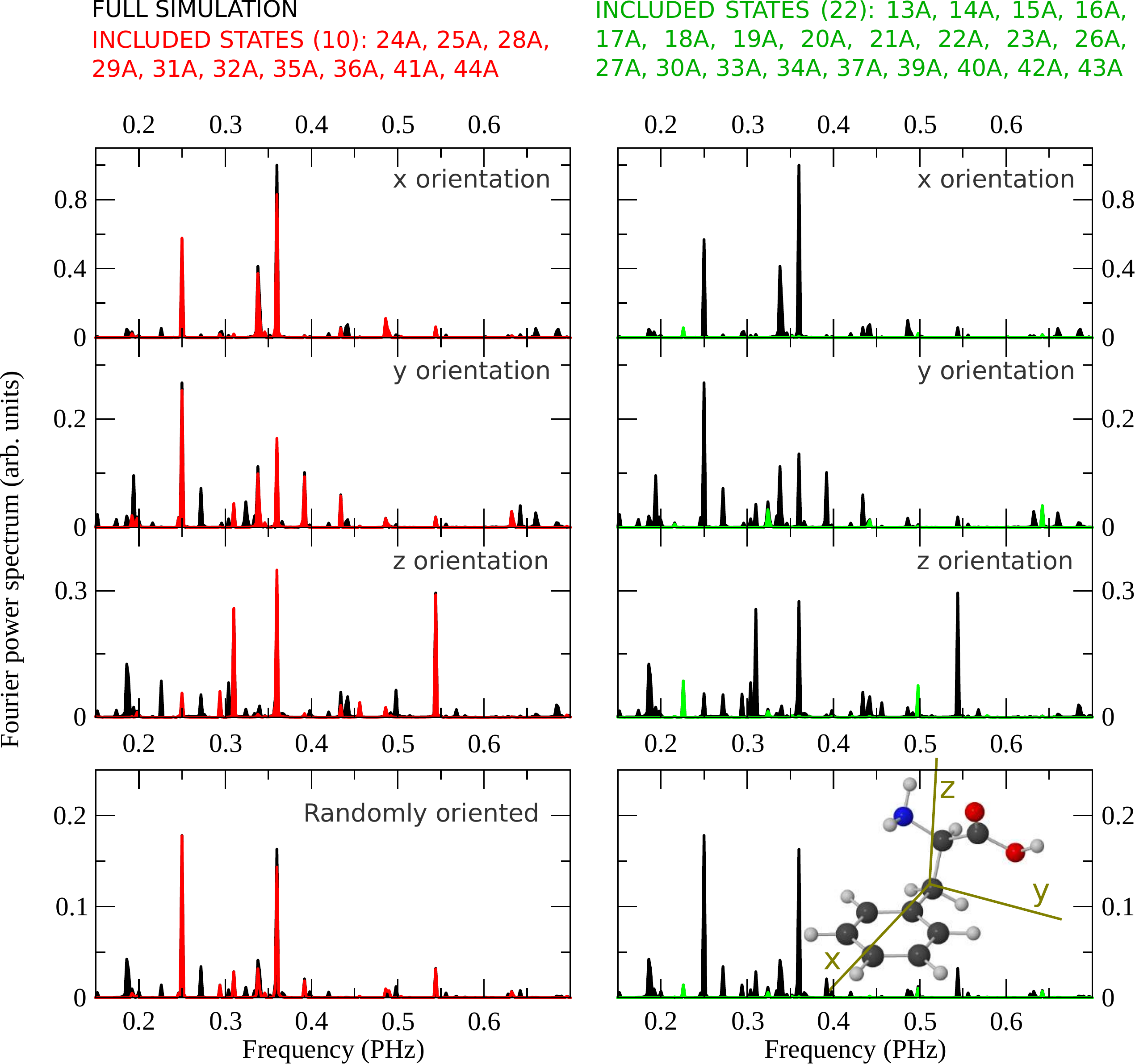}
\caption{Fourier power spectra of the hole density integrated over the amino group of phenylalanine:
results from the full calculation (black lines in all panels),
from a calculation in which only the ionic states resulting from removing an electron from the 24A, 25A, 28A, 29A, 31A, 32A, 35A, 36A, 41A, and 44A orbitals are included (red lines on top of the black lines, left panels),
and from a calculation in which all ionic states but the above mentioned ones are included (green curves on top of the black lines, right panels).}
\label{fig_FPS_partialSimulation_phe}
\end{figure}

It is clear from the Fourier analysis presented here that the hole density not only provides information about energy spacings between different quantum states responsible of the electronic beatings, but also probes the specific dynamics generated by the attosecond pulse.
This is further illustrated in fig. \ref{fig_FPS_barcode} for the case of phenylalanine, in which the calculated Fourier spectrum is compared with a similar one containing all possible energy spacings with an equal weight.
The dynamics of the electronic wave packet generation by the attosecond pulse is responsible for the fact that only a few beatings are observed.
These depend on the dipole matrix elements (i.e., the ionizing transition induced by the XUV attosecond pulse) and the interference between the different amplitudes, which are imprinted in the time evolution of the hole density.\vspace{2 mm}

To illustrate that the observed dynamics in the amino group can almost be entirely explained in terms of some of the ionic states that are populated by the attosecond XUV pulse, we have carried out calculations for phenylalanine in which only ionic states resulting from removing an electron from the 24A, 25A, 28A, 29A, 31A, 32A, 35A, 36A, 41A, and 44A orbitals are included in the free propagation of the electronic wave packet.
Conversely, we have also performed calculations in which all 1h states but the above mentioned are included.
The results of these two calculations are shown and compared with the full calculations in fig. \ref{fig_FPS_partialSimulation_phe}.
As can be seen, the full spectrum is almost entirely reproduced by only including the above ten states.
In contrast, the dynamics resulting from excluding these states is almost inexistent.

\subsection{Fourier analysis on different atoms}\label{section_fourier_atoms}

\begin{figure}
\centering
\includegraphics[width=\textwidth]{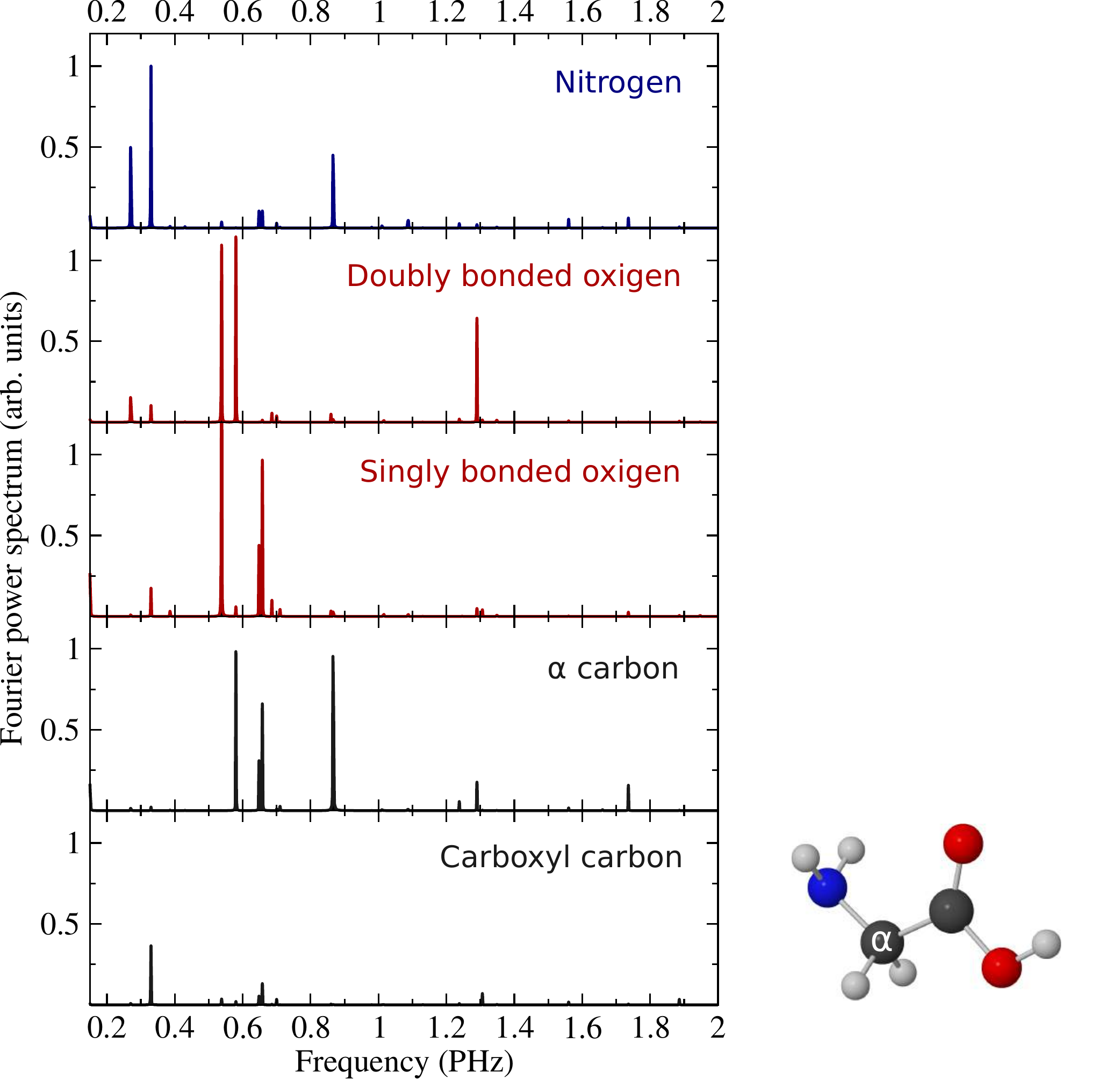}
\caption{Fourier power spectra of the calculated hole density integrated over various atoms of glycine for the case of randomly oriented molecules.
In order to obtain well resolved peaks in frequency, the hole density has been evaluated up to 500 fs.}
\label{fig_FPS_atoms_gly}
\end{figure}

\begin{figure}
\centering
\includegraphics[width=\textwidth]{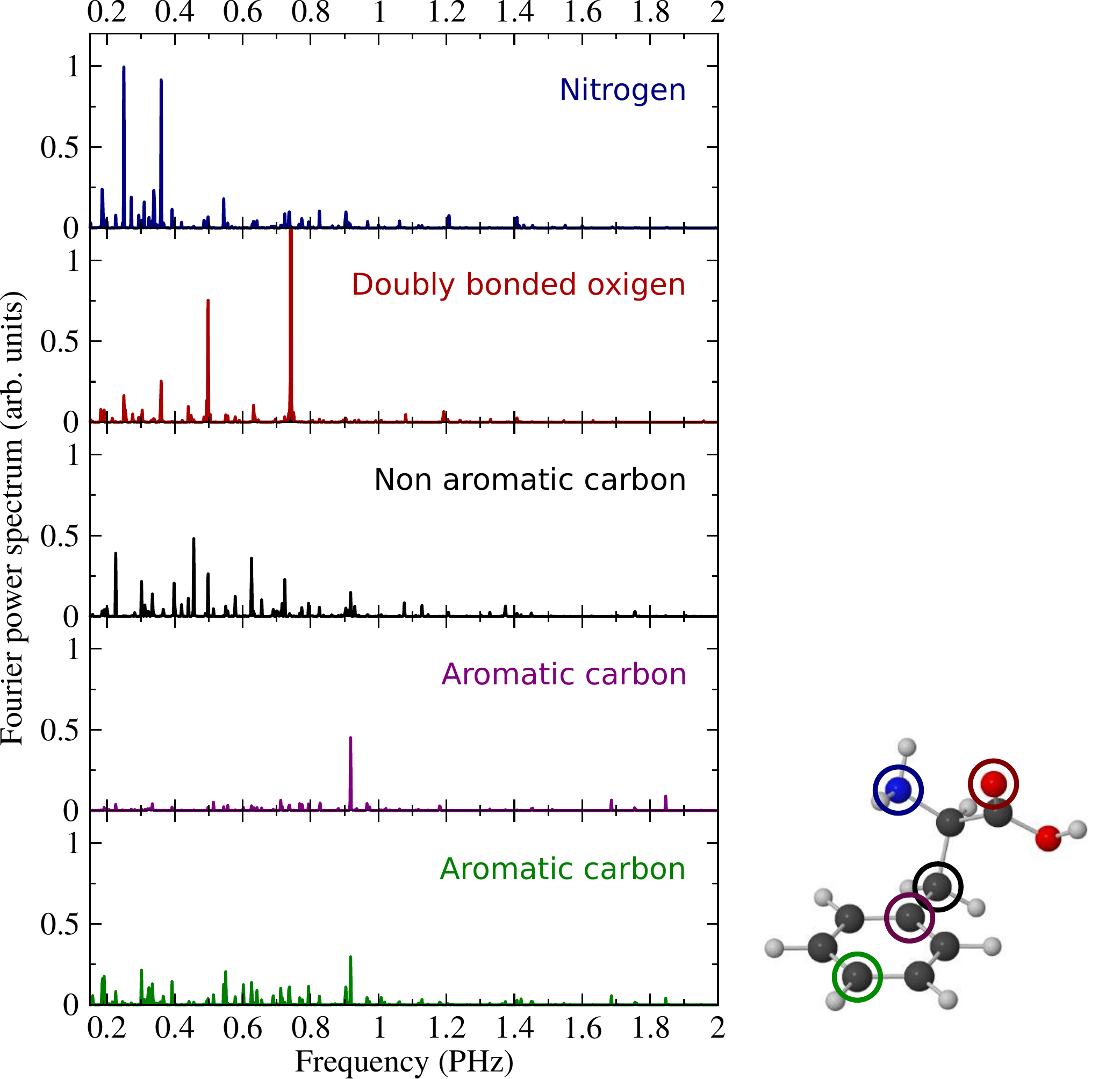}
\caption{Same as \ref{fig_FPS_atoms_gly} for phenylalanine.}
\label{fig_FPS_atoms_phe}
\end{figure}

\begin{figure}
\centering
\includegraphics[width=\textwidth]{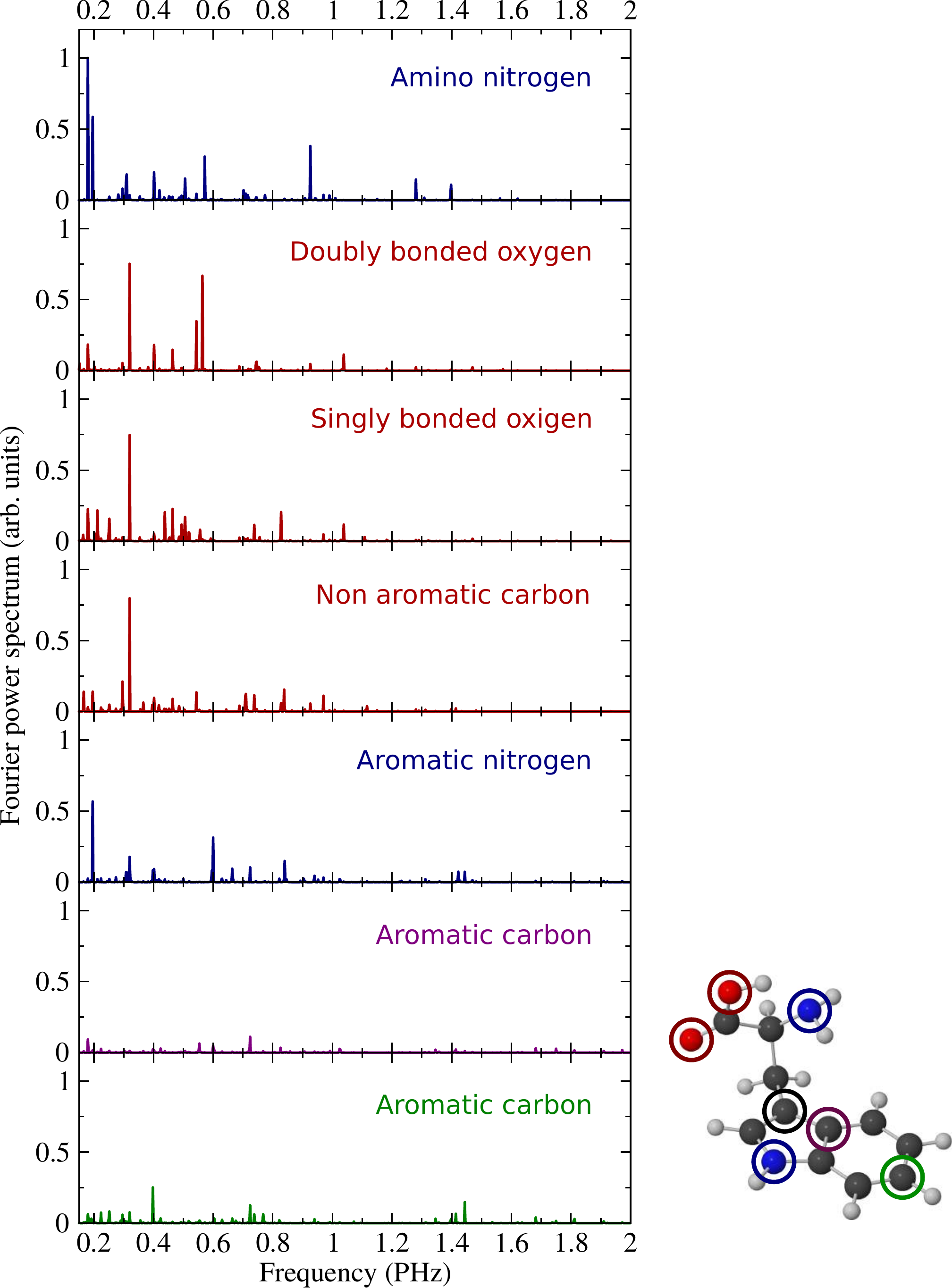}
\caption{Same as \ref{fig_FPS_atoms_gly} for tryptophan.}
\label{fig_FPS_atoms_trp}
\end{figure}

Charge fluctuations on the amino group are most likely to be responsible for the observed beatings in the experimental results shown in \ref{fig_fragmentationYield_phe}.
In order to perform a complete analysis of the charge modulations on different sites of the molecule, we have performed a Fourier analysis also on different atoms of the three amino acids.
Figs. \ref{fig_FPS_atoms_gly}, \ref{fig_FPS_atoms_phe} and \ref{fig_FPS_atoms_trp} show the Fourier power spectra of the hole density integrated over various atoms of the three molecules. 
As expected, charge fluctuations occur with different frequencies on different atoms because they are due to interferences between different pairs of states. 
Indeed, it is a requirement for a beating to be observed on a specific site (atom) of the molecule that the two states involved contain a significant part of their hole localized on that particular site.

In the case of phenylalanine, beating frequencies in agreement with the experimental observations were observed when the charge density was integrated around the nitrogen atom in amino group.
The hole densities at different positions do not show clear and clean frequency components, with the exception of the doubly bonded O atom in the carboxyl group. 
We note that the VIS/NIR probe pulse is not locally absorbed only by the amino group, but also by other sites of the molecule.
However, the simulations indicate that the periodic modulations observed in the experiment are mainly related to the absorption of the probe pulse by the amino group.

\subsection{Comparison with the experiment: Gabor profiles}

\begin{figure}
\centering
\includegraphics[scale=0.23]{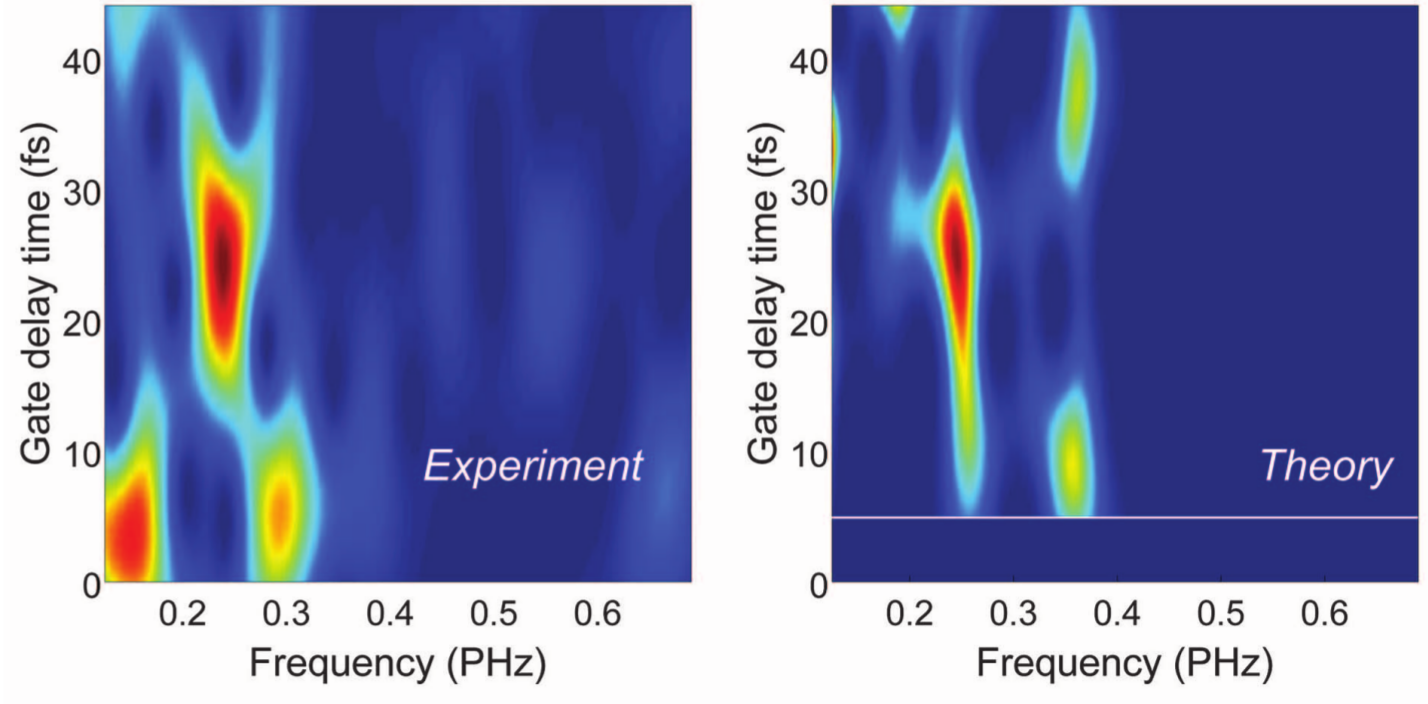}
\caption{Fourier analysis of charge dynamics.
Spectrograms calculated for the measured data of \ref{fig_fragmentationYield_phe} (left panel) and for the calculated hole density integrated over the amino group of phenylalanine (right panel), shown in fig \ref{fig_snapshots_phe}.
The sliding window Fourier transforms have been calculated by using a Gaussian window function $g(t-t_d) = e^{-(t-t_d)^2 / t_0^2}$, with $t_0 = 10$ fs and peak at $t_d$ (gate delay time).
The theoretical spectrogram (right panel) was calculated considering an experimental temporal resolution of $3$ fs.}
\label{fig_gabor_phe}
\end{figure}

The hole density on the amino group has been analyzed by using a sliding-window Fourier transform which, at the expense of frequency resolution, shows frequency and time information on the same plot.
The same procedure has been applied to the fragmentation yield of doubly-charged immonium presented in fig. \ref{fig_fragmentationYield_phe}.
Fig. \ref{fig_gabor_phe} shows the resulting spectrograms in a temporal window up to 45 fs.
The theoretical spectrogram presents a dominant peak around 0.25 PHz, which forms in about 15 fs and vanishes after about 35 fs, in close agreement with the results of the Fourier analysis of the experimental data.
A higher frequency component is visible around 0.36 PHz in the delay intervals below $15$ fs and above $30$ fs.
At short delays, this component favorably compares with the experimental observation of the frequency peak around 0.30 PHz in the same window of pump-probe delays.
The good agreement between between theory and experiment strongly supports the interpretation of the measured data in terms of charge migration.
The temporal evolution of the main Fourier components is a consequence of the complex interplay among several beating processes initiated by the broadband excitation pulse.
Despite the agreement with the experimental results, we cannot exclude that the nuclear dynamics, which are not included in the simulations, also play a role in the temporal evolution of the measured oscillation frequencies.

The good agreement between theory and experiment is rather remarkable in light of the fact that simulations do not take into account the interaction of the VIS/NIR probe pulse.
The fact that the effects of the probe pulse are not included in the simulations can explain why the calculated intensities of the different beatings differ from the experimental ones.
We note that the beating frequencies have been observed experimentally even though the initial hole density is highly delocalized.
An important result of the simulations is that the measured beating frequencies originate from charge dynamics around the amino group.
This leads to the conclusion that the periodic modulations measured in the experiment are mainly related to the absorption of the probe pulse by the amino group.
The mechanism that makes the probe pulse sensitive specifically to the charge density on this group is still not well understood.
Moreover, we observe that, in spite of the large number of potential frequency beatings associated to the wave packet motion induced by the attosecond pulse, only a few ones manifest in the experiment, thus reducing the impact of the modulations introduced by the probe pulse in the analysis of the wave packet motion.
These results can be seen as the first experimental confirmation that attosecond pulses and techniques are essential tools for understanding of dynamical processes on a temporal scale that is relevant for the evolution of crucial microscopic events at the heart of the macroscopic biological response of molecular complexes.

\subsection{What about molecular conformation?}
\label{section_conformation}

\begin{figure}
\centering
\includegraphics[width=\textwidth]{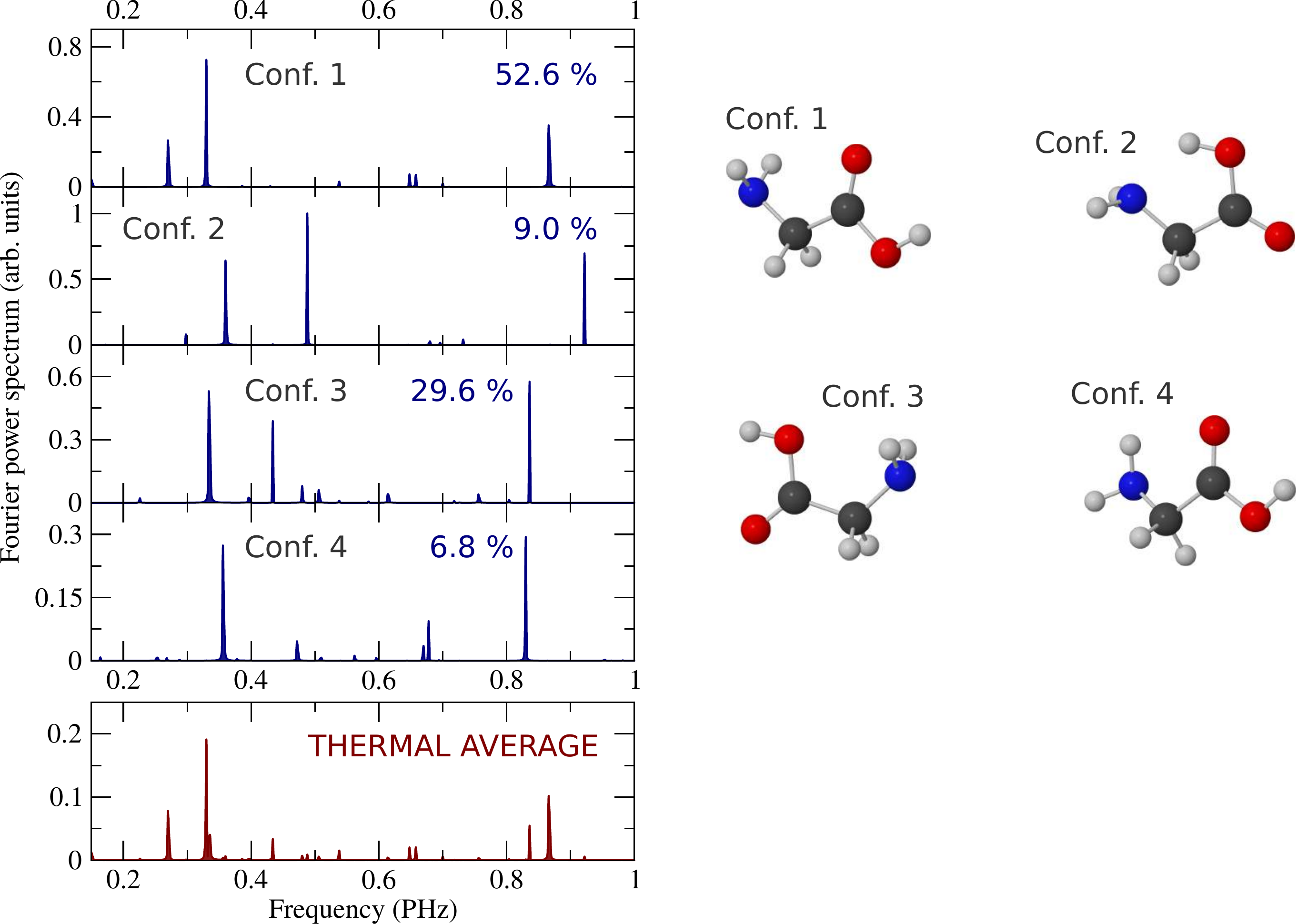}
\caption{Fourier power spectra of the hole density on the amino group of the most abundant conformers of glycine at 430 K.
The relative populations have been calculated assuming a Boltzman distribution.
The lower panel shows the averaged results and the corresponding geometries are shown in the right.}
\label{fig_FPS_confs_gly}
\end{figure}

\begin{figure}
\centering
\includegraphics[width=\textwidth]{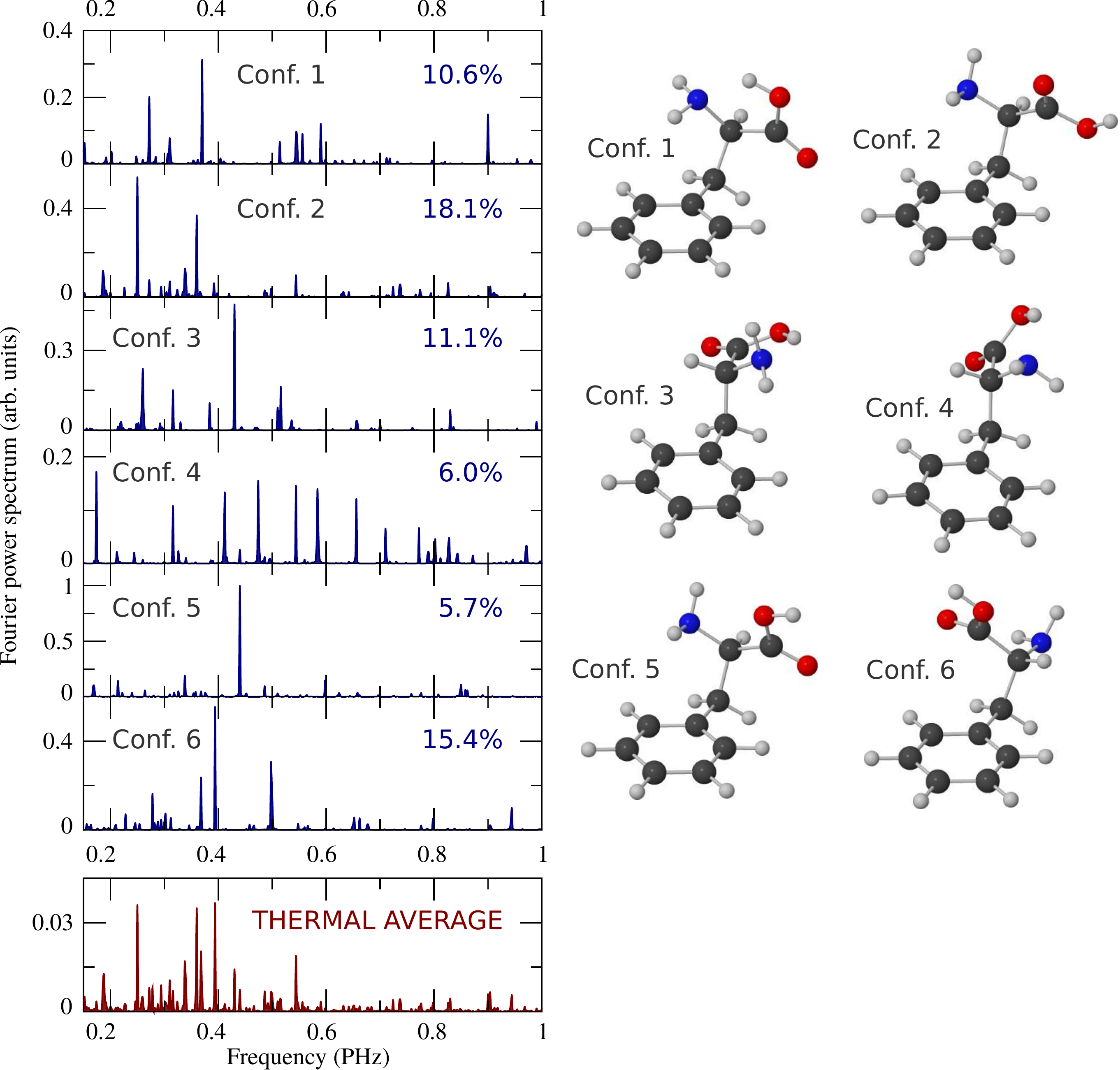}
\caption{Same as fig. \ref{fig_FPS_confs_gly} for the most stable conformers of phenylalanine, rescaled with their relative weights for a better illustration.
Their relative populations according to Huang \emph{et al.} \cite{HuangTHEOCHEM2006} are indicated.}
\label{fig_FPS_confs_phe}
\end{figure}

\begin{figure}
\centering
\includegraphics[width=\textwidth]{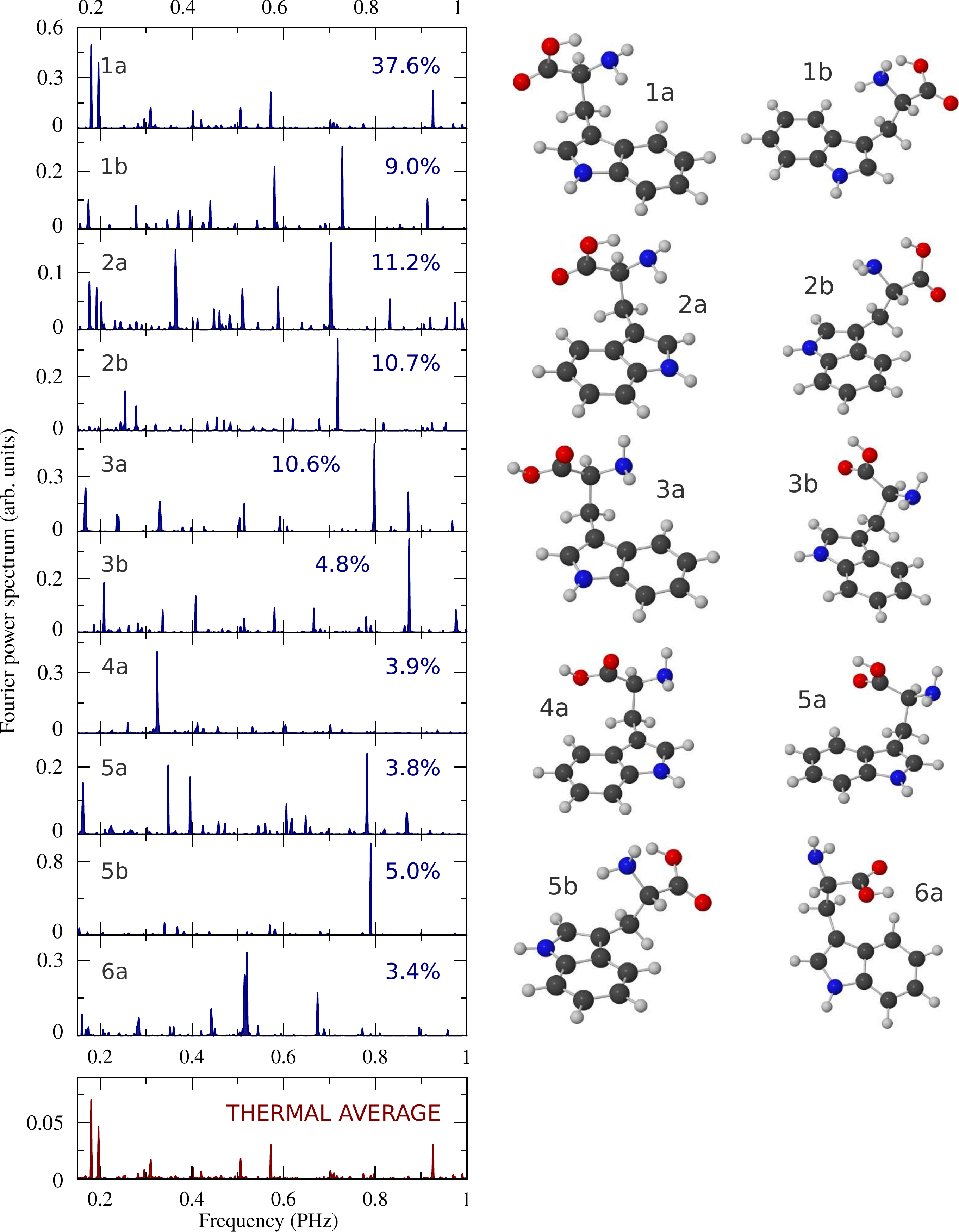}
\caption{Same as fig. \ref{fig_FPS_confs_gly} for the most stable conformers of tryptophan.
The relative populations have been calculated assuming a Boltzman distribution.}
\label{fig_FPS_confs_trp}
\end{figure}

It is well known that amino acids exist in many conformations as a result of their structural flexibility.
Typically, the energy barrier to interconversion between different conformers is small, of the order of a few kcal/mol, so that, even at room temperature, thermal energy is sufficient to induce conformational changes.
Theoretical investigations have shown that such changes can affect the charge migration process \cite{KuleffCP2007}.
In the case of phenylalanine, 37 conformers have been found by ab initio calculations \cite{HuangTHEOCHEM2006}, with a conformational distribution that depends on the temperature.
However, in the experiment presented in fig. \ref{fig_fragmentationYield_phe}, performed at an average temperature of about 430 K, only six conformers are substantially present \cite{HuangTHEOCHEM2006}.
Figs. \ref{fig_FPS_confs_gly}, \ref{fig_FPS_confs_phe} and \ref{fig_FPS_confs_trp} show the most stable conformations of glycine, phenylalanine and tryptophan at 430K.
The relative populations of the conformers of phenylalanine have been calculated using statistical mechanics methods based on DFT/B3LYP quantum chemistry calculations of the geometries, energies, vibrational frequencies and rotational constants \cite{HuangTHEOCHEM2006}.
In the case of glycine \cite{CsaszarJACS1992,NevilleJACS1996} and tryptophan \cite{SnoekPCCP2001}, the relative populations were estimated assuming Boltzman distributions.\vspace{2 mm}

The ultrafast temporal evolution of the wave packet generated by the attosecond pump pulse has also been calculated for the most stable conformers of the tree amino acids.
The corresponding Fourier power spectra are shown in figs. \ref{fig_FPS_confs_gly}, \ref{fig_FPS_confs_phe} and \ref{fig_FPS_confs_trp}, together with the results for the thermal average obtained by taking into account the estimated populations (given in figs. \ref{fig_FPS_confs_gly}, \ref{fig_FPS_confs_phe} and \ref{fig_FPS_confs_trp}).
We have found that, although the precise frequencies of the relevant peaks in the calculated Fourier spectra depend on the particular conformer, the spectrum of the most abundant conformer is very similar to that of the thermal average.
In the case of the most populated conformers of phenylalanine (and therefore for the averaged results), the frequencies at which the dominant peaks appear are in good agreement with those observed experimentally.

\section{Time-evolution of the full electron system}
\label{section_fullSystem}

\begin{figure}
\centering
\includegraphics[width=\textwidth]{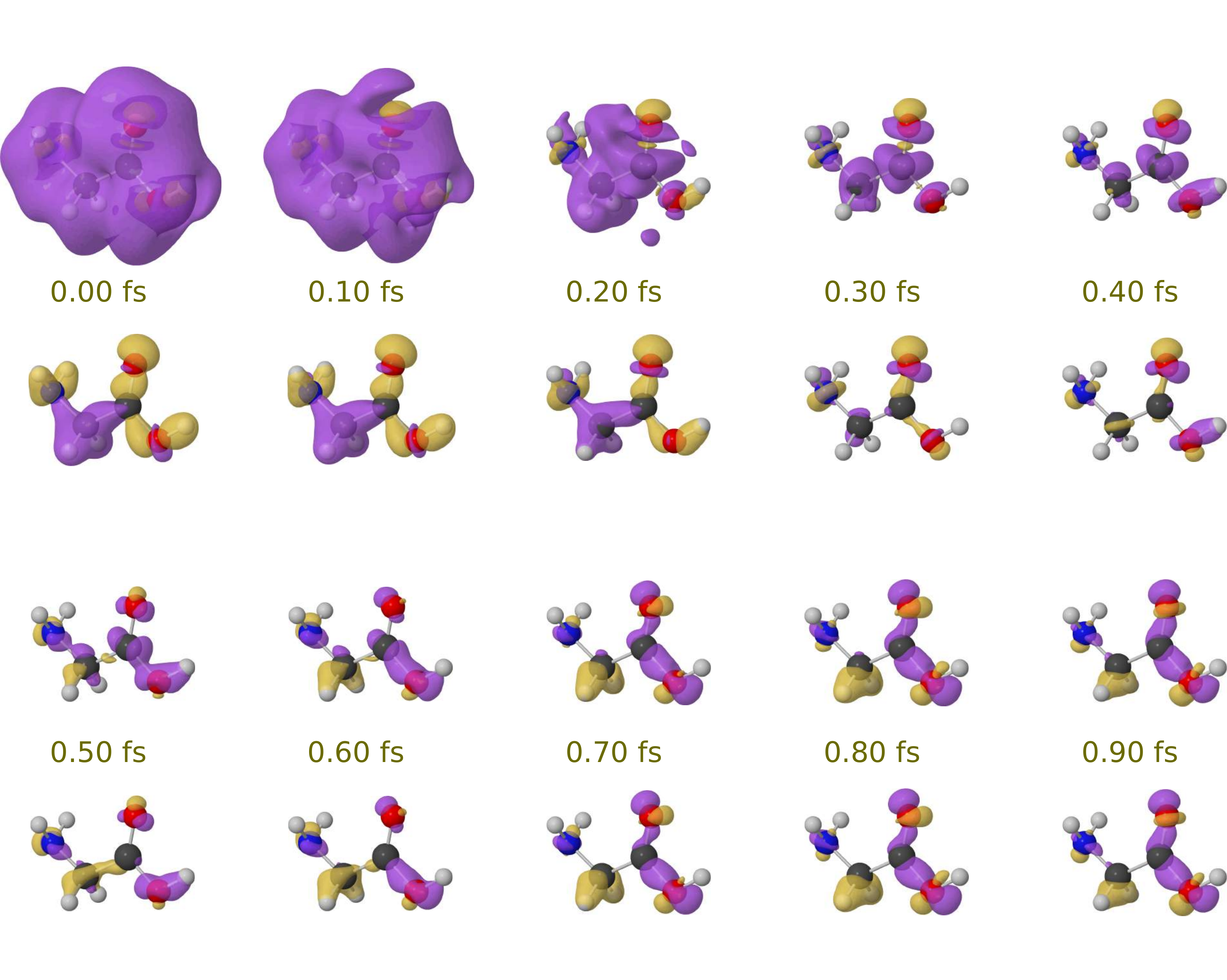}
\caption{Relative variation of the electron density of glycine with respect to its time-averaged value as a function of time.
Rows 1 and 3: dynamics of the full-system, containing all electrons.
Rows 2 and 4: dynamics of the residual ion, calculated in terms of its reduced density matrix.
Isosurfaces of the relative electron density are shown for cutoff values of $10^{-3}$ a.u. (purple) and $-10^{-3}$ (yellow).
Time is with reference to the end of the XUV pulse.}
\label{fig_snapshots_fullSystem}
\end{figure}

The evaluation of the hole density generated upon ionization in terms of the reduced density matrix of the ionic subsystem has allowed to describe the ultrafast response of the parent ion to attosecond ionization.
In order to analyze the time-evolution of the electronic wave packet created in the continuum, one needs to evaluate the electron density of the full system, including the photoelectron.
This can be done by making use of eq. \ref{elecDens_NelecFinal}, which retrieves the electron density in terms of the the wave packet coefficients (see eq. \ref{WFexpansion}) and the corresponding bound and continuum orbitals.
We have analyzed the time-evolution of the electron density of the glycine molecule upon interaction with an attosecond pulse similar to that used in the experiment illustrated in fig. \ref{fig_fragmentationYield_phe}.
Fig. \ref{fig_snapshots_fullSystem} (rows 1 and 3) shows snapshots of the relative variation of the electron density of the full system for different times, from right after the interaction with the pulse to up to 1fs.
For comparison, snapshots of the relative electron density of the residual ion, calculated using eq. \ref{elecDens_ion} and the reduced density matrix of the ionic subsystem (eq. \ref{elecDens_ion}), are also shown in fig. \ref{fig_snapshots_fullSystem} (rows 2 and 4).
For analysis purposes, we have chosen a bigger cuttoff value ($10^{-3}$ a.u.) to generate the isosurfaces of the electron density than that employed in fig. \ref{fig_snapshots_gly} ($10^{-4}$ a.u.).
Note that positive values of the relative electron density (purple surfaces in fig. \ref{fig_snapshots_fullSystem}) correspond to negative values of the relative hole density (purple surfaces in fig. \ref{fig_snapshots_gly}).\vspace{2 mm}

We can see in fig. \ref{fig_snapshots_gly} that at $t=0$ the molecule is surrounded by a bulky electronic cloud due to the presence of the photoelectron.
This cloud gradually vanishes as the photoelectron wave is emitted and, at $t=0.7$ fs, the electron density of the full system is identical to that of the residual ion.
These results justify the analysis of the ultrafast charge redistributions occurring in the system in terms of the electron density of the residual ion.

%% file: Chapters/Conclusions.tex
\chapter*{Conclusions}
\label{Conclusions}
\addcontentsline{toc}{part}{Conclusions}
\fancyhead[LE]{}
\fancyhead[RO]{\fontsize{11pt}{11pt}\selectfont Conclusions}

The advent of ultrabright light sources based on synchrotron radiation and ultrafast pulses based on high harmonic generation (HHG) opened the door to image molecular structures and to monitor and even control electron and nuclear dynamics in their intrinsic time scales.
In this quest, theoretical modelling and numerical simulations have been probed to be fundamental in the design of new experimental schemes and to understand the complex outcomes of those.
Methods based on the Density Functional Theory (DFT) provide an excellent compromise between accuracy and computational effort.
In this work, we have employed the static-exchange DFT method to evaluate the electronic structure of molecules (from CO or F$_2$ up to large amino acids) in order to predict and to understand recent experiments performed in two well-defined contexts: using third-generation light sources and attosecond pulses generated via HHG. 
The development of high-resolution third-generation synchrotron facilities has enabled the investigation of vibrationally-resolved inner-shell photoionization in small molecules made of first-row atoms of the periodic table.
In chapter \ref{chapter4} and in appendices \ref{appendix1}, \ref{appendix2}, \ref{appendix3}, \ref{appendix4} and \ref{appendix5} we have presented C 1s and B 1s vibrationally resolved photoionization cross sections of CO, CF$_4$ and BF$_3$, which have been evaluated using the static-exchange and the time-dependent DFT methods including the nuclear motion at the Born-Oppenheimer level.
By comparing our results with photoelectron spectra measured at SOLEIL and Spring-8 synchrotron light sources, we have found clear evidence of non-Franck-Condon effects that take place upon photoabsorption. 
When the de Broglie wave length of the photoelectron is comparable to the dimensions of the molecular target, i.e., for photon energies of the order of a few hundreds of eVs, the undulatory nature of the electron manifests as diffraction interferences by the surrounding atomic centers.
These interferences are imprinted in the ratios between vibrationally resolved cross sections ($\nu-$ratios) as clear oscillations as a function of the photoelectron momentum.
As a proof-of-principle, we have used the C 1s photoelectron spectra of CO to determine the internuclear distance of the neutral molecule and the bond contraction accompanying C 1s ionization.
This is a surplus of photoelectron spectroscopy with respect to more conventional spectroscopic techniques, which usually can only provide structural information of neutral molecular species.
Near the photoionization threshold, where the continuum waves are sensitive to the details of the molecular potentials, the cross sections show sharp features due to the presence of shape resonances.
From the theoretical calculations, and the angular momentum decompositions, we have also confirmed the existence of confinement effects as those previously found in more simple molecules.
In the future, molecules such as BF$_3$, CF$_4$ or CO could provide an interesting workbench to study the photoelectron scattering phenomenon in a time-resolved (pump-probe) scheme using novel ultrahigh intensity photon sources like seeded FELs, which now become available.\vspace{2 mm}

We have also investigated deviations from the Franck-Condon approximation in the photoionization of molecular fluorine, by considering electron ejection from different molecular orbitals using the static-exchange DFT method.
For the outermost shells, we have included the nuclear degrees of freedom, which has allowed us to describe dissociative and non-dissociative ionization channels.
The photoionization cross sections show an oscillatory behavior as a function of the photoelectron momentum, which is the result of the coherent emission from two equivalent centers.
The observed interference patterns are similar to those already explained by Cohen and Fano \cite{CohenPR1966} in the early sixties. 
These interferences can be described by using very simple expressions that account for both the ratio between total photoionization cross sections associated to different orbitals and to the branching ratios between vibrationally resolved cross sections associated to the same electronic ionic state.
Both effects were demonstrated in previous work \cite{CantonPNAS2011} in other diatomic molecules, so the present work reinforces the general validity of those findings.
The fact that there is a non-negligible probability of dissociation accompanying valence-ionization could be exploited experimentally to obtain molecular frame photoelectron angular distributions by using multi-coincidence techniques, since detection of the charged atomic fragments provides information about the orientation of the molecule at the instant of ionization.\vspace{2 mm}

Since the generation of the first isolated attosecond pulses at the very beginning of the present century, tracing ultrafast phenomena using attosecond pump-probe techniques became a reality.
The observation of charge migration in complex molecular structures and, in particular, in biologically relevant systems, is one of the main targets of attosecond science.
Although charge migration was predicted in the late nineties by L. S. Cederbaum and coworkers, experimental evidence or reliable simulations of the electronic wave packet that could be created by an attosecond pulse were inexistent.
In this work, we have investigated the electronic response of molecules of biological relevance, the amino acids glycine, phenylalanine and tryptophan, to attosecond ionization. 
We have found that attosecond pulses can induce ultrafast charge fluctuations over large regions of these complex molecules on a temporal scale that is much shorter than the vibrational response of the system.
The work presented here differs from most previous theoretical work on charge migration, where charge dynamics is initiated by removing an electron from a given molecular orbital. 
Here, the use of the static-exchange DFT method in the framework of time-dependent first-order perturbation theory has allowed us to evaluate the actual wave packet generated upon attosecond ionization.
Because of the broad energy spectrums of attosecond pulses, in a realistic experiment electrons are coherently emitted from various (many) molecular orbitals and thus the initial hole is highly delocalized along the molecular skeleton.
The application of isolated attosecond pulses to prompt ionization of the amino acid phenylalanine has allowed the subsequent detection of ultrafast dynamics on a sub-4.5 femtosecond temporal scale, which is shorter than the vibrational response of the molecule.
The good agreement with our numerical simulations of the temporal evolution of the electronic wave packet created by the attosecond pulse strongly supports the interpretation of the experimental data in terms of charge migration resulting from ultrafast electron dynamics preceding any nuclear rearrangement.

%% file: Chapters/Conclusiones.tex
\chapter*{Conclusiones}
\label{Conclusiones}
\addcontentsline{toc}{part}{Conclusiones}
\fancyhead[LE]{}
\fancyhead[RO]{\fontsize{11pt}{11pt}\selectfont Conclusiones}

La llegada de fuentes de luz ultra-brillantes basadas en radiaci\'on sincrotr\'on y de pulsos l\'aser ultra-cortos basados en la generaci\'on de altos arm\'onicos (HHG, del ingl\'es ``high harmonic generation'') han abierto la puerta a la visualizaci\'on de estructuras moleculares y al seguimiento e incluso el control de din\'amica electr\'onica y nuclear en sus escalas de tiempo intr\'insecas.
El modelado te\'orico y las simulaciones num\'ericas han demostrado jugar un papel fundamental en el dise\~no de nuevos experimentos as\'i como en la comprensi\'on de los complejos resultados procedentes de los mismos.
Los m\'etodos basados en la Teor\'ia del Funcional de la Densidad (DFT, del ingl\'es ``Density Functional Theory'') presentan un excelente compromiso entre precisi\'on y esfuerzo computacional.
En este trabajo hemos empleado el m\'etodo ``static-exchange DFT'' para evaluar la estructura electr\'onica de mol\'eculas (desde CO o F$_2$ hasta grandes amino \'acidos) con el objetivo de predecir y comprender experimentos recientes realizados bajo condiciones bien diferenciadas: utilizando fuentes de luz de tercera generaci\'on y pulsos ultra-cortos basados en HHG.
El desarrollo de instalaciones sincrotr\'on de \'ultima generaci\'on ha permitido la investigaci\'on de procesos de fotoionizaci\'on de capa interna en mol\'eculas peque\~nas, constituidas por \'atomos del primer periodo de la tabla peri\'odica, con resoluci\'on vibracional.
En el cap\'itulo \ref{chapter4} y en los ap\'endices \ref{appendix1}, \ref{appendix2}, \ref{appendix3}, \ref{appendix4} y \ref{appendix5} hemos presentado las secciones eficaces de fotoionizaci\'on de capa interna de las mol\'eculas CO, CF$_4$ y BF$_3$, calculadas mediante el m\'etodo ``static-exchange DFT'' y su versi\'on dependiente del tiempo, incluyendo el movimiento nuclear a nivel de Born-Oppenheimer.
La comparaci\'on de nuestros resultados te\'oricos con los espectros fotoelectr\'onicos medidos en los sincrotrones SOLEIL y Spring-8, nos ha permitido investigar desviaciones de la aproximaci\'on Franck-Condon en procesos de ionizaci\'on de capa interna.
Cuando la longitud de onda de de Broglie del fotoelectr\'on es comparable a las dimensiones del sistema, es decir, cuando su \mbox{energ\'ia} cin\'etica es del orden de unos cientos de eVs, la naturaleza ondulatoria del fotoelectr\'on produce interferencias de difracci\'on intramolecular. 
Estas interferencias quedan impresas en los ratios entre secciones eficaces resueltas vibracionalmente ($\nu-$ratios), produciendo claras oscilaciones en funci\'on del momento del fotoelectr\'on.
Como prueba de concepto, hemos utilizado espectros fotoelectr\'onicos de la mol\'ecula de CO para determinar la distancia internuclear de la especie neutra y la contracci\'on del enlace que tiene lugar tras la extracci\'on un electr\'on del orbital 1s del \'atomo de carbono.
Cerca del umbral de ionizaci\'on, donde las funciones del continuo son m\'as sensibles a los detalles de los potenciales moleculares, las secciones eficaces presentan estructuras agudas debido a la presencia de resonancias de forma.
La descomposici\'on en momento angular ha permitido confirmar la existencia de efectos de confinamiento electr\'onico similares a los que ya se hab\'ian observado previamente en mol\'eculas m\'as simples.
En el futuro, mol\'eculas como BF$_3$, CF$_4$ or CO podr\'ian constituir un punto de referencia para el estudio del fen\'omeno de difracci\'on fotoelectr\'onica mediante el uso de t\'ecnicas bombeo-sonda utilizando pulsos ultra-intensos de l\'aseres de electrones libres.\vspace{2 mm}

Tambi\'en hemos investigado el origen de las desviaciones de la aproximaci\'on de Franck-Condon que tienen lugar en la fotoionizaci\'on de la mol\'ecula de fl\'uor, considerando emisi\'on fotoelectr\'onica desde distintos orbitales.
La inclusi\'on del movimiento nuclear en el estudio de las capas m\'as externas nos ha permitido describir canales de fotoionizaci\'on disociativos y no disociativos.
Como consecuencia de un proceso de emisi\'on multic\'entrica, las secciones eficaces de fotoionizaci\'on presentan claras oscilaciones en funci\'on del momento del fotoelectr\'on.
Los patrones de interferencia observados son similares a aquellos para los que Cohen y Fano \cite{CohenPR1966} fueron capaces de dar una explicaci\'on te\'orica en los a\~nos 60.
Estas interferencias pueden ser descritas de forma cualitativa mediante el uso de expresiones sencillas que permiten calcular ratios entre secciones eficaces totales correspondientes a diferentes canales i\'onicos as\'i como entre secciones eficaces resueltas vibracionalmente asociadas al mismo canal.
Estos efectos fueron previamente observados \cite{CantonPNAS2011} en mol\'eculas diat\'omicas m\'as sencillas, as\'i que este trabajo refuerza la validez general de esos hallazgos.
El hecho de que exista una probabilidad no despreciable de que la mol\'ecula de fl\'uor disocie tras ser ionizada desde sus capas de valencia podr\'ia ser explotado para medir distribuciones angulares del fotoelectr\'on mediante el uso de t\'ecnicas de coincidencia m\'ultiple, ya que la detecci\'on de fragmentos at\'omicos cargados proporciona informaci\'on sobre la orientaci\'on de la mol\'ecula en el instante de la ionizaci\'on.\vspace{2 mm}

La generaci\'on de pulsos de attosegundos a comienzos de este siglo ha permitido la observaci\'on y el seguimiento de procesos ultra-r\'apidos mediante el uso de t\'ecnicas bombeo-sonda.
La observaci\'on de migraci\'on de carga en estructuras moleculares complejas y, en particular, en mol\'eculas de relevancia biol\'ogica, es uno de los objetivos fundamentales de la ciencia de attosegundos.
A pesar de que los procesos de migraci\'on de carga fueron predichos a finales de la d\'ecada de los noventa por Lorenz S. Cederbaum y sus colaboradores, no exist\'ian pruebas experimentales s\'olidas ni estudios te\'oricos del paquete de ondas electr\'onico generado por un pulso de attosegundos.
En este trabajo hemos investigado la respuesta electr\'onica de mol\'eculas de relevancia biol\'ogica, los amino \'acidos glicina, fenilalanina y tript\'ofano, a la ionizaci\'on ultra-r\'apida.
Hemos descubierto que los pulsos de attosegundos pueden inducir fluctuaciones de carga ultra-r\'apidas, en una escala de tiempo que precede a la respuesta vibracional del sistema.
Nuestro trabajo es distinto a otros estudios te\'oricos previos, donde la din\'amica se inicia arrancando un electr\'on de un determinado orbital molecular.
El uso del m\'etodo ``static-exchange DFT'' en el marco de teor\'ia de perturbaciones a primer orden ha permitido evaluar el paquete de ondas electr\'onico generado por un pulso de attosegundos.
Debido sus anchos espectros energ\'eticos, los pulsos de attosegundos son capaces de emitir fotoelectrones desde varios (muchos) orbitales moleculares de forma coherente, generando huecos que est\'an completamente deslocalizados sobre el esqueleto molecular.
La reciente aplicaci\'on de pulsos de attosegundos aislados a la ionizaci\'on del amino\'acido fenilalanina ha permitido la detecci\'on experimental de din\'amica puramente electr\'onica.
El buen acuerdo con nuestras simulaciones num\'ericas de la evoluci\'on temporal del paquete de ondas electr\'onico apoya firmemente la interpretaci\'on de los resultados experimentales en t\'erminos de migraci\'on de carga ultra-r\'apida que precede al movimiento nuclear.